\documentclass[a4paper,11pt]{article}
\usepackage{jheppub}
\usepackage[T1]{fontenc}
\usepackage{subfig}
\usepackage{multirow}

\title{\boldmath Non-Conformal Behavior of Holographic Entanglement Measures}
\author[a]{M. Ali-Akbari,}
\author[b]{M. Asadi}
\author[a]{and B. Amrahi}
\affiliation[a]{Department of Physics, Shahid Beheshti University, Tehran, Iran}
\affiliation[b]{IPM, School of Particles and Accelerators, P.O. Box 19395-5531, Tehran, Iran}
\emailAdd{m\_aliakbari@sbu.ac.ir}
\emailAdd{m\_asadi@ipm.ir}
\emailAdd{b\_amrahi@sbu.ac.ir}

\abstract{We evaluate the holographic entanglement entropy, HEE, holographic mutual information, HMI, and holographic entanglement of purification, EoP, in a non-conformal model at zero and finite temperature. In order to find the analytical results we consider some specific regimes of the  parameter space. We find that the non-conformal effects decrease the redefined HEE and increase the redefined HMI and EoP in the all studied regimes. On the other side, the  temperature effects increase (decrease) the redefined HEE (HMI) in the all studied regimes while it has no definite effect on the redefined EoP. 
Finally, from the information point of view, we find that in the vicinity of the phase transition the zero temperature state is more favorable than the finite temperature one.}

\begin{document} 
\maketitle
\flushbottom

\section{Introduction}
The gauge/gravity duality is a conjectured relation between  a classical gravity theory  and a strongly coupled gauge theory \cite{Maldacena:1997re,Witten:1998qj}. This duality has been utilized to study various areas of physics such as condensed matter, quantum information theory and the quark-gluon plasma \cite{Natsuume:2014sfa,CasalderreySolana:2011us,Hartnoll:2009sz,Camilo:2016kxq}. A significant example of the gauge/gravity duality is the AdS/CFT correspondence which states that type IIB string theory on ($d+1$)-dimensional AdS spacetimes is dual to $d$-dimensional superconformal field theories.  The gauge/gravity duality provides a powerful tool to investigate various physical quantities in the quantum information theory utilizing their corresponding gravity duals. Some of these quantities are entanglement entropy, mutual information and entanglement of purification. Although it may be difficult to calculate these quantities in the field theory, they are relatively simple to calculate on the gravity side. Surprisingly, the gauge/gravity duality is extended to the more general cases such that the field theories which are not conformal. There are many different families of non-conformal theories which one can study the effect of the non-conformality on their physical observables \cite{Attems:2016ugt,Pang:2015lka,Rahimi:2016bbv}.

Entanglement entropy (EE) is a significant non-local quantity which measures the quantum entanglement between subsystem $A$ and its complement $\bar{A}$. It has been shown that the EE in the QFTs contains short-distance divergence which satisfies an area law and  thus the EE is a scheme-dependent quantity \cite{Bombelli:1986rw,Srednicki:1993im}. Although the EE has simple definition in the QFT but it is very difficult to compute this quantity using QFT techniques \cite{Holzhey:1994we,Calabrese:2004eu,Casini:2009sr,Hertzberg:2012mn,Rosenhaus:2014ula,Rosenhaus:2014zza,Iso:2021vrk,Iso:2021dlj,Sorkin:2012sn,Chen:2020ild}. Fortunately, the gauge/gravity duality has made this problem easier and the EE has a simple holographic dual. For a spatial region $A$ in the boundary field theory, the EE of $A$ corresponds to the area of minimal surface, called Ryu-Takayanagi surface extended in the bulk whose boundary coincides with the boundary of $A$, known as RT prescription \cite{Ryu:2006bv,Ryu:2006ef}. Using this prescription, many studies have been done in the literature to better understand the EE, for example, see \cite{Casini:2011kv,Myers:2012ed,Lokhande:2017jik,Rahimi:2018ica,Fischler:2012ca,Ben-Ami:2014gsa,Pang:2014tpa,Kundu:2016dyk,Ebrahim:2020qif,Buniy:2005au}.

 When the total system is described by mixed state $\rho_{AB}$, the EE is not a suitable quantity to measure the full correlation between two subsystems $A$ and $B$. In this case, there are known quantities in quantum information theory which measure total, classical and quantum, correlation between subsystems $A$ and $B$. One of the most famous and important of these quantities is the mutual information (MI) which measures the total correlation between subsystems A and B which is defined as $I(A,B)=S_A+S_B+S_{A\cup B}$, where $S_X$ is the EE of the region $X$ \cite{Casini:2004bw,Wolf:2007tdq,Fischler:2012uv,Allais:2011ys,Hayden:2011ag,MohammadiMozaffar:2015wnx,Asadi:2018ijf,Ali-Akbari:2019zkf}. The MI is a finite quantity since the divergent pieces in the EE cancel out and the subadditivity property of the EE also guarantees that MI is always non-negative. When the subsystem $B$ is the complement of $A$, $B=\bar{A}$, the MI is proportional to the EE. Although the EE is dominated by the thermal entropy in the high temperature limit, it has been shown that the MI obeys an area law in the same limit \cite{Wolf:2007tdq,Fischler:2012uv}.

The entanglement of purification (EoP) is another quantity which measures the total correlation between two disjoint subsystems $A$ and $B$ for mixed state $\rho_{AB}$ \cite{arXiv:quant-ph/0202044v3,arXiv:quant-ph/1502.01272}. In general, by enlarging the Hilbert space we can purify a mixed state. In fact, purification refers to the fact that every mixed state acting on the Hilbert spaces can be viewed as the reduced state of some pure states. The EoP is defined by the minimum of the EE between subsystems $AA'$ and $BB'$ against all possible purifications where $A'$ and $B'$ are arbitrary. Obviously, when two subsystems $A$ and $B$ are complementary to each other, the EoP between two subsystems reduces to the EE of subsystem $A$. It is very difficult to compute the EoP in QFT, however it has been computed by numerical lattice calculation in QFT \cite{Bhattacharyya:2018sbw,Caputa:2018xuf,Camargo:2020yfv}. It has been conjectured that the EoP is holographically dual to the minimal cross-section of entanglement wedge $E_w$ of $\rho_{AB}$, which satisfies the basic properties of the EoP \cite{Takayanagi:2017knl,Nguyen:2017yqw}. $E_w$ of $\rho_{AB}$ is reduced to the holographic entanglement entropy (HEE), when the total system $A\cup B$ is pure. In the literature, other correlation measures such as reflected entropy, odd entropy and logarithmic negativity are discussed and all of them are connected with the entanglement wedge cross-section \cite{Dutta:2019gen,Tamaoka:2018ned,Kudler-Flam:2018qjo}. Various aspects of the EoP are discussed in the literature, for example, see \cite{Yang:2018gfq,Espindola:2018ozt,Bao:2017nhh,Umemoto:2018jpc,Bao:2018gck,Bao:2018fso,Bhattacharyya:2019tsi,Liu:2019qje,Jokela:2019ebz,BabaeiVelni:2019pkw,BabaeiVelni:2020wfl,Amrahi:2020jqg,Amrahi:2021lgh, Sahraei:2021wqn,Guo:2019azy,Ghodrati:2019hnn,Du:2019emy,Guo:2019pfl,Bao:2019wcf,Harper:2019lff,Camargo:2021aiq,DiNunno:2021eyf,Saha:2021kwq,Lin:2020yzf,Liu:2020blk,Fu:2020oep, Huang:2019zph,Jain:2020rbb,Ghodrati:2021ozc}

In this paper we consider a four dimensional QCD-like model, which is non-conformal, at zero and finite temperature. The zero temperature model is dual to modified AdS$_5$ (MAdS) background and the finite temperature one is dual to a modified AdS$_5$ black hole (MBH) \cite{Andreev:2006ct,Andreev:2006eh}. There is a thermal phase transition, in the MBH background,  which occurs at the point that the position of the horizon approaches to the lower bound of the minimum value of the radial coordinate. 

The rest of the paper is organized as follows. Section \ref{section2} includes a brief review of the MAdS and MBH backgrounds. In section \ref{section3}, after a short review on HEE, we compute and study this quantity in our model for a strip entangling subsystem at both zero and finite temperature. In order to obtain the analytical results we use the systematic expansion up to some specific order of the expansion parameters and consider some specific limits of the QFTs which we call high energy limit.
Since we can not obtain the analytical results in the low energy limit, we do not study this limit. Section \ref{section4} will be devoted to study the MI at zero and finite temperature. Using the analytical results obtained for the HEE, we reach analytical expressions for the MI corresponding to the two strip entangling subsystems in the some specific limits. Section \ref{section5} is reserved for study the EoP at both zero and finite temperature. At zero temperature, we consider two favorable limits called high and intermediate energy and obtain analytical results in these limits. At finite temperature, we can get analytical expressions for the EoP in the four regimes which are low and intermediate temperature limits at high  and at intermediate energy. We will conclude in section \ref{section6} with the discussion of our results. We present the full details of our calculations for the HEE in appendix \ref{Appendix1}. Appendices \ref{Appendix2} and \ref{Appendix3} include the numerical constants and some of the calculations for the MI and EoP, respectively.
\section{Backgrounds}\label{section2}
We are interested in studying the EE, MI and EoP in the (3+1)-dimensional non-conformal field theories using the framework of holography. Therefore, we start with a holographic background in five dimensions, which is called MAdS and its black hole version MBH. These backgrounds are dual to QCD-like theories at zero and finite temperature, respectively. MAdS background described by the following metric \cite{Andreev:2006ct}
\begin{align}\label{MAdS}
ds^2=\frac{r^2}{R^2}g(r)\left(-dt^2+d\vec{x}^2+\frac{R^4}{r^4}dr^2\right) , \ \ \ \ \ g(r)=e^{\frac{r_c^2}{r^2}},\ \ \ \ \ \ r_c=R^2\sqrt{\frac{c}{2}},
\end{align}
and the MBH background is described by \cite{Andreev:2006eh}
\begin{align}\label{MBH}
ds^2=\frac{r^2}{R^2}g(r)\left(-f(r)dt^2+d\vec{x}^2+\frac{R^4}{r^4f(r)}dr^2\right) , \ \ \ \ \ f(r)=1-\frac{r_H^4}{r^4},
\end{align}
where $\vec{x}\equiv(x,y,z)$. $r$ is the radial coordinate, $R$ is the asymptotic AdS$_5$ radius and $r_H$ is the location of the horizon. The QCD-like model lives on the boundary at $r\rightarrow \infty$. $r_c$ is an lower bound on the minimum value of the radial coordinate which is related to the deformation parameter $c$ as $r_c=R^2\sqrt{\frac{c}{2}}$. From the trajectory of $\rho$ meson the value of $c$ is fixed and almost equals to $0.9$ GeV$^2$ \cite{Andreev:2006eh}. We assign an energy scale $\Lambda_c$ to $r_c$ which is defined $\Lambda_c\equiv\sqrt{c}$. Since $c$ has (energy)$^2$ dimension, $\Lambda_c$ has (energy) dimension. In the limit of $c\rightarrow 0$, the background \eqref{MAdS} becomes AdS$_5$ and $r$ is not bounded \cite{Andreev:2006ct}. Also in this limit, the background \eqref{MBH} becomes AdS$_5$ black hole. In other words, in the $c\rightarrow 0\ (\Lambda_c \rightarrow 0)$ limit we have a conformal field theory. The thermodynamics properties of these backgrounds are studied in \cite{Lezgi:2020bkc} and some thermodynamics parameters such as entropy, density and pressure are expressed in terms of non-conformal parameters.
\section{Entanglement entropy}\label{section3}
The entanglement entropy is one of the most important quantities which measure the quantum entanglement among different degrees of freedom of a quantum mechanical systems \cite{Casini:2009sr,Horodecki:2009zz}. In fact, EE has emerged as a valuable tool to probe the physical information in quantum systems. To define the EE, we consider a quantum system which is described by the pure state $\vert \psi \rangle$ and density matrix $\rho=\vert \psi \rangle\langle \psi \vert$. By dividing the total system into two subsystems $A$ and its complement $\bar{A}$, the total Hilbert space becomes factorized $\mathcal{H}_{tot}=\mathcal{H}_A\otimes \mathcal{H}_{\bar{A}}$. The reduced density matrix for the subsystem $A$, $\rho_A$, is obtained by tracing out the degrees of freedom in the complementary subsystem, $\rho_A=Tr_{\bar{A}}(\rho)$. The entanglement between subsystems $A$ and $\bar{A}$ is measured by the entanglement entropy which is a non-local quantity and defined as the Von-Neumann entropy of the reduced density matrix $\rho_A$
\begin{align}
S_A=-Tr(\rho_A \log \rho_A).
\end{align}
In the framework of the gauge/gravity duality, the RT-proposal is a remarkably simple prescription to compute entanglement entropy in terms of a geometrical quantity in the bulk \cite{Ryu:2006bv,Ryu:2006ef}. According to the RT-prescription, the HEE is given by
\begin{align}\label{HEE}
S_A=\frac{\rm{Area}(\Gamma_A)}{4G_N^{(d+2)}},
\end{align}
where $G_N^{(d+2)}$ is the $(d+2)$-dimensional Newton constant. $\Gamma_A$ is a codimension-2 minimal hypersurface in the bulk (RT-surface) whose boundary $\partial \Gamma_A$ coincides with the boundary of region $A$, $\partial \Gamma_A=\partial A$. We also require $\Gamma_A$ to be homologous to region $A$. It can be shown that the EE for two disjoint subsystems $A$ and $B$ satisfies the so-called strong subadditivity condition $S_A+S_B\geqslant S_{A\cup B}+S_{A\cap B}$ \cite{Headrick:2007km}. Using the RT-prescription the HEE is calculated in the AdS soliton background and it is shown that the HEE decreases under the tachyon condensations \cite{Nishioka:2006gr}. In \cite{Klebanov:2007ws,Pakman:2008ui} the HEE is used to determine the phase structure of the confining field theories. Using QFT techniques the dependence of the EE on the renormalized mass is studied in the scalar field theory and it is shown that the area law that exists in the free field theory persists in the interacting theory through  the mass renormalization \cite{Hertzberg:2012mn}. In \cite{Rosenhaus:2014ula} it is shown that the EE for a general theory can be expressed in terms of a spectral function of this theory. Also the EE has been studied perturbatively for a CFT perturbed by a relevant operator and a deformed planar entangling surface \cite{Rosenhaus:2014zza}. The renormalization of the UV divergences and the non-Gaussianity of the vacuum which are two important issues of the EE in the intacting field theory are studied in \cite{Iso:2021vrk,Iso:2021dlj}. In \cite{Sorkin:2012sn}, a covariant definition of the EE is proposed in terms of the spacetime two-point correlation function for Gaussian theories. This formulation developed for a non-Gaussian scalar field theory in \cite{Chen:2020ild}.
 In this section we would like to study the HEE in a non-conformal field theory at zero, low and high temperature.
\begin{figure}
\centering
\includegraphics[scale=0.4]{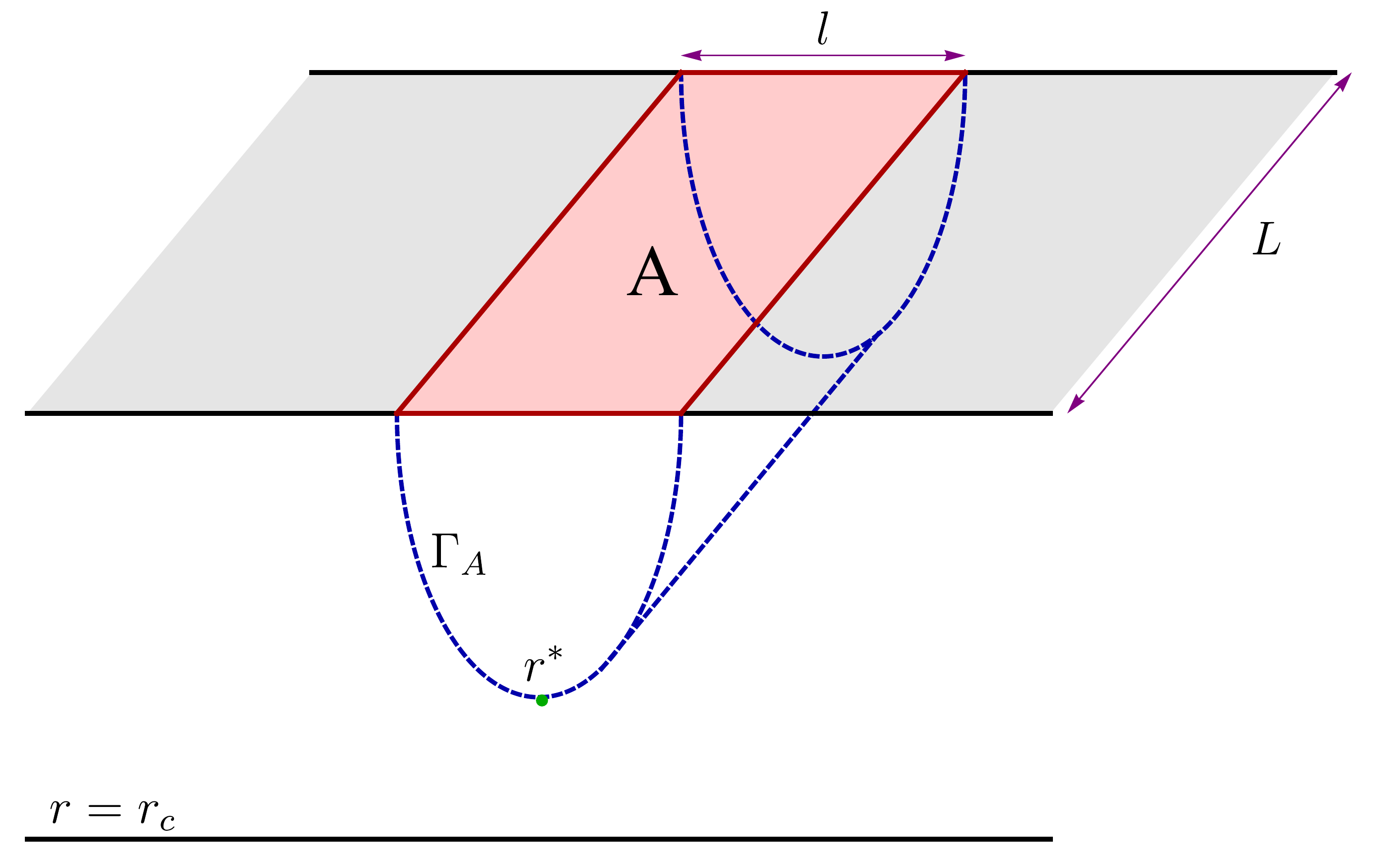} 
\caption{A simplified sketch of a strip region $A$ with width $l$ and length $L$. $\Gamma_A$ is the RT-surface of the region $A$, $r^*$ is the turning point of this surface and $r_c$ is the lower bound on the radial coordinate which is related to the deformation parameter $c$.}
\label{EE}
\end{figure}

\begin{figure}
\centering
\subfloat[]{\includegraphics[scale=0.2]{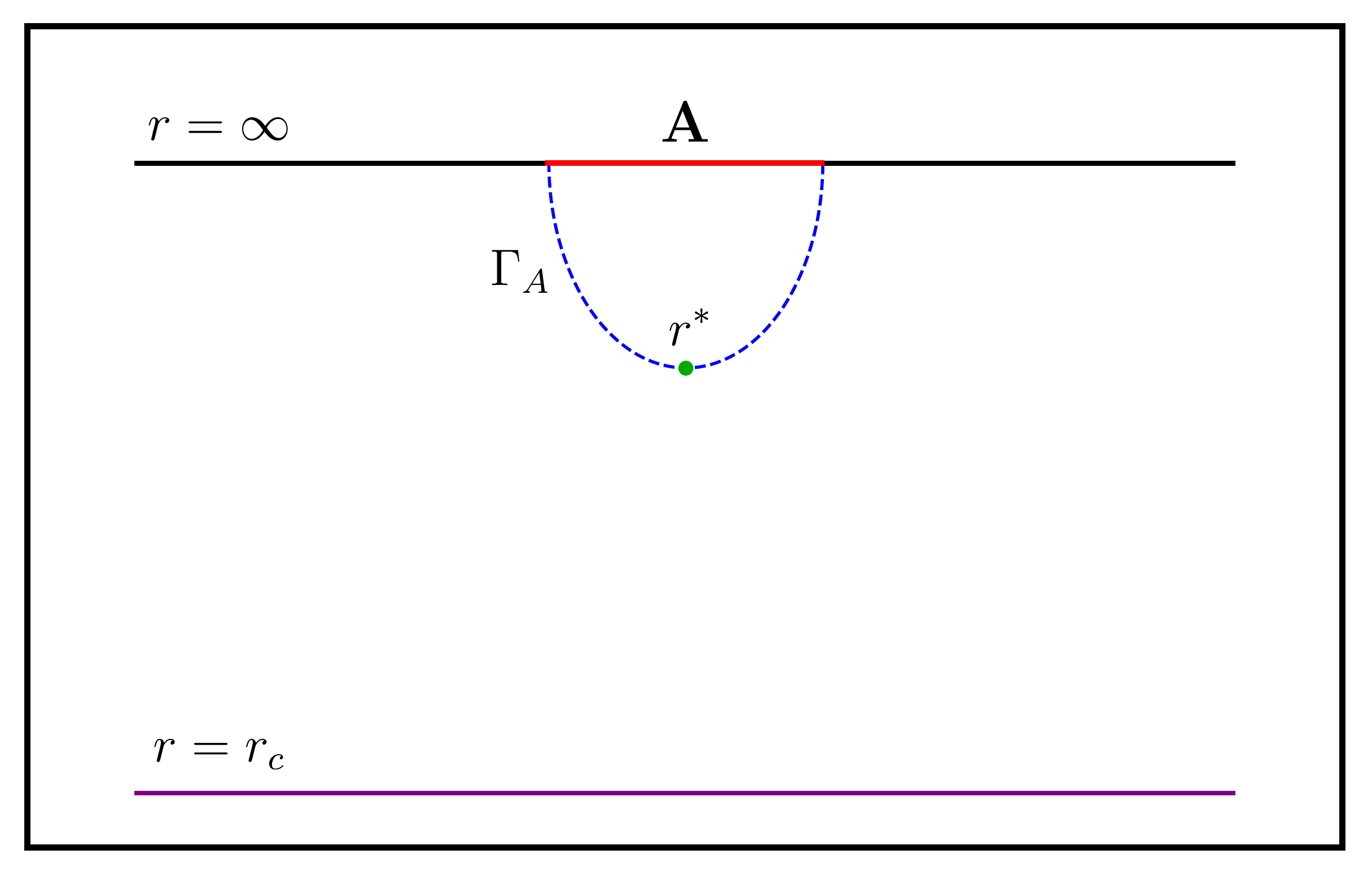}\label{aa}}
\subfloat[]{\includegraphics[scale=0.2]{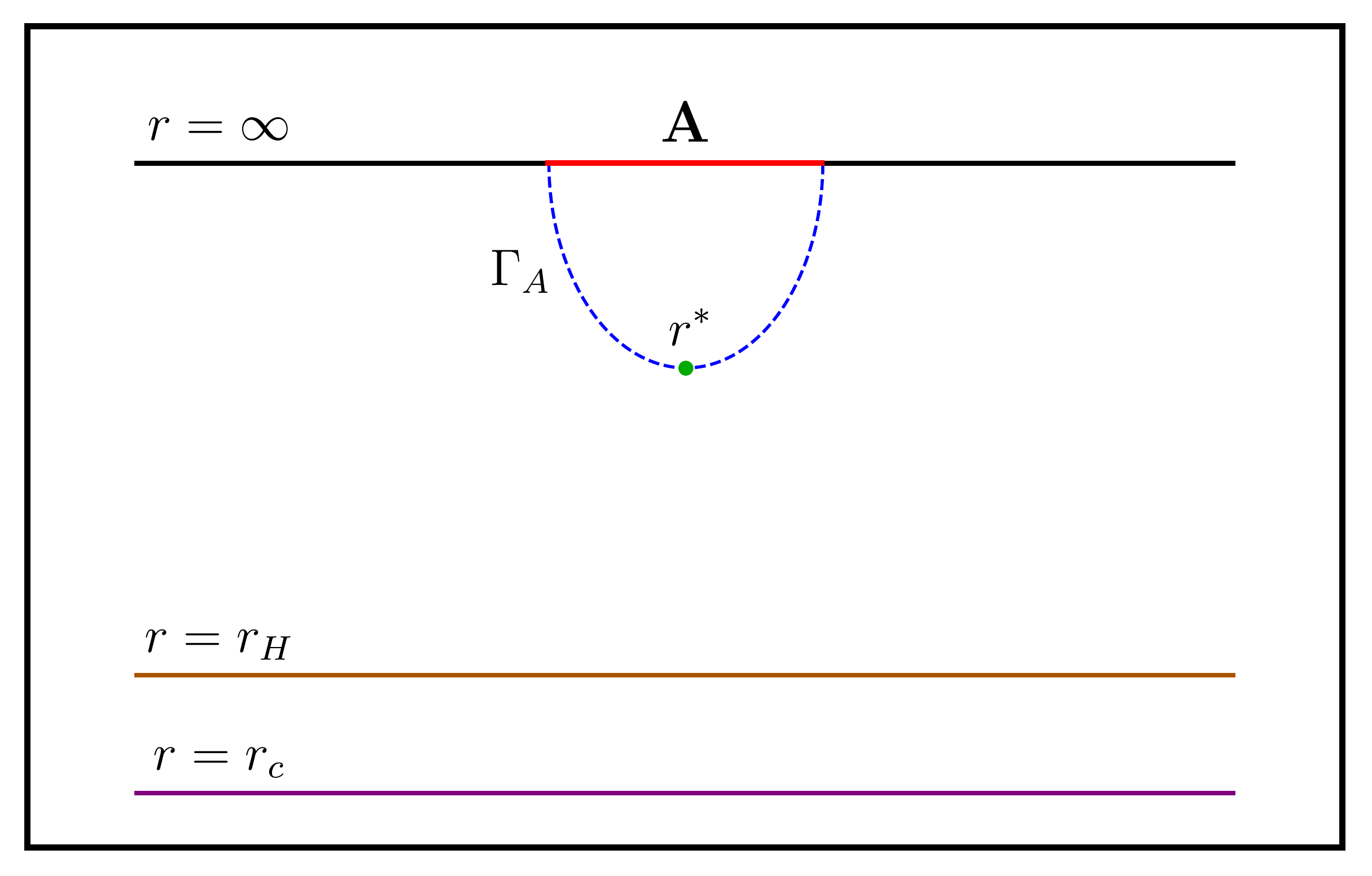}\label{bb}}
\subfloat[]{\includegraphics[scale=0.20]{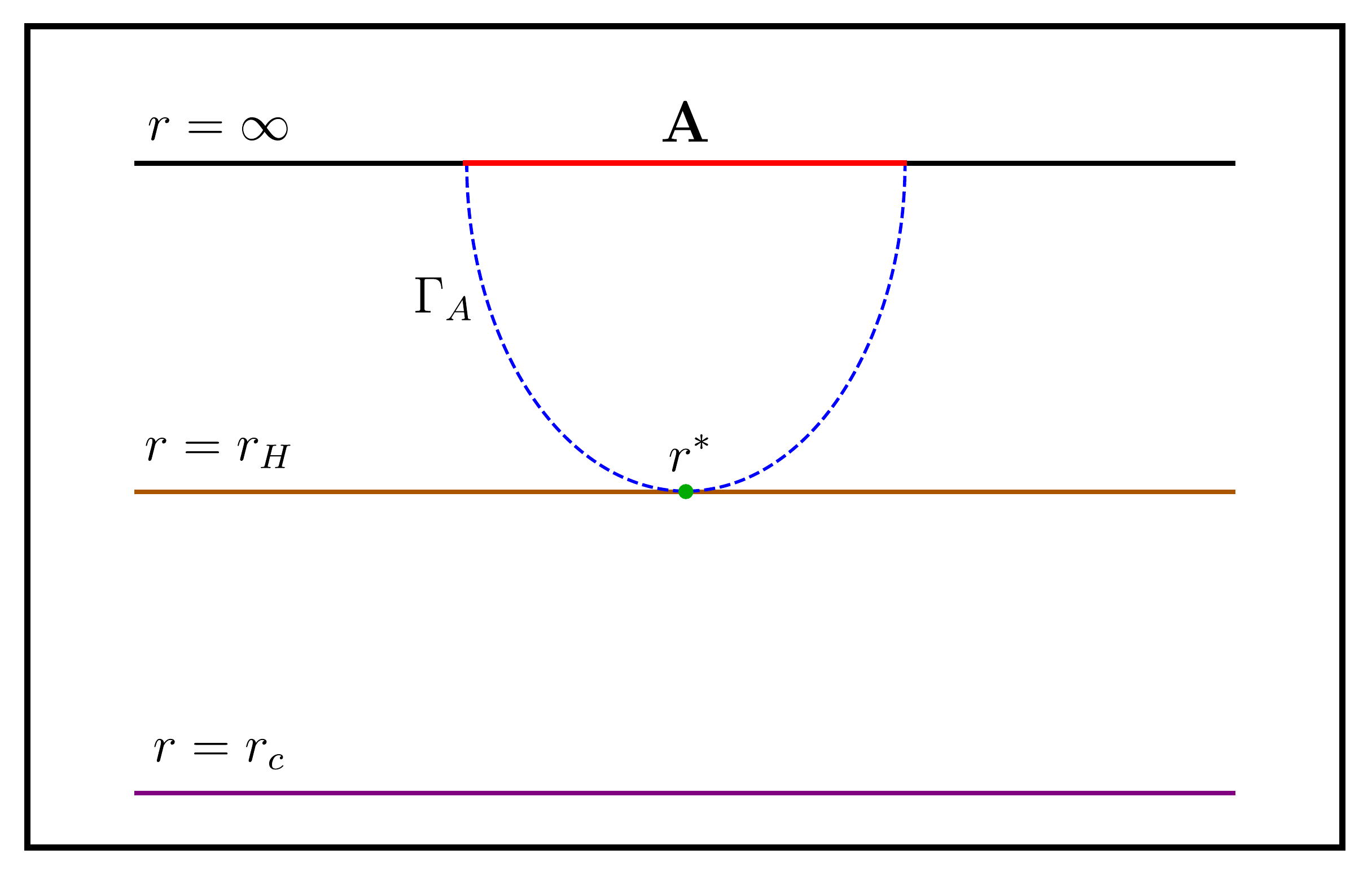}\label{cc}}
\caption{(a): The high energy limit i.e. $l\Lambda_c\ll 1$ or equivalently $r_c\ll r^*$ at the zero temperature. (b): The high energy limit at the low temperature i.e. $lT\ll 1$ or equivalently $r_H\ll r^*$. (c): The high energy limit at the high temperature i.e. $lT\gg 1$ or equivalently $r^*\rightarrow r_H$.}
\label{EEhighE}
\end{figure}

We want to compute the HEE for a rectangular strip with width $l$ and length $L(\rightarrow \infty)$, depicted in figure \ref{EE}, specified by
\begin{align}\label{config}
-\frac{l}{2}&\leq x(r) \leq \frac{l}{2},\ \ \ \ \ \ \ \ \ -\frac{L}{2}\leq y\ \&\ z \leq \frac{L}{2}.
\end{align}
In order to perform the computations analytically we need to focus on the some specific limits of the non-conformal model at zero and finite temperature which are dual to the backgrounds \eqref{MAdS} and \eqref{MBH}, respectively. We consider the high energy limit ($l\Lambda_c\ll 1$) at zero ($T=0$), low ($lT\ll 1$) and high ($lT\gg 1$) temperature where the temperature is given by $T=\frac{r_H}{R^2\pi}$ \cite{Andreev:2006eh}. In the high energy limit, the energy scale corresponding to the subsystem $A$ should be very larger than the energy scale $\Lambda_c$  i.e. $\Lambda _c\ll \frac{1}{l}$. In the MAdS background \eqref{MAdS} the high energy limit equals to $r_c\ll r^*$  and in the MBH background \eqref{MBH} the low and high temperature limits at high energy are equivalent to $r_H \ \& \ r_c\ll r^*$ and $r_c\ll r^* \rightarrow r_H$, respectively, where $r^*$ is the turning point of the RT-surface $\Gamma_A$. In figure \ref{EEhighE} we show these regimes schematically. Hereafter, we assume the AdS radius to be one.
\subsection{HEE at zero temperature}\label{sectionzeroEE}
Using the RT-prescription eq.\eqref{HEE} and background \eqref{MAdS}, the corresponding area surface for the configuration in eq.\eqref{config}, $\mathcal{A}$, is given by
\begin{align}\label{areaa}
\mathcal{A}=2L^2\int_{r^*}^\infty  r e^{\frac{3}{2}\left(\frac{r_c}{r}\right)^2}\sqrt{1+r^4x'(r)^2} dr.
\end{align}
Since there is no explicit $x(r)$ dependence in eq.\eqref{areaa}, the corresponding Hamiltonian is constant and one can easily obtain
\begin{align}\label{xdif}
x'(r)=\pm \frac{{r^*}^3}{r^5}e^{\frac{3}{2}\left(\left(\frac{r_c}{r^*}\right)^2-\left(\frac{r_c}{r}\right)^2\right)}\left[1-\left(\frac{r^*}{r}\right)^6e^{3\left(\left(\frac{r_c}{r^*}\right)^2-\left(\frac{r_c}{r}\right)^2\right)}\right]^{-\frac{1}{2}},
\end{align}
where the constant is chosen to be $\frac{\sqrt{r^6 e^{3\left(\frac{r_c}{r}\right)^2}-{r^*}^6 e^{3(\frac{r_c}{r^*})^2}}}{r^2}$. Plugging eq.\eqref{xdif} back into eq.\eqref{areaa}, we find
\begin{align}\label{HEE1}
\mathcal{A}=2{r^*}^2L^2\int _{r^*\epsilon}^1 u^{-3} e^{\frac{3}{2}(\frac{r_c}{r^*})^2u^2}\left[1-u^6e^{3(\frac{r_c}{r^*})^2(1-u^2)}\right]^{-\frac{1}{2}}du,
\end{align}
where $u=\frac{r^*}{r}$ and $\epsilon$ is an ultraviolet cut off. By integrating the differential equation \eqref{xdif}, the relation between $l$ and $r^*$ is given by
\begin{align}\label{lengthh}
l=\frac{2}{r^*}\int_0^1 u^3 e^{\frac{3}{2}\left(\frac{r_c}{r^*}\right)^2(1-u^2)}\left[1-e^{3\left(\frac{r_c}{r^*}\right)^2(1-u^2)}u^6\right]^{-\frac{1}{2}} du.
\end{align}
Unfortunately eqs. \eqref{HEE1} and \eqref{lengthh} can not be analytically solved. Therefore, we need to use the following binomial expansion \cite{Fischler:2012ca,Fischler:2012uv}
\begin{align}\label{expansionn}
(1+x)^{-r}&=\sum\limits_{n=0}^{\infty}(-1)^n\binom{r+n-1}{n}x^n , \ \ \ \ \ \ \ |x|<1,
\end{align}
where $x$ and $r$ are real numbers and $r>0$. By identifying $x=-e^{3\left(\frac{r_c}{r^*}\right)^2(1-u^2)}u^6$ and using the facts that $0<u<1$ and $r_c<r^*$, one can easily see that $\vert x\vert<1$ and hence the sum is well-defined. Using eq.\eqref{expansionn}, we can write eqs. \eqref{lengthh} and \eqref{HEE1} as follows
\begin{align}\label{Lengthh1}
l=\frac{2}{r^*}\sum\limits_{n=0}^{\infty}\frac{\Gamma(n+\frac{1}{2})}{\sqrt{\pi}\Gamma(n+1)}\int_0^1 u^{6n+3} e^{3(n+\frac{1}{2})\left(\frac{r_c}{r^*}\right)^2(1-u^2)} du,
\end{align}
\begin{align}\label{HEE2}
\mathcal{A}=2{r^*}^2L^2\sum\limits_{n=0}^{\infty}\frac{\Gamma(n+\frac{1}{2})}{\sqrt{\pi}\Gamma(n+1)}\int_{r^*\epsilon}^1 u^{6n-3}e^{(3n+(\frac{3}{2}-3n)u^2)(\frac{r_c}{r^*})^2} du.
\end{align}
In order to find the $\mathcal{A}$ as a function of $l$, we should solve eq.\eqref{Lengthh1} for $r^*$ and then substitute back this in eq.\eqref{HEE2}. Since we can not analytically solve the eq.\eqref{Lengthh1} to find $r^*$ as a function of  $l$, we need to focus on the high energy limit ($l\Lambda _c\ll 1$). Note that, in the low energy limit, i.e. $l\Lambda _c\gg 1$, or equivalently $r^*\rightarrow r_c$ we do not reach the analytical results and then we have to neglect studying this limit.
\subsubsection{High Energy limit}
In the section \ref{sectionzeroEE}, we saw that in order to find $\mathcal{A}$ in terms of $l$, we must consider the high energy limit. This limit can be interpreted in terms of the bulk data as $r_c \ll r^*$, see figure \ref{aa} . Hence the RT-surface $\Gamma$ is restricted to be near the boundary. The boundary is AdS and we should get the conformal entanglement entropy as the leading term. The non-conformal effects appear as sub-leading terms corresponding to the deviations from the boundary geometry. These effects are small and hence can be computed perturbatively. By expanding eq.\eqref{Lengthh1} up to the 4th order in $\frac{r_c}{r^*}$ we obtain
\begin{align}\label{lengthhh2}
l=\frac{2}{r^*}\left[ a_1+a_2 \left(\frac{r_c}{r^*}\right)^2+a_3 \left(\frac{r_c}{r^*}\right)^4\right] , \ \ \ \ \ \ \ \ a_1, a_2, a_3>0,
\end{align}
where numerical coefficients $a_1$, $a_2$ and $a_3$ are given by eq.\eqref{qeoff1}  in appendix \ref{Appendix1}. Solving eq.\eqref{lengthhh2} perturbatively for $r^*$ and doing some calculations, we finally reach the following expression for the HEE (for more details of the calculations, see appendix \ref{Appendix1})
\begin{align}\label{HEEz}
S=\frac{1}{4G_N^{(5)}}\left(\frac{L}{\epsilon }\right)^2-\frac{3}{8G_N^{(5)}}\Lambda_c^2 L^2\log (\Lambda_c \epsilon)+S_{finite}(l,l\Lambda_c) ,
\end{align}
where $S_{finite}(l,l\Lambda_c)$ is the finite part of the HEE which is given by
\begin{align}
S_{finite}(l,l\Lambda_c)=\frac{1}{4G_N^{(5)}}\frac{L^2}{l^2}\bigg[ \kappa_1 &+\left(\kappa_2+\frac{3}{2}\log(l\Lambda_c) \right) (l \Lambda_c)^2\cr
&+\kappa_3(l\Lambda_c)^4\bigg], \ \ \ \ \ \ \ \ \ \kappa _1<0, \ \kappa_2 , \kappa_3>0,
\end{align}
where $\kappa_1$, $\kappa_2$ and $\kappa_3$ are numerical coefficients which are given by eq.\eqref{Kappa} in Appendix \ref{Appendix1}. The first term in eq.\eqref{HEEz} is a divergent term in the limit of $\epsilon\rightarrow 0$ which appears in the AdS background.  From the second term in eq.\eqref{HEEz} we observe that there is a logarithmic divegence because of non-conformality.  Since the underlying field theory is conformal, we expect that the dimensionless parameter $l\Lambda_c$ appears. Therefore, we redefined $S_{finite}$ as follows 
\begin{align}\label{HEEhighh1}
\hat{\mathcal{S}}_{finite}(l\Lambda_c)\equiv\frac{4G_N^{(5)}S_{finite}(l,l\Lambda_c)}{L^2\Lambda_c^2}=\frac{1}{(l\Lambda_c)^2}\bigg[& \kappa_1 +\left(\kappa_2+\frac{3}{2}\log(l\Lambda_c) \right) (l \Lambda_c)^2\cr
&+\kappa_3(l\Lambda_c)^4\bigg], \ \ \ \ \kappa _1<0, \kappa_2 , \kappa_3>0,
\end{align}
where $\hat{\mathcal{S}}_{finite}(l\Lambda_c)$ is the redefined $S_{finite}(l,l\Lambda_c)$. By this redefinition, the limit of $l\Lambda_c \rightarrow 0$ become meaningful and intuitive. The first term in eq.\eqref{HEEhighh1} is the contribution of the AdS boundary corresponds to the entanglement entropy in conformal field theory. Obviously, this term is negative. The other two terms in eq.\eqref{HEEhighh1} are the non-conformal effects. In the second term the logarithmic term is the dominant term which is always negative in the high energy limit. This shows that the non-conformal effects decrease $\hat{\mathcal{S}}_{finite}(l\Lambda_c)$ which  is in complete agreement with \cite{Rahimi:2016bbv}.  We are not worried about the logarithmic term. Although, this term is large for small $l\Lambda_c$, it can still be considered as a perturbation compared to the first term.
\subsection{HEE at finite temperature}
In this section we want to study the thermal behavior of the entanglement entropy in the non-conformal field theory. Therefore we calculate the HEE in the MBH background \eqref{MBH}.
Using eqs. \eqref{HEE} and \eqref{MBH} and doing some calculation similar to the previous section, we reach the following expressions for $l$ and $\mathcal{A}$
\begin{align}\label{LengthhT}
l=\frac{2}{r^*}\int_0^1 u^3 e^{\frac{3}{2}\left(\frac{r_c}{r^*}\right)^2(1-u^2)}\left[1-e^{3\left(\frac{r_c}{r^*}\right)^2(1-u^2)}u^6\right]^{-\frac{1}{2}}\left[1-(\frac{r_H}{r^*})^4u^4\right]^{-\frac{1}{2}} du,
\end{align}
\begin{align}\label{HEET}
\mathcal{A}=2{r^*}^2L^2\int _{r^*\epsilon}^1 u^{-3} e^{\frac{3}{2}(\frac{r_c}{r^*})^2u^2}\left[1-u^6e^{3(\frac{r_c}{r^*})^2(1-u^2)}\right]^{-\frac{1}{2}}\left[1-(\frac{r_H}{r^*})^4u^4\right]^{-\frac{1}{2}} du,
\end{align}
where again $u=\frac{r^*}{r}$ and $\epsilon$ is an ultraviolet cut off. Unfortunately, the above integrals can not be solved analytically. Therefore, in order to calculate eqs. \eqref{LengthhT} and \eqref{HEET}, we use eq.\eqref{expansionn} by identifying $x=-\frac{r_H^4}{r^4}u^4$ and $x=e^{3\left(\frac{r_c}{r^*}\right)^2(1-u^2)}u^6$ and we obtain
\begin{align}\label{lengthhT1}
l=\frac{2}{r^*}\sum\limits_{m=0}^{\infty}\sum\limits_{n=0}^{\infty}\frac{\Gamma(n+\frac{1}{2})\Gamma(m+\frac{1}{2})}{\pi \Gamma(n+1)\Gamma(m+1)}\left(\frac{r_H}{r^*}\right)^{4m}\int_0^1 u^{6n+4m+3} e^{3(n+\frac{1}{2})\left(\frac{r_c}{r^*}\right)^2(1-u^2)} du,
\end{align}
\begin{align}\label{HEET1}
\mathcal{A}=2{r^*}^2L^2\sum\limits_{n=0}^{\infty}\sum\limits_{m=0}^{\infty}\frac{\Gamma(n+\frac{1}{2})\Gamma(m+\frac{1}{2})}{\pi\Gamma(n+1)\Gamma(m+1)}(\frac{r_H}{r^*})^{4m}\int _{r^*\epsilon}^1u^{6n+4m-3}e^{(3n+(\frac{3}{2}-3n)u^2)(\frac{r_c}{r^*})^2} du.
\end{align}
One can check that $\vert x\vert < 1$ for any allowable values of background parameters, that is $0<u<1$, $r_H< r$ and $r_c < r^*$ and hence the sums are convergent.
Now, we should solve eq.\eqref{lengthhT1} for $r^*$ and use this in eq.\eqref{HEET1} to get $\mathcal{A}$ in terms of $l$. It is impossible to solve eq.\eqref{lengthhT1} analytically and we only consider some orders of expansion in different regimes such as low temperature ($lT  \ll 1$) and high temperature ($l T  \gg 1$), in the high energy limit ($l\Lambda _c\ll 1$).
\subsubsection{High energy and low temperature}\label{EE-highT}
Here, we focus on the limit of low temperature i.e $lT \ll 1$ at the high energy $l\Lambda _c\ll 1$. This regime can be interpreted in terms of bulk parameters as $r_H\ll r^*$ and $r_c\ll r^*$, see figure \ref{bb}. In this regime, the leading contribution comes from the AdS boundary and finite temperature and non-conformal corrections are sub-leading terms which are small and hence we can perturbatively do the calculations. We expand eq.\eqref{lengthhT1} up to 4th order in $\frac{r_c}{r^*}$ and $\frac{r_H}{r^*}$ and finally have
\begin{align}\label{LengthhhLT}
l=\frac{2}{r^*}\Bigg\lbrace &a_1+b_1\left(\frac{r_H}{r^*}\right)^4+\left[a_2+b_2\left(\frac{r_H}{r^*}\right)^4\right]\left(\frac{r_c}{r^*}\right)^2\cr
&+\left[a_3+b_3\left(\frac{r_H}{r^*}\right)^4\right]\left(\frac{r_c}{r^*}\right)^4\Bigg\rbrace ,\ \ \ \ \ \ \ b_1,b_2,b_3>0,
\end{align}
where numerical coefficients $b_1$, $b_2$ and $b_3$ are given by eq.\eqref{qeoff3} in appendix \ref{Appendix1}. Solving eq.\eqref{LengthhhLT} perturbatively for $r^*$ and replacing $r^*$ in the eq.\eqref{HEET1}, we reach  (see appendix \ref{Appendix1} for more details of calculation)
\begin{align}\label{HEEloww}
\tilde{\mathcal{S}}_{finite}(l\Lambda_c,lT)&\equiv\frac{4G_N^{(5)}S_{finite}(l,l\Lambda_c,lT)}{L^2\Lambda_c T}\cr
&=\frac{1}{(l \Lambda_c)(lT)}\Big\lbrace \kappa_1 +\bar{\kappa}_1(lT)^4+\left[\kappa_2+\frac{3}{2}\log(l\Lambda_c) +\bar{\kappa}_2(lT)^4\right] (l \Lambda_c)^2\cr
&+\left[\kappa_3 +\bar{\kappa}_3(lT)^4\right] (l\Lambda_c)^4\Big\rbrace , \ \ \ \bar{\kappa}_1,\kappa_2,\kappa_3 >0,\kappa_1,\bar{\kappa}_2,\bar{\kappa}_3<0, \ \ \ \ \ 
\end{align}
where $\tilde{\mathcal{S}}_{finite}(l\Lambda_c,lT)$ is the redefined $S_{finite}(l,l\Lambda_c,lT)$ at low temperature and numerical coefficients  $\bar{\kappa}_1$, $\bar{\kappa}_2$ and $\bar{\kappa}_3$ are given by eq.\eqref{kappabar} in appendix \ref{Appendix1}. Since the underlying field theory is conformal we expect the dimensionless parameters $l\Lambda_c$ and $lT$ to appear and therefore we redefined $S_{finite}(l,l\Lambda_c,lT)$. By this redefinition we can take more easily the limits of $l\Lambda_c\rightarrow 0$ and $lT\rightarrow 0$. The first two terms are the known results corresponding to the pure AdS and AdS black hole HEE, respectively, and the next two terms are the thermal and non-conformal corrections. Since $\bar{\kappa}_1$ is a positive constant the second term is always positive and hence the thermal fluctuations increase $\tilde{\mathcal{S}}_{finite}$. This behavior is due to the increase in the number of degrees of freedom because of the thermal fluctuations. In the third term the logarithmic term is dominant and always negative in the high energy limit. Therefore, the non-conformal effects decrease $\tilde{\mathcal{S}}_{finite}$.  From eq.\eqref{HEEloww} we observe that if we fix $lT$ ($l\Lambda_c$) and increases $l\Lambda_c$ ($lT$), then $\tilde{\mathcal{S}}_{finite}(l\Lambda_c,lT)$ will increase. This result can be clearly seen in figure \ref{N-EE} where we plot $\tilde{\mathcal{S}}_{finite}$ as a function of $l\Lambda_c$ ($lT$) for fixed value of $lT$ ($l\Lambda_c$). Below, we represent the results corresponding to the mentioned limits:
\begin{figure}
\centering
\includegraphics[scale=0.38]{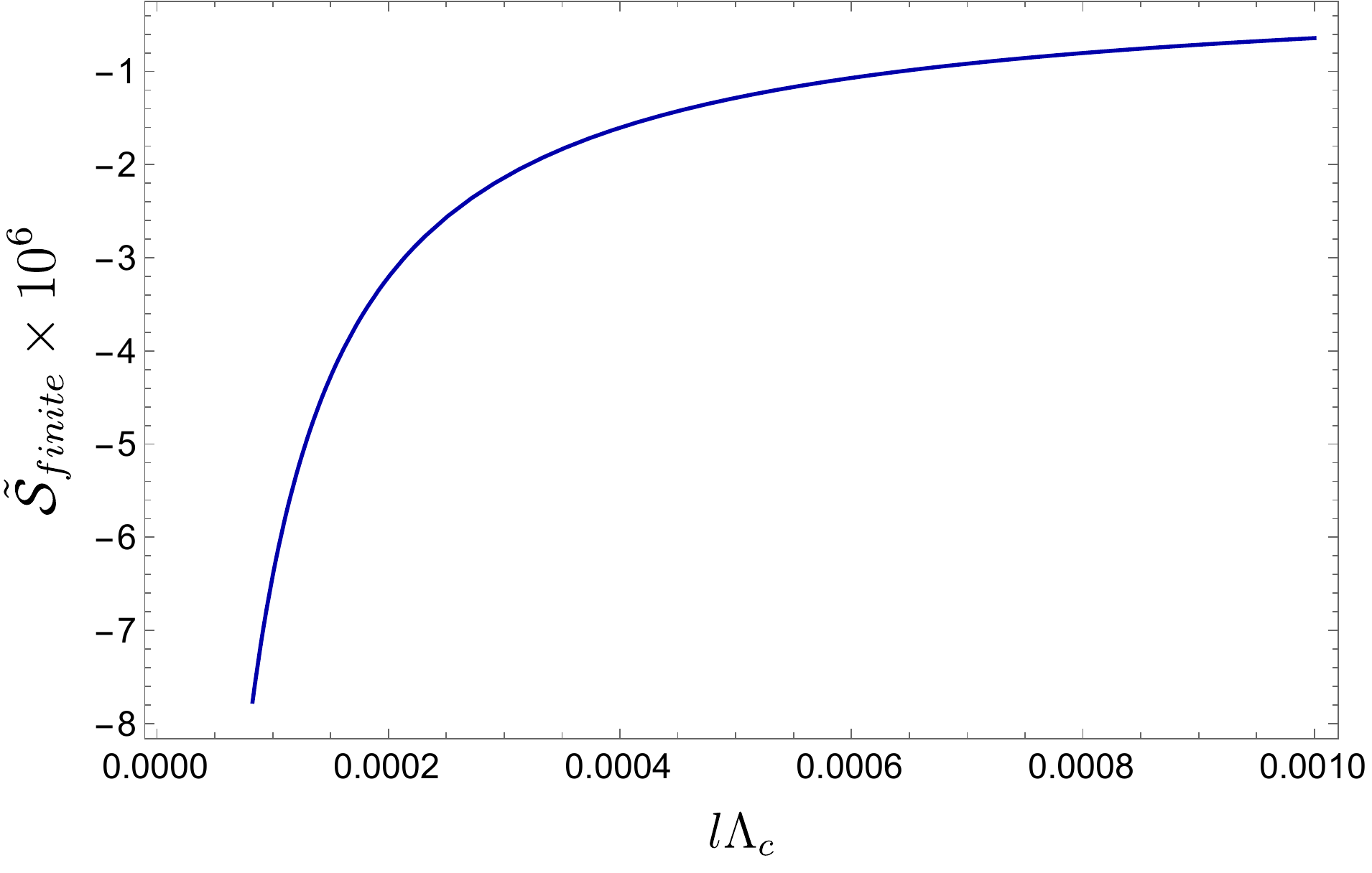} \ \
\includegraphics[scale=0.38]{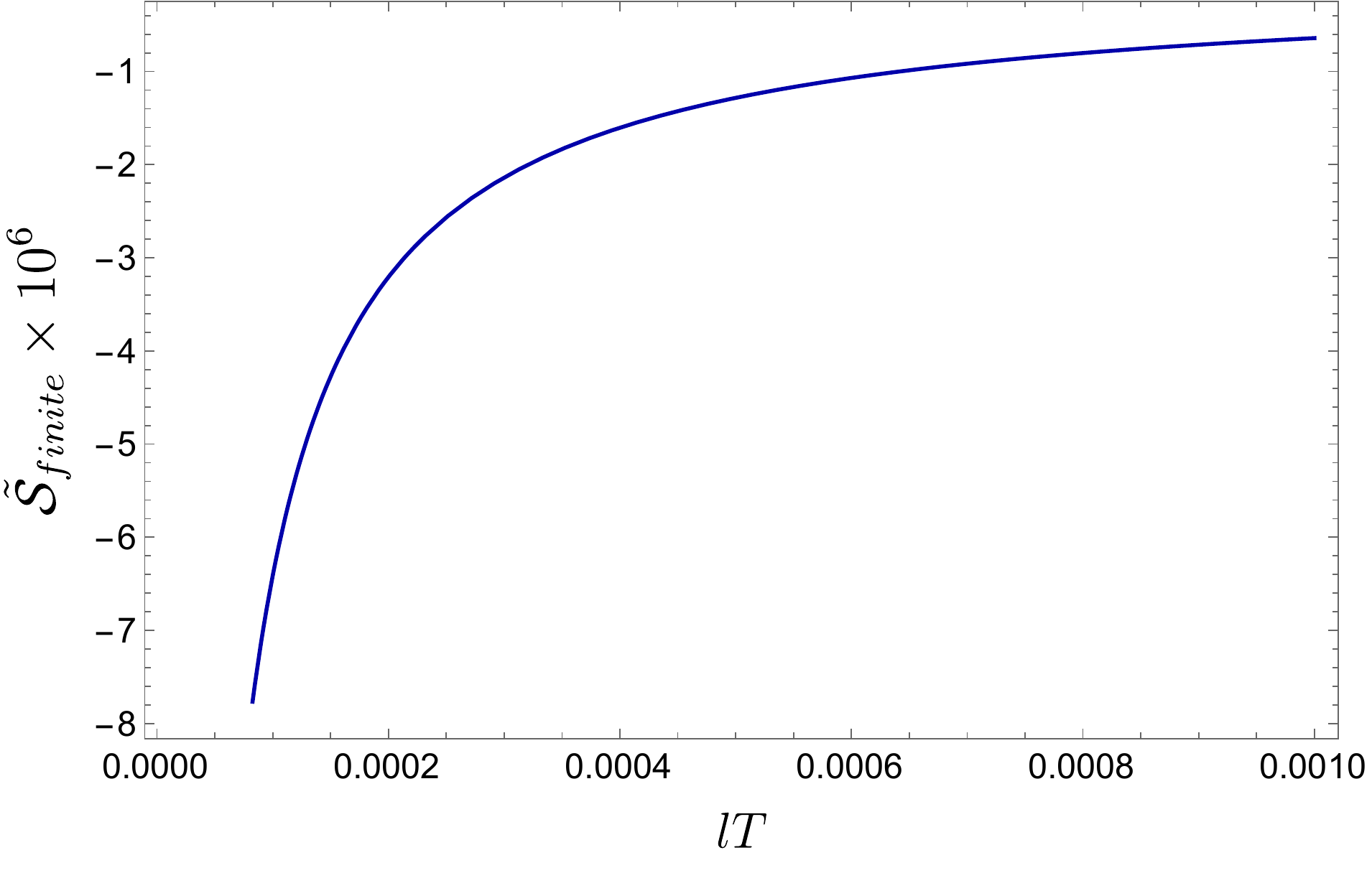}
\caption{Left: $\tilde{\mathcal{S}}_{finite}(l\Lambda_c,lT)$ in terms of $l\Lambda_c$ for fixed $lT=0.0005$ in the low temperature limit at high energy. Right:  $\tilde{\mathcal{S}}_{finite}(l\Lambda_c,lT)$ in terms of $lT$ for fixed $l\Lambda_c=0.0005$ in the low temperature limit at high energy. }
\label{N-EE}
\end{figure}
\begin{itemize}
\item $\Lambda_c\neq 0$ and $T=0$: In this case, we reproduce the results obtained for MAdS background eq.\eqref{HEEhighh1}. The leading term corresponds to the entanglement entropy in the conformal field theory and the non-conformal effects, the terms including $\kappa_2+\frac{3}{2}\log(l\Lambda_c)$ and $\kappa_3$, appear as the sub-leading terms. These effects are negative and therefore the non-conformality decrease the entanglement between our considered subsystems. 
\item $\Lambda_c=0$ and $T\neq 0$: In this case, we reproduce the previous results obtained for AdS black hole, see \cite{Fischler:2012ca,BabaeiVelni:2019pkw}. The leading term corresponds to the zero temperature HEE and the finite temperature correction appears as the sub-leading term. Since the constant $\bar{\kappa }_1$ is positive, the thermal fluctuations increase $\tilde{\mathcal{S}}_{finite}$.
\item $\Lambda_c= 0$ and $T=0$:  We reach the previous results obtained for pure AdS$_5$, which is corresponding to the HEE in zero
temperature conformal field theory. Obviously, this term is negative.
\end{itemize}
There is a phase transition point at $r_H=r_c$ \cite{Andreev:2006eh}. Therefore we are interested in studying the HEE in the limit of $r_H \rightarrow r_c$ or equivalently $T \rightarrow \frac{\Lambda_c}{ \sqrt{2}\pi}$ , which we call it the transition limit. If we take the transition limit from eq.\eqref{HEEloww}, up to the second order in $l\Lambda_c$, we obtain the following expression
\begin{align}\label{HEEtra}
\tilde{\mathcal{S}}_{finite}(l\Lambda_c ,lT)\bigg\vert _{T \rightarrow \frac{\Lambda_c}{ \sqrt{2}\pi}}=\frac{\sqrt{2}\pi}{(l \Lambda_c)^2}\left\lbrace \kappa_1+ \left(\kappa_2+\frac{3}{2}\log(l\Lambda_c)\right) (l \Lambda_c)^2+ \left(\kappa_3+\frac{\bar{\kappa}_1}{4\pi ^4}\right) (l\Lambda_c)^4\right\rbrace .
\end{align}
For fixed $l\Lambda_c$, one can compare $\hat{\mathcal{S}}_{finite}(l\Lambda_c)$ at the zero temperature eq.\eqref{HEEhighh1} and $\tilde{\mathcal{S}}_{finite}(l\Lambda_c,lT)$ at low temperature one in the transition limit eq.\eqref{HEEtra}. To do this, we subtract eq.\eqref{HEEhighh1} from eq.\eqref{HEEtra} and we get 
\begin{align}\label{EEtransition}
\frac{\tilde{\mathcal{S}}_{finite}(l\Lambda_c ,lT)}{\sqrt{2}\pi} \bigg\vert _{T \rightarrow\frac{\Lambda_c}{ \sqrt{2}\pi}} -\hat{\mathcal{S}}_{finite}(l\Lambda_c)  =\frac{\bar{\kappa}_1}{4\pi^4 }(l\Lambda_c)^2 >0 .
\end{align}
This result shows that near the transition point, the subsystem $A$ and its complement $\bar{A}$ are less entangled at zero temperature. As we know, the EE describes the amount of information loss because of integrating out the subsystem $\bar{A}$. The higher the EE, the more information we lose. From the information point of view, we would like to define a favorable state such that the subsystems $A$ and $\bar{A}$ are less entangled. Therefore, from eq.\eqref{EEtransition}, we find that near the transition point the state at zero temperature is the favorable one. 

Note that we define the favorable state here in the point of view of the amount of information loss. However, from resource theory point of view, the definition of favorable state can be different. For example from the resource theory of entanglement we know that if the local operations and classical communication (LOCC) are allowed opperations, then the entangled states can be regarded as a valuable resource \cite{Horodecki:2009zz,Chitambar,Gour,Tejada}.
\subsubsection{High energy and high temperature}\label{EE-highT}
In this subsection we consider the high temperature limit i.e. $Tl\gg1$ or equivalently $r^*\rightarrow r_H$, at high energy ($r_c\ll r^*$), see figure \ref{cc}. In this case, the turning point of the extremal surface $\Gamma$ approaches the horizon and the leading contribution comes from the near horizon background.  We expand eqs. \eqref{lengthhT1} and \eqref{HEET1}  up to second order in $\frac{r_c}{r^*}$ and take the integrals. we get
\begin{align}\label{LengthTH}
l=\frac{2}{r^*}\sum\limits_{n=0}^{\infty}\sum\limits_{m=0}^{\infty}\frac{\Gamma(n+\frac{1}{2})\Gamma(m+\frac{1}{2})}{\pi\Gamma(n+1)\Gamma(m+1)}(\frac{r_H}{r^*})^{4m}\left[L_1+L_2(\frac{r_c}{r^*})^2\right],
\end{align}
\begin{align}\label{HEEhighT1}
\mathcal{A}=\frac{L^2}{\epsilon^2}-3r_c^2L^2\log(r^*\epsilon)+\mathcal{A}',
\end{align}
where $\mathcal{A}'$ contains the terms which do not include $\epsilon$ 
\begin{align}\label{HEETH}
\mathcal{A}'=2{r^*}^2L^2\sum\limits_{n=0}^{\infty}\sum\limits_{m=0}^{\infty}\frac{\Gamma(n+\frac{1}{2})\Gamma(m+\frac{1}{2})}{\pi\Gamma(n+1)\Gamma(m+1)}(\frac{r_H}{r^*})^{4m}\left[D_1+D_2(\frac{r_c}{r^*})^2\right],
\end{align}
and $L_1$, $L_2$, $D_1$ and $D_2$ are given by
\begin{align}
L_1=&\frac{1}{4 m+6 n+4}, \ \ \ \ \ \ \ L_2=\frac{3}{2} (2 n+1) \left(\frac{1}{4 m+6 n+4}-\frac{1}{4 m+6 n+6}\right),\cr
D_1=&\frac{1}{4 m+6 n-2} , \ \ \ \ \ \ \ D_2=\frac{3 n}{2 (2 m+3 n-1)}+\frac{3-6 n}{8 m+12 n}.
\end{align}
In order to write $\mathcal{A}'$ in terms of $l$, we do the calculations order by order up to $(\frac{r_c}{r^*})^2$ and check the convergence of the resulting infinite series.
\begin{itemize}
\item Up to $(\frac{r_c}{r^*})^0$:  In this case, $r_c$ is not appear and it is expected that we reach the results corresponding to AdS black hole in the high temperature limit \cite{Fischler:2012ca,BabaeiVelni:2019pkw}. We consider the first terms in brackets in eqs. \eqref{LengthTH} and \eqref{HEETH} which include $L_1$ and $D_1$, respectively and sum over $m$. We obtain
\begin{align}\label{LengthTH1}
l=\frac{2}{r^*}\sum\limits_{n=0}^{\infty}\frac{1}{4n+1}\frac{\Gamma(n+\frac{1}{2})\Gamma(\frac{2n+2}{3})}{\Gamma(n+1)\Gamma(\frac{4n+1}{6})}(\frac{r_H}{r^*})^{4n},
\end{align}
\begin{align}\label{HEETH1}
\mathcal{A}'=2{r^*}^2L^2\sum\limits_{n=0}^{\infty}\frac{1}{4n-2}\frac{\Gamma(n+\frac{1}{2})\Gamma(\frac{2n+2}{3})}{\Gamma(n+1)\Gamma(\frac{4n+1}{6})}(\frac{r_H}{r^*})^{4n}.
\end{align}
Using eqs. \eqref{LengthTH1} and \eqref{HEETH1} and doing some calculations, we get 
\begin{align}
\mathcal{A}'=2{r^*}^2L^2\Bigg\lbrace & \frac{lr^*}{2}-\frac{3\sqrt{\pi}\Gamma(\frac{2}{3})}{2\Gamma(\frac{1}{6})}+\sum\limits_{n=1}^{\infty}\frac{3}{(4n-2)(4n+1)}\frac{\Gamma(n+\frac{1}{2})\Gamma(\frac{2n+2}{3})}{\Gamma(n+1)\Gamma(\frac{4n+1}{6})}(\frac{r_H}{r^*})^{4n}\bigg\rbrace .
\end{align}
It is easy to see that in the large $n$ limit the above infinite series behaves as $\frac{1}{n^2}(\frac{r_H}{r^*})^{4n}$, so we can safely consider $r^*\rightarrow r_H$ limit. Hence, the final result in the high temperature limit becomes
\begin{align}\label{HEETH2}
\mathcal{S}_{HT}^{(0)}(lT)\equiv\frac{4G_N^{(5)}S_{finite}(l,lT)}{L^2T^2}= \left[F_1+\pi^3 (lT)\right], \ \ \ \ \ \ \ F_1<0,
\end{align}
where $\mathcal{S}_{HT}^{(0)}(lT)$ is the redefined $S_{finite}(l,lT)$ in the high temperature limit up to the zero order of $\frac{r_c}{r^*}$ and $F_1$ is a numerical constant given by eq.\eqref{F2} in appendix \ref{Appendix1}. The second term in eq.\eqref{HEETH2} is the dominant term in $S_{finite}(l,lT)$ which scales with the volume of the rectangular strip, $lL^2$, and the first term in $S_{finite}(l,lT)$ is area dependent, $L^2$. Therefore, the first term corresponds to the entanglement entropy between the strip region and its complement while the second term corresponds to the thermal entropy.
\item Up to $(\frac{r_c}{r^*})^2$: In this case, we add the non-conformal effects to the previous one by considering the second terms in brackets  in eqs. \eqref{LengthTH} and \eqref{HEETH}. we rewrite eq.\eqref{HEETH} as follows
\begin{align}
\mathcal{A}'=&2{r^*}^2L^2\sum\limits_{n=0}^{\infty}\sum\limits_{m=0}^{\infty}\frac{\Gamma(n+\frac{1}{2})\Gamma(m+\frac{1}{2})}{\pi\Gamma(n+1)\Gamma(m+1)}(\frac{r_H}{r^*})^{4m}\cr
&\times\left[L_1+L_2(\frac{r_c}{r^*})^2+(D_1-L_1)+(D_2-L_2)(\frac{r_c}{r^*})^2\right]
\end{align}
Performing the summation over $m$, then the above equation leads to
\begin{align}\label{HEETHH}
\mathcal{A}'=&{r^*}^3L^2l+2{r^*}^2L^2\left[-\frac{3\sqrt{\pi}\Gamma(\frac{2}{3})}{2\Gamma(\frac{1}{6})}+\sum\limits_{n=1}^{\infty}\frac{3}{(4n-2)(4n+1)}\frac{\Gamma(n+\frac{1}{2})\Gamma(\frac{2n+2}{3})}{\Gamma(n+1)\Gamma(\frac{4n+1}{6})}(\frac{r_H}{r^*})^{4n}\right]\cr
&+2{r^*}^2L^2B(\frac{r_c}{r^*})^2,
\end{align}
where $B$ is given by
\begin{align}\label{B}
B=B_0&+\sum\limits_{n=0}^{\infty}\sum\limits_{m=1}^{\infty}\frac{\Gamma(n+\frac{1}{2})\Gamma(m+\frac{1}{2})}{\pi\Gamma(n+1)\Gamma(m+1)}(\frac{r_H}{r^*})^{4m}\cr
&\times\Bigg[\frac{3 n}{4m+6 n-2)}+\frac{3-6 n}{8 m+12 n}
-\frac{\frac{3}{2}(2 n+1)}{4 m+6 n+4}+\frac{\frac{3}{2}(2 n+1)}{4 m+6 n+6}\Bigg].
\end{align}
$B_0$ is a constant comes from $m=0$ and given by eq.\eqref{B0} in appendix \ref{Appendix1}. The convergence of the series in the second term of eq.\eqref{HEETHH} was seen in the previous case. In order to study the convergence of the series in eq.\eqref{B} at high temperature, $r^*\rightarrow r_H$, we sum over $n$ and obtain
\begin{figure}
\centering
\includegraphics[scale=0.38]{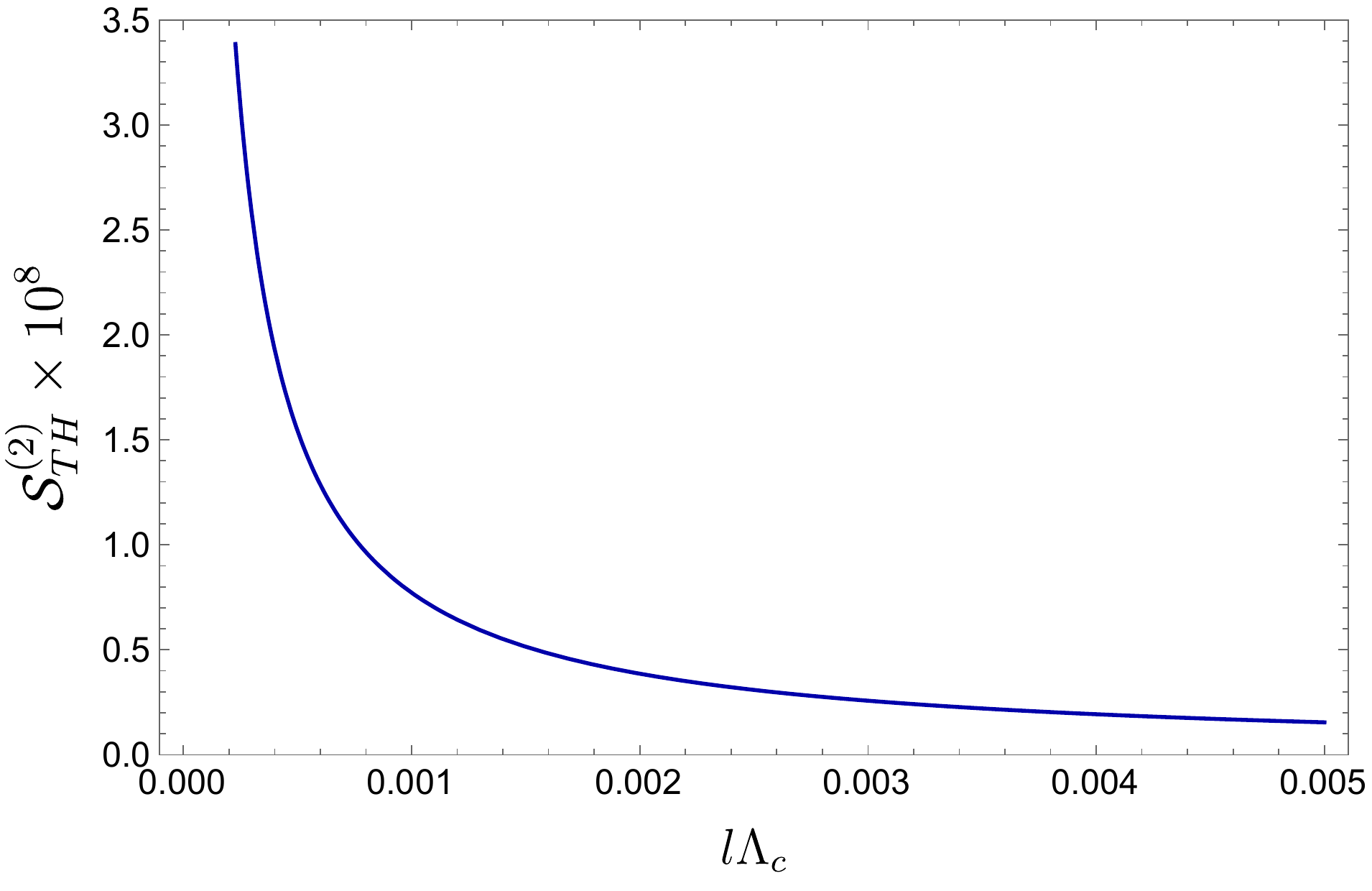} \ \
\includegraphics[scale=0.38]{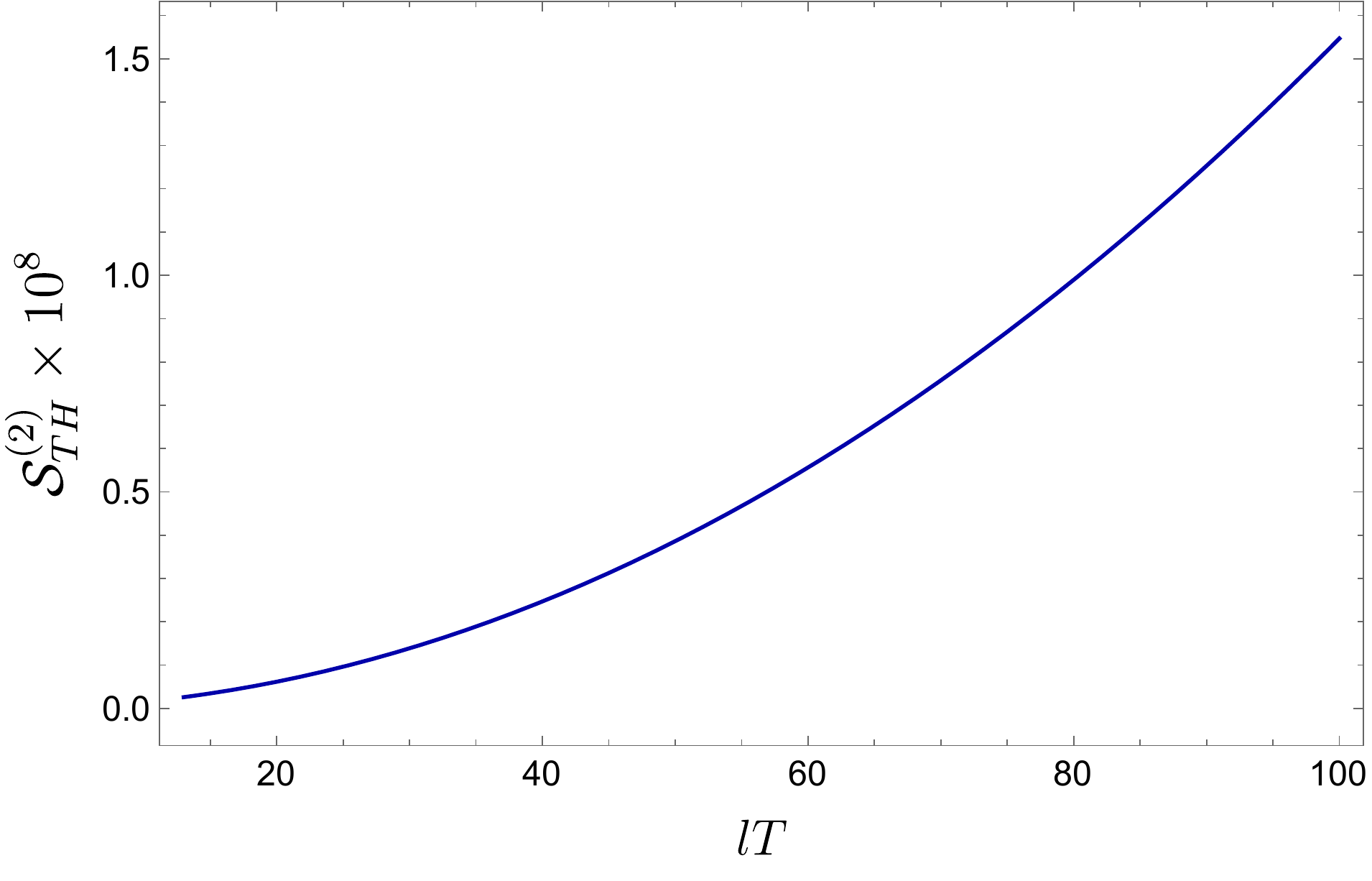}
\caption{Left: $\mathcal{S}_{HT}^{(2)}(lT,l\Lambda_c)$ in terms of $l\Lambda_c$ for fixed $lT=50$ in the high temperature limit at high energy. Right:  $\mathcal{S}_{HT}^{(2)}(lT,l\Lambda_c)$ in terms of $lT$ for fixed $l\Lambda_c=0.002$ in the high temperature limit at high energy. }
\label{EE-HT}
\end{figure}
\begin{align}
B&=B_0+\sum\limits_{m=1}^{\infty}\frac{3\Gamma(m+\frac{1}{2})}{8\sqrt{\pi}\Gamma(m+1)}(\frac{r_H}{r^*})^{4m}\cr
&\times\Bigg[\frac{1}{m}\left(\, _4F_3\left(\frac{1}{2},\frac{2 m}{5}+\frac{4}{5},\frac{2 m}{3}-\frac{1}{3},\frac{2 m}{3};\frac{2 m}{5}-\frac{1}{5},\frac{2 m}{3}+\frac{2}{3},\frac{2 m}{3}+1;1\right)\right)\cr
&-\frac{1}{(m+1) (2 m+3)}\left(\, _3F_2\left(\frac{3}{2},\frac{2 m}{3}+\frac{2}{3},\frac{2 m}{3}+1;\frac{2 m}{3}+\frac{5}{3},\frac{2 m}{3}+2;1\right)\right)\Bigg]. \ \ \ \ \
\end{align}
It is easy to see that in the large $m$ limit, the first series in the brackets behaves as $\frac{1}{m^\frac{3}{2}}(\frac{r_H}{r^*})^{4m}$ and the second series in the brackets behaves as $\frac{1}{m^\frac{5}{2}}(\frac{r_H}{r^*})^{4m}$. Therefore, both of these series are convergent at $r^*\rightarrow r_H$ and we can safely take the mentioned limit. If we add the contribution of the second term in eq.\eqref{HEEhighT1} to the finite part of the HEE, then we will finally reach the following expression
\begin{align}\label{HEEhighT}
\mathcal{S}_{HT}^{(2)}(l\Lambda_c,lT)&\equiv\frac{4G_N^{(5)}S_{finite}(l,l\Lambda_c,lT)}{L^2\Lambda_c T}\cr
&=\frac{1}{(l\Lambda_c )(lT)}\bigg\lbrace F_1(lT)^2+\pi^3 (lT)^3\cr
& \ \ \ \ \ \ \ \ \ \ \ \ \ \ \ \ \ \ +\left[F_2+\frac{3}{2}\log(l\Lambda_c )\right](l\Lambda_c )^2\bigg\rbrace, \ \ \ F_1<0, F_2>0,\ \ \ \ \
\end{align}
\begin{scriptsize}
•
\end{scriptsize}where  $S_{HT}^{(2)}(l\Lambda_c,lT)$ is the redefined  $S_{finite}(l,l\Lambda_c,lT)$ and $F_2$ is positive and given by eq.\eqref{F1} in appendix \ref{Appendix1}. The first two terms in eq.\eqref{HEEhighT} correspond to $S_{HT}^{(0)}$ in the AdS black hole in the high temperature limit. The non-conformal effect appears in third term which is very small with respect to the first two terms. The logarithmic term in the braket is the dominant term and is always negative in the high energy limit. Hence the non-conformal effect decreases the $S_{HT}^{(2)}(l\Lambda_c,lT)$ with respect to $S_{HT}^{(0)}(lT)$ at high temperature in the high energy limit which is in complete agreement with \cite{Rahimi:2016bbv}. In figure \ref{EE-HT} we plot $S_{HT}^{(2)}(l\Lambda_c,lT)$ as a function of $l\Lambda_c$ and $lT$ for fixed value of $lT$  and $l\Lambda_c$, respectively. From this figure, we observe that if we fix $l\Lambda_c$ and increase $lT$, then $S_{HT}^{(2)}(l\Lambda_c,lT)$ will increase. But in the fixed $lT$, $S_{HT}^{(2)}(l\Lambda_c,lT)$ has different behavior and decreases by increasing $l\Lambda_c$.
\end{itemize}
\section{Mutual information}\label{section4}
When the boundary entangling region is made by two disjoint subsystems, an important quantity to study is the mutual information which is a quantity that is derived from entanglement entropy. The
definition of mutual information between two disjoint subsystems $A$ and $B$ is given by
\begin{figure}
\centering
\includegraphics[scale=0.39]{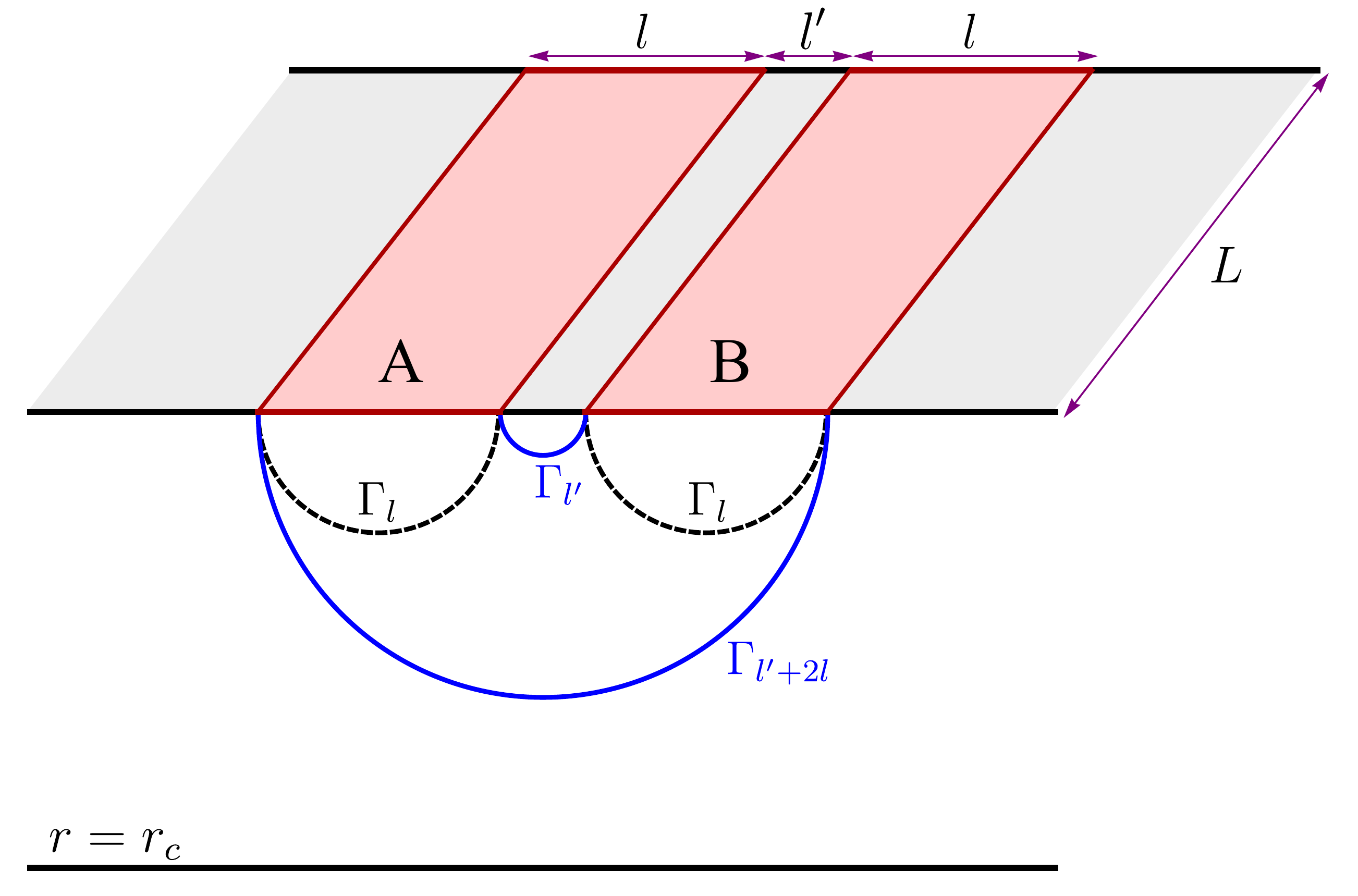}
\caption{A simplified sketch of two strip regions $A$ and $B$ with equal size $l$ which are separated by the distance $l'$. When $l'$ is small enough, the minimal surface of $A\cup B$ are given by the union of blue curves and when  $l'$ is large enough, the minimal surface of $A\cup B$ are given by the union of black-dashed curves.}
\label{MIandEE}
\end{figure}
\begin{align}
I(A,B)=S_A+S_B-S_{A\cup B},
\end{align}
where $S_A$, $S_B$ and $S_{A\cup B}$ denote the entanglement entropy of the region $A$, $B$ and $A\cup B$, respectively. The mutual information measures the total correlation between the two sub-systems, including both classical and quantum correlations \cite{Groisman}. From the definition, it is clear that the mutual information is a finite quantity since the divergent pieces in the entanglement entropy cancel out and the subadditivity of the entanglement entropy, $S_A+S_B\geq S_{A\cup B}$, guarantees that $I(A,B)\geq 0$. We consider the two disjoint subsystems both rectangular strips of size $l$ which are separated by the distance $l'$ on the boundary, see figure \ref{MIandEE}. One can easily follow the RT-prescription to compute the HEE of the individual sub-systems $A$ and $B$. For computing $S_{A\cup B}$, we have two configurations. When the separation distance is large enough, the two subsystems $A$ and $B$ are completely disentangled and we have $S_{A\cup B}=S_A+S_B=2S(l)$. In this case, the mutual information vanishes $I(A,B)=0$. On the other hand, when $l'$ is small enough, the two subsystems $A$ and $B$ are entangled and $S_{A\cup B}$ is given by $S_{A\cup B}=S(l')+S(l'+2l)$. In this case, we have the non-zero mutual information $I(A,B)>0$. Therefore, one can assume that there would be a critical separation distance $x_d$ where the mutual information exhibits a phase transition from positive values to zero at this distance. To summarize, the holographic mutual information (HMI) of two disjoint subsystems is given by
\begin{align}\label{HMI}
I(A,B)=\Bigg\lbrace
\begin{array}{lr}
2S(l) -S(2l+l')-S(l') \ \ l'<x_d   & \\ 
0 \ \ \ \ \ \ \ \ \ \ \ \ \ \ \ \ \ \ \ \ \ \ \ \ \ \ \ \ \ \ \ \ \ l'\geqslant x_d   & 
\end{array} .
\end{align}
We will use this relation to discuss the behavior of HMI in MAdS \eqref{MAdS} and MBH \eqref{MBH} backgrounds. 
\subsection{HMI at zero temperature; High energy limit}
In this section, we would like to study the HMI for MAdS background \eqref{MAdS}. Since the HMI is a linear combination of HEE we consider the special limit that the HEE can be computed analytically. Considering the high energy limit i.e. $l\Lambda _c \ \& \ l'\Lambda_c \ll 1$ or equivalently $r_c\ll r^*_l \ \& \ r^*_{l'}$ and using eqs. \eqref{HEEhighh1} and \eqref{HMI}, we get
\begin{align}\label{HMIhigh}
\hat{\mathcal{I}}(l\Lambda_c ,l'\Lambda_c )&\equiv\frac{4G_N^{(5)} I(l,l',l\Lambda_c,l'\Lambda_c)}{L^2\Lambda_c^2}\cr
&=\kappa _1\left[\frac{2}{(l\Lambda_c)^2}-\frac{1}{(l'\Lambda_c)^2}-\frac{1}{(l'\Lambda_c+2l\Lambda_c)^2}\right]+\frac{3}{2}\log\left(\frac{(l\Lambda_c)^2}{(l'\Lambda_c)(l'\Lambda_c +2l\Lambda_c)} \right)\cr
&-2\kappa _3(l'\Lambda _c+l\Lambda _c)^2, \ \ \ \ \ \ \ \ \ \ \kappa _1<0,\kappa _3>0,
\end{align}
where $\hat{\mathcal{I}}(l\Lambda_c ,l'\Lambda_c )$ is the redefined $I(l,l',l\Lambda_c,l'\Lambda_c)$ which is given by eq.\eqref{I-highE} in appendix \ref{Appendix2}.  By this redefinition, the limits of $l\Lambda\rightarrow 0$ and $l'\Lambda_C\rightarrow 0$ become meaningful. The first term in eq.\eqref{HMIhigh}, including $\kappa_1$, is the contribution of the AdS boundary corresponds to the mutual information between two subsystems in conformal field theory \cite{Fischler:2012uv} and the other terms are the non-conformal effects. We focus on the limit of $l\Lambda_c\gg l'\Lambda_c$. Threfore, the second term which is the dominant term in the non-conformal effects is always positive and hence the non-conformality increases $\hat{\mathcal{I}}(l\Lambda_c ,l'\Lambda_c )$. From the above equation, we observe that by increasing $l\Lambda_c$ ($l'\Lambda_c$) $\hat{\mathcal{I}}(l\Lambda_c ,l'\Lambda_c )$ increase (decrease) for fixed $l'\Lambda_c$ ($l\Lambda_c$) in the high energy limit.
\subsection{HMI at finite temperature}
Here, we investigate the thermal behavior of HMI in the MBH background \eqref{MBH}. In order to reach the analytical results, we consider the low ($lT\ \&\ l'T \ll 1$), intermediate  ($l'T  \ll 1\ll  lT$)  and high ($lT\ \&\ l'T \gg 1$) temperature, in the high energy limit ($l\Lambda _c\ \&\ l'\Lambda _c\ll 1$).
\subsubsection{Low temperature at high energy}
In the low temperature limit i.e. $lT \ \&\ l'T \ll 1$ or equivalently  $r_H \ll {r^*}_{l'}\ \&\ r^*_{l'+2l}$, at high energy, we use eqs. \eqref{HEEloww} and \eqref{HMI} and obtain
\begin{align}\label{HMIlow}
\tilde{\mathcal{I}}(l\Lambda_c,l'\Lambda_c ,lT,l'T)&\equiv\frac{4G_N^{(5)} I(l,l',l\Lambda_c,l'\Lambda_c ,lT,l'T)}{L^2\Lambda_c T}\cr
&=\kappa _1\left[\frac{2}{(l\Lambda_c)(lT)}-\frac{1}{(l'\Lambda_c)(l'T)}-\frac{1}{(l'\Lambda_c+2l\Lambda_c)(l'T+2lT)}\right]\cr
&-2\bar{\kappa}_1\frac{(l'T+lT)^3}{(l'\Lambda_c+l\Lambda_c)}+\frac{3(l\Lambda_c)}{2(lT)}\log\left(\frac{(l\Lambda_c)^2}{(l'\Lambda_c)(l'\Lambda_c +2l\Lambda_c)} \right)\cr
&-2\kappa _3\frac{(l'\Lambda_c+l\Lambda_c)^3}{(l'T+lT)},
 \ \ \ \ \ \ \ \ \kappa _1<0, \bar{\kappa}_1,\kappa _3>0,
\end{align}
where $\tilde{\mathcal{I}}(l\Lambda_c,l'\Lambda_c ,lT,l'T)$ is the redefined $I(l,l',l\Lambda_c,l'\Lambda_c ,lT,l'T)$ which is given by eq.\eqref{I-lowT} in appendix \ref{Appendix2}. By this redefinition the limits $l\Lambda_c \rightarrow 0$, $lT\rightarrow 0$, $l'\Lambda_c \rightarrow 0$ and $l'T\rightarrow 0$ become meaningful and intuitive. The first two terms, including $\kappa _1$ and $\bar{\kappa}_1$, are the known results corresponding to the pure AdS and AdS black hole, respectively and the next two terms indicate the non-conformal effect. Since $\bar{\kappa}_1$ is positive constants the second term is always negative and hence thermal fluctuations decrease $\tilde{\mathcal{I}}(l\Lambda_c,l'\Lambda_c ,lT,l'T)$. Therefore, thermal fluctuations reduce the total correlation between the subsystems. From eq.\eqref{HMIlow} one can see that $\tilde{\mathcal{I}}(l\Lambda_c,l'\Lambda_c ,lT,l'T)$ decreases by increasing the temperature. At low temperature, this behavior is seen in \cite{Bernigau} which study the MI for thermal free fermions.  The third term is the dominant term in the non-conformal effects. We focus on the limit of $l\Lambda_c\gg l'\Lambda_c$. In this limit the logarithmic term is positive and therfere the non-conformal effects increase $\tilde{\mathcal{I}}(l\Lambda_c,l'\Lambda_c ,lT,l'T)$. In the following, we represent the results corresponding to above limits.
\begin{itemize}
\item $\Lambda_c\neq 0$ and $T=0$: In this case, we reproduce the results obtained for MAdS background \eqref{HMIhigh}. The leading contribution comes from the AdS boundary which corresponds to the HMI between our considered subsystems and the non-conformal effects appear as the sub-leading terms. These effects are positive and therefore the non-conformality increases $\hat{\mathcal{I}}(l\Lambda_c ,l'\Lambda_c )$. 
\item $\Lambda_c=0$ and $T\neq 0$: In this case, we reproduce the previous results obtained for AdS black hole, see \cite{Fischler:2012uv}. The leading term corresponds to the zero temperature HMI and the finite temperature correction appears as the sub-leading term. The constant $\bar{\kappa}_1$ is positive and therefore, the thermal fluctuations decrease $\tilde{\mathcal{I}}(lT,l'T)$.
\item $\Lambda_c= 0$ and $T=0$:  We reach the previous results obtained for pure AdS$_5$ which corresponds to the HMI in the zero temperature conformal field theory \cite{Fischler:2012uv}. Obviously, this term is positive.
\end{itemize}
We want to study the $\tilde{\mathcal{I}}(l\Lambda_c,l'\Lambda_c ,lT,l'T)$ near the transition point and hence we take the transition limit i.e. $T\rightarrow\frac{\Lambda_c}{ \sqrt{2}\pi}$ of eq.\eqref{HMIlow}. Up to the second order in $l\Lambda_c$ and $l'\Lambda_c$ we obtain the following expression
\begin{align}\label{HMItra}
 \tilde{\mathcal{I}}(l\Lambda_c,l'\Lambda_c ,lT,l'T)&\vert _{T \rightarrow \frac{\Lambda_c}{ \sqrt{2}\pi}}=\sqrt{2}\pi\Bigg\lbrace \kappa _1\left[\frac{2}{(l\Lambda_c)^2}-\frac{1}{(l'\Lambda_c)^2}-\frac{1}{(l'\Lambda_c+2l\Lambda_c)^2}\right]\cr
&+\frac{3}{2}\log\left(\frac{(l\Lambda_c)^2}{(l'\Lambda_c)(l'\Lambda_c +2l\Lambda_c)} \right)-2(\kappa _3+\frac{\bar{\kappa}_1}{4\pi^4}) (l'\Lambda _c+l\Lambda _c)^2\Bigg\rbrace .
\end{align}
We fix $l \Lambda_c$ and $l'\Lambda_c$ and compare the zero temperature $\hat{\mathcal{I}}(l\Lambda_c,l'\Lambda_c)$ eq.\eqref{HMIhigh} with the finite temperature $\tilde{\mathcal{I}}(l\Lambda_c,l'\Lambda_c ,lT,l'T)$ in transition limit eq.\eqref{HMItra}
\begin{align}\label{HMItran}
\frac{\tilde{\mathcal{I}}(l\Lambda_c,l'\Lambda_c ,lT,l'T)}{\sqrt{2}\pi}\bigg\vert _{T \rightarrow \frac{\Lambda_c}{ \sqrt{2}\pi}}-\hat{\mathcal{I}}(l\Lambda_c,l'\Lambda_c)=-\frac{\bar{\kappa}_1}{4\pi^4} (\Lambda _cl'+\Lambda _cl)^2<0.
\end{align}
From the above equation it is seen that, near the transition point, the subsystems $A$ and $B$ have more information in common  at zero temperature than the low temperature. When the two subsystems $A$ and $B$ are more correlated, we can get more information about the subsystem $A$ through subsystem $B$ and hence the state at zero temperature is the favorable one.
\subsubsection{Intermediate temperature at high energy}
In this subsection we study another interesting limit called intermediate temperature limit which is defined by $l'T  \ll 1\ll lT $ or equivalently $r_H\ll r^*_{l'}$ and $r^*_{l'+2l}\to r_H$. We consider this limit at high energy $l\Lambda _c\ \&\ l'\Lambda _c\ll 1$. Using eqs. \eqref{HEEloww}, \eqref{HEEhighT} and \eqref{HMI}, we reach the following expression
\begin{align} \label{MIintT}
\tilde{\mathcal{I}}(l'\Lambda_c ,l'T)&\equiv\frac{4G_N^{(5)} I(l',l'\Lambda_c ,l'T)}{L^2\Lambda_c T}\cr
&=\frac{1}{(l'\Lambda_c)(l'T)}\bigg\lbrace -\kappa_1+F_1(l'T)^2 -\pi^3(l'T)^3-\bar{\kappa}_1 (l'T)^4\cr
& +\left[F_3-\frac{3}{2}\log(l'\Lambda_c)-\bar{\kappa}_2(l'T)^4\right](l'\Lambda_c )^2 -\left[\kappa_3+\bar{\kappa}_3(l'T)^4\right] (l'\Lambda_c )^4\bigg\rbrace , \cr
&\ \ \ \ \ \ \ \ \ \ \ \ \ \ \ \ \ \ \kappa_1,\bar{\kappa}_2,\bar{\kappa}_3,F_1<0, \ \bar{\kappa}_1,\kappa_3, F_3>0, \ \
\end{align}
where $\tilde{\mathcal{I}}(l'\Lambda_c ,l'T)$ is the redefined $I(l',l'\Lambda_c ,l'T)$ which is given by eq.\eqref{I-intT} and $F_3$ is a numerical constant given by eq.\eqref{F3} in appendix \ref{Appendix2}. It is observed that $\tilde{\mathcal{I}}(l'\Lambda_c ,l'T)$ does not depend on the length of the subsystems $l$ which is in complete agreement with the results obtained in \cite{Fischler:2012uv}. The logarithmic term and the term including $F_3$ which are the non-conformal effects are always positive and therefore the non-conformality increases $\tilde{\mathcal{I}}(l'\Lambda_c ,l'T)$. In the limit of $l'\Lambda_c\rightarrow 0$, we reproduce the results obtained for the AdS black hole \cite{Fischler:2012uv}. Since the  term including $F_1$ is proportional to the area of entangling region the HMI obeys an area law even at intermediate temperature. This term is negative and hence temperature effects decrease $\tilde{\mathcal{I}}(l'\Lambda_c ,l'T)$.  We want to study the HMI in the limit of $l'\rightarrow 0$ which corresponds to the case that the two subsystems $A$ and $B$ touch each other. By taking $l'\rightarrow 0$ limit from eq.\eqref{I-intT} we reach the following expression
\begin{align}\label{I(l'=0)}
I(l',l'\Lambda_c ,l'T)\bigg\vert_{l'\rightarrow 0}=\frac{1}{4G_N^{(5)}}\bigg\lbrace & -\kappa_1\left(\frac{L}{l'}\right)^2-\frac{3}{2}\log(l'\Lambda_c)(L\Lambda_c)^2\cr
&+F_1(LT)^2+F_3(L\Lambda_c )^2\bigg\rbrace , \ \ \ \ \ \ \ \kappa_1,F_1<0, F_3>0.
\end{align}
The first term in the above equation is the leading term which obeys an area law divergence in the limit $l'\rightarrow 0$. The second term has a logarithmic divergence because of non-conformality. The third and forth terms are finite sub-leading terms which scale with the area of the strip region $L^2$ times to the $T^2$ and $\Lambda_c^2$, respectively. In the subsection \ref{EE-highT} we observe that at the high temperature there is a thermal entropy contribution to the entanglement entropy which scales with the volume of the strip region $lL^2$. The thermal entropy contribution is not appear in eq.\eqref{I(l'=0)} and hence this equation only measure pure quantum entanglement \cite{Kundu:2016dyk,Ebrahim:2020qif,Fischler:2012uv}.
\subsubsection{High temperature at high energy}
 In the high temperature limit we have $l'T \gg 1$ or equivalently $l' \gg \frac{\pi}{r_H}$. As we discussed at the beginning of this section, for very large separation distances the two subsystems are completely disentangled and hence the HMI is identically zero in this regime. In other words, for large separation distance, the RT-surface corresponding to the $A\cup B$ is equal to the union of the RT-surfaces corresponding to the two subregions $A$ and $B$ and hence the HMI vanishes for this case.
\section{Entanglement of purification}\label{section5}
Entanglement of purification is an important quantity which measure the total (quantum and classical) correlation between two disjoint subsystems in the mixed states. Consider a bipartite system described by a mixed state and density matrix $\rho_{AB}$. We can always purify this mixed state by enlarging its Hilbert space as $\mathcal{H}_A\otimes \mathcal{H}_B\rightarrow \mathcal{H}_A\otimes \mathcal{H}_B \otimes \mathcal{H}_{A'}\otimes \mathcal{H}_{B'}$ such that the total density matrix in enlarged Hilbert space $\rho_{AA'BB'}$ is given by $\rho_{AA'BB'}=\vert \psi_{AA'BB'}\rangle\langle\psi_{AA'BB'}\vert$. Such a pure state is called a purification of $\rho_{AB}$ if we have
\begin{align}
\rho_{AB}=Tr_{A'B'}\left(\vert \psi_{AA'BB'}\rangle\langle\psi_{AA'BB'}\vert\right).
\end{align}
\begin{figure}\label{fig2}
\centering
\includegraphics[width=80 mm]{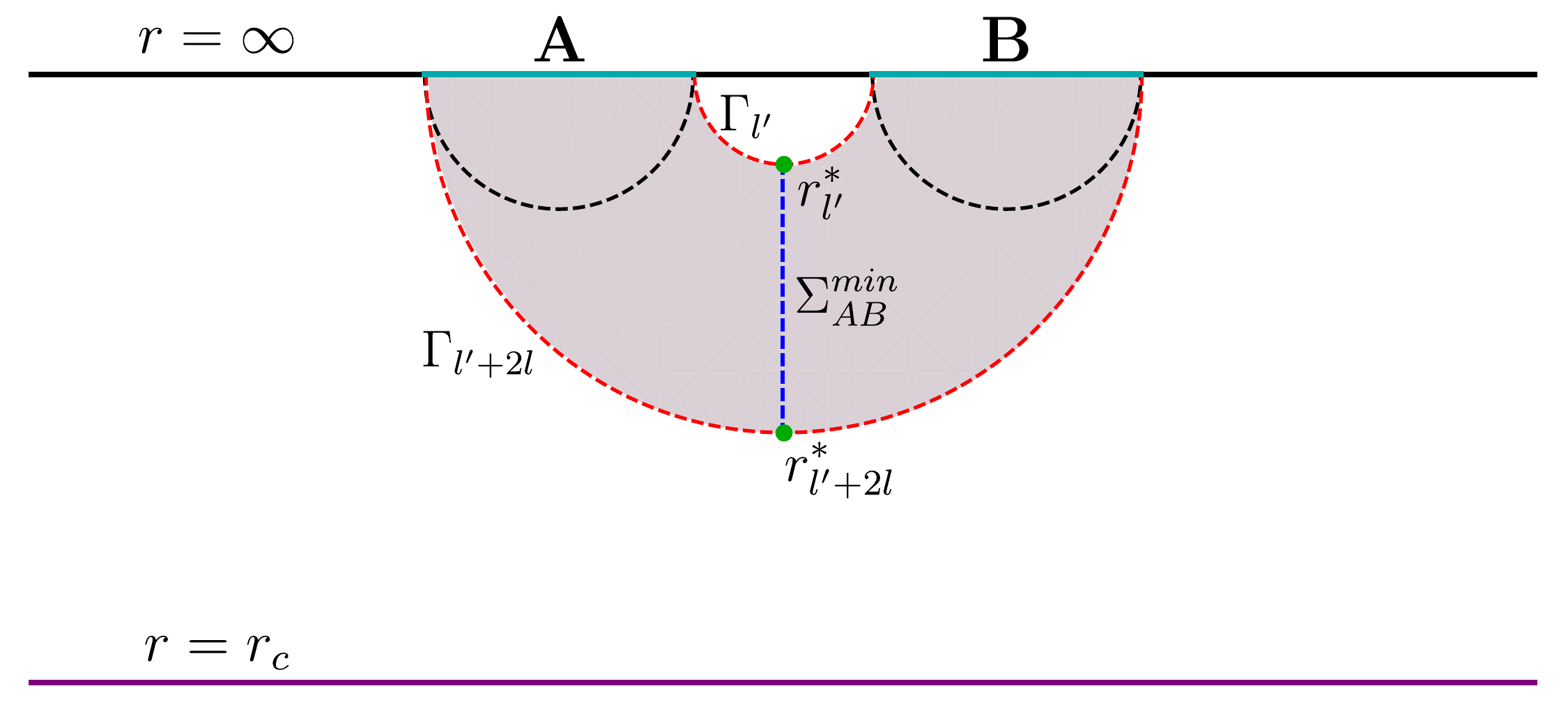}
\caption{The gray region shows the entanglement wedge dual to $\rho_{AB}$. The minimal surfaces, RT-surfaces, are denoted by $\Gamma$, the dashed curves.}
\label{fig2}
\end{figure}
Obviously there exist infinite ways to purify $\rho_{AB}$. The EoP is defined by minimizing the entanglement entropy $S_{AA'}$ over all purifications of $\rho_{AB}$ \cite{arXiv:quant-ph/0202044v3}
\begin{align}
E_p(\rho_{AB})=\underset{\vert \psi _{AA'BB'}\rangle}{\rm{min}}(S_{AA'}),
\end{align}
where $S_{AA'}$ is the entanglement entropy corresponding to the density matrix $\rho_{AA'}$ and $\rho_{AA'}={\rm{Tr}}_{BB'}\big[\left(|\psi\rangle_{ABA'B'}\right) \left({}_{ABA'B'}\langle\psi|\right)\big]$.

In the context of the gauge/gravity duality, it has been conjectured  that the EoP is dual to the entanglement wedge cross-section $E_w$ of $\rho_{AB}$ which is defined by \cite{Takayanagi:2017knl, Nguyen:2017yqw}
\begin{align}\label{EWCS}
E_w(\rho_{AB})=\frac{{\rm{Area}}(\Sigma_{AB}^{min})}{4G_N^{(d+2)}}.
\end{align}
where $\Sigma_{AB}^{min}$ is the minimal surface in the entanglement wedge $E_w(\rho_{AB})$, separating two regions of A and B, that ends on the RT-surface of $A\cup B$, the blue-dashed line in figure \ref{fig2}.
As a result, we have \cite{Takayanagi:2017knl, Nguyen:2017yqw}
\begin{align}\label{eop}
E_p(\rho_{AB})\equiv E_w(\rho_{AB}).
\end{align}
In \cite{Ghodrati:2021ozc} the EoP is used to probe the phase structure of the QCD and confining backgrounds. We would like to study the EoP in a non-conformal field theory at zero, low and high temperature. In order to compute the EoP, we use the holographic prescription given in \cite{Takayanagi:2017knl, Nguyen:2017yqw}. We consider two parallel strips with equal widths $l$ extended along $y$ and $z$ directions with length $L(\rightarrow \infty)$ and separated by a distance $l'$, see figure \ref{fig2}. Indeed, in this case $\Sigma_{AB}^{min}$ runs along the radial direction and connects the turning points of the minimal surfaces $\Gamma_{l'}$ and $\Gamma_{l'+2l}$. In order to obtain the analytical results, we focus in the some specific limits of the our models such as high ($l'\Lambda_c \ \& \ l\Lambda_c \ll 1$) and intermediate ($l'\Lambda_c \ll 1\ll l\Lambda_c $) energy at zero ($T=0$), low ($l'T \ \& \ lT \ll 1$) and intermediate ($l'T \ll 1\ll lT$) temperature. In the MAdS background \eqref{MAdS} the high and intermediate energy limits are equivalent to $r_c\ll r^*_{l'} \ \& \ r^*_{l'+2l}$ and $r_c\ll r^*_{l'}\ \& \ r^*_{l'+2l}\rightarrow r_c$, respectively and in the MBH background \eqref{MBH} the low temperature limit at high and intermediate energy are given in terms of balk parameters as ($r_c \ \& \ r_H \ll r^*_{l'} \ \& \ r^*_{l'+2l}$) and ($r_c \ \& \ r_H \ll r^*_{l'}$ and $r_H\ll r^*_{l'+2l}\rightarrow r_c$), respectively and the intermediate temperature at high and intermediate energy are given in terms of balk parameters as ($r_c \ \& \ r_H \ll r^*_{l'}$ and $r_c\ll r^*_{l'+2l}\rightarrow r_H$) and ($r_c \ \& \ r_H \ll r^*_{l'}$ and $r^*_{l'+2l}\rightarrow r_c =r_H$), respectively. These regimes are depicted in figure \ref{limitsT1}.
\subsection{EoP at zero temperature}
At zero temperature, using background \eqref{MAdS} and eqs. \eqref{EWCS} and \eqref{eop} we obtain
\begin{figure}
\centering
\subfloat[]{\includegraphics[scale=0.2]{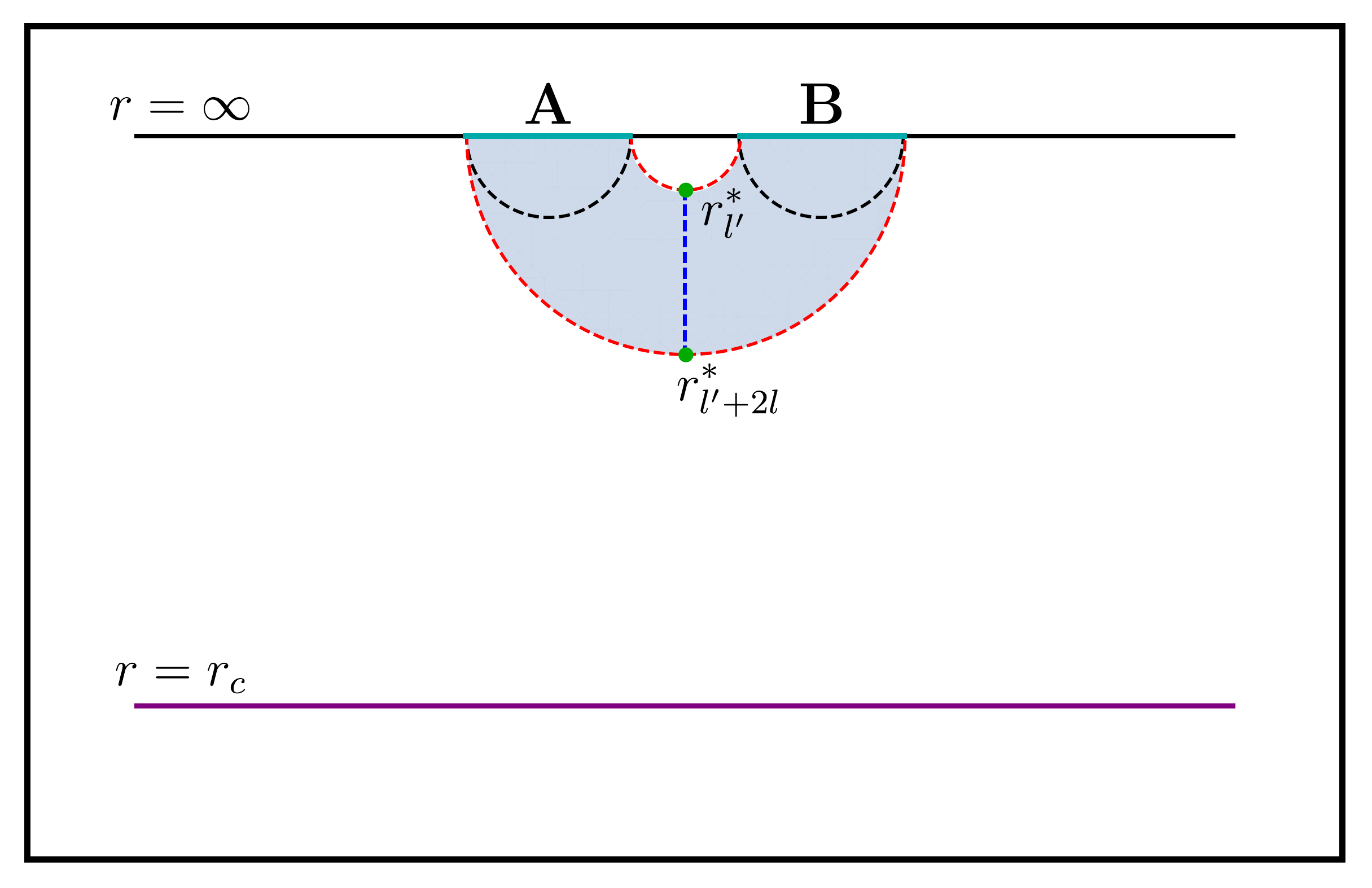}\label{a}}
\subfloat[]{\includegraphics[scale=0.2]{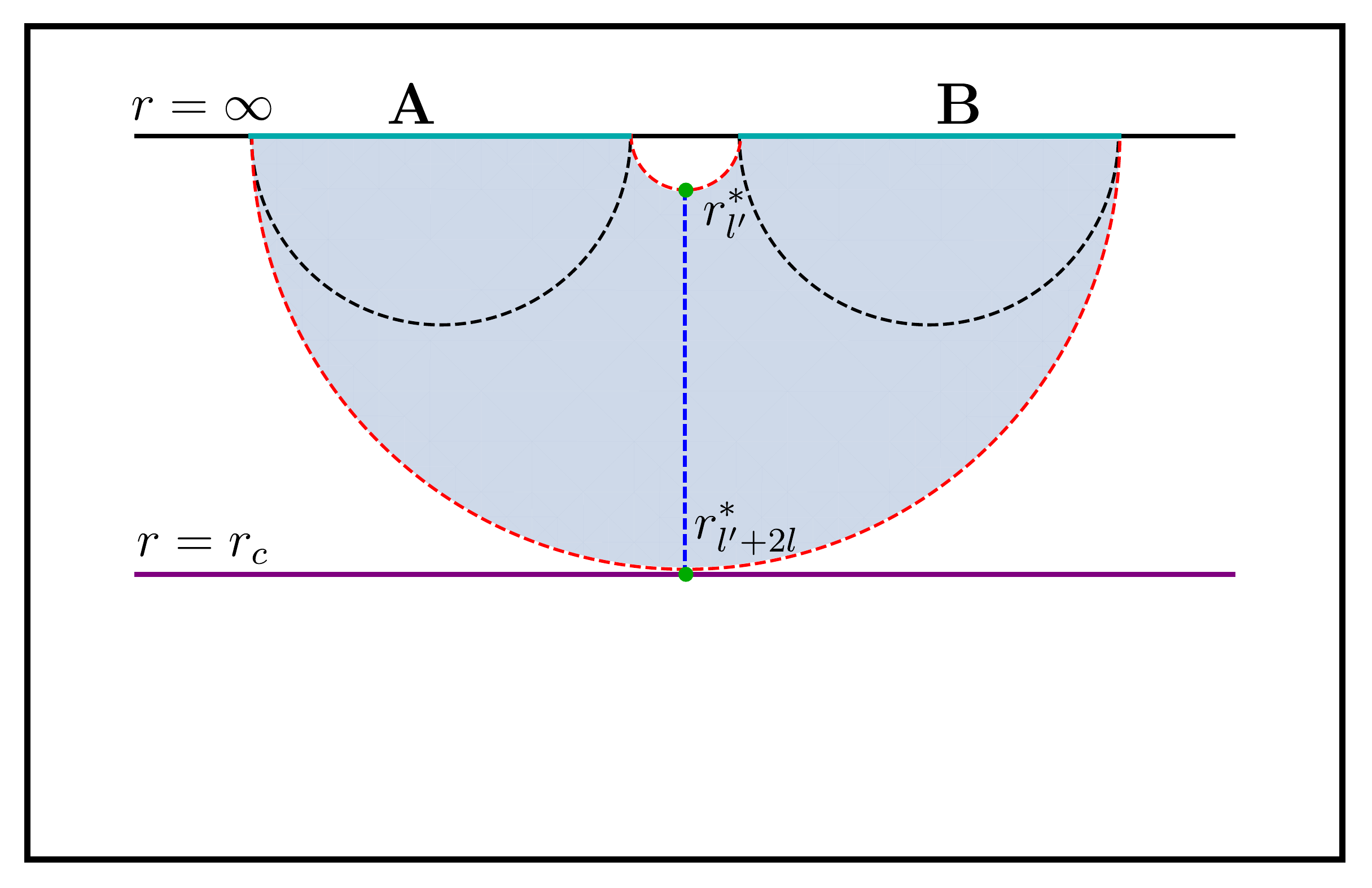}\label{b}}
\subfloat[]{\includegraphics[scale=0.2]{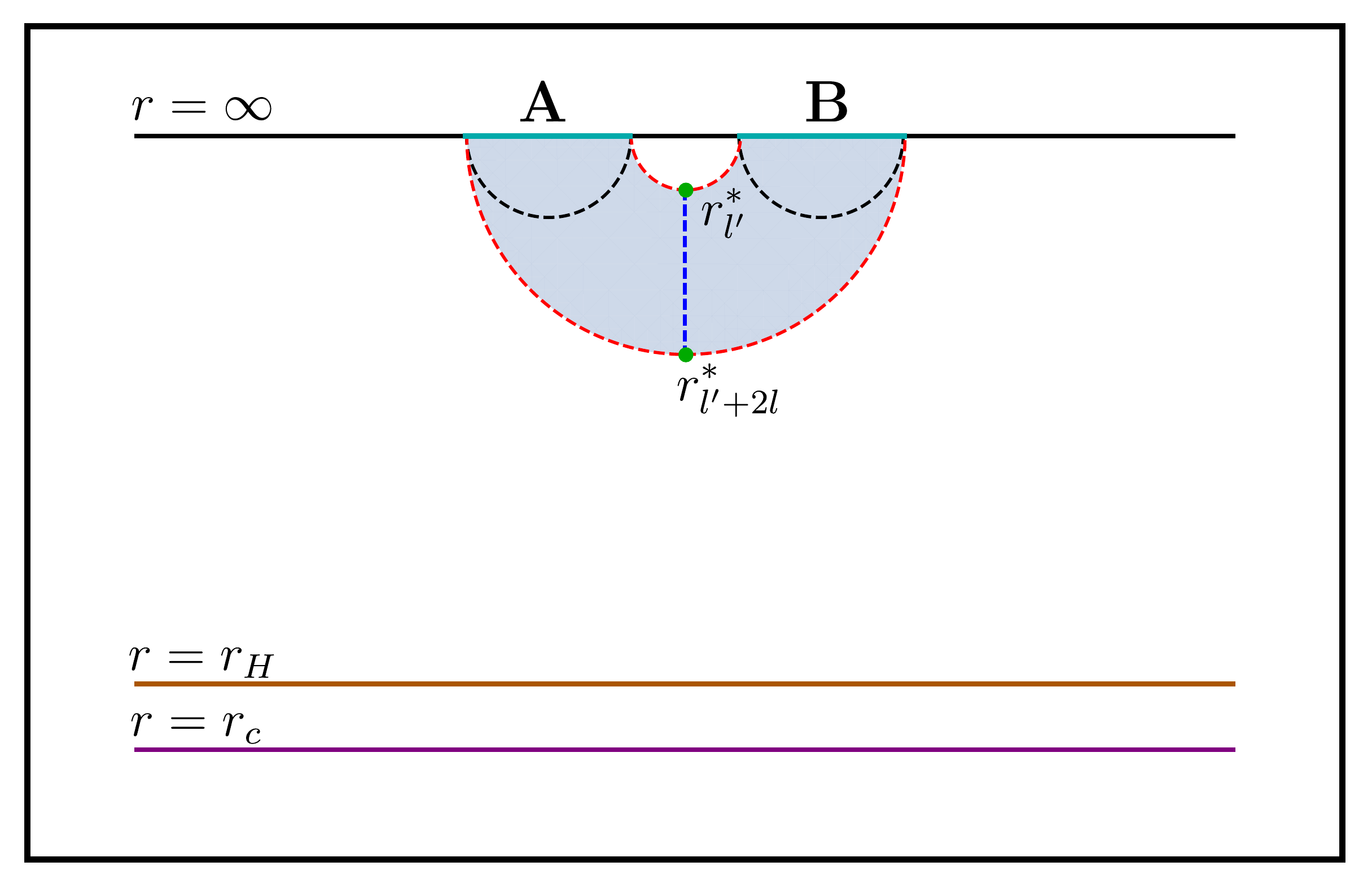}\label{c}}\\
\subfloat[]{\includegraphics[scale=0.2]{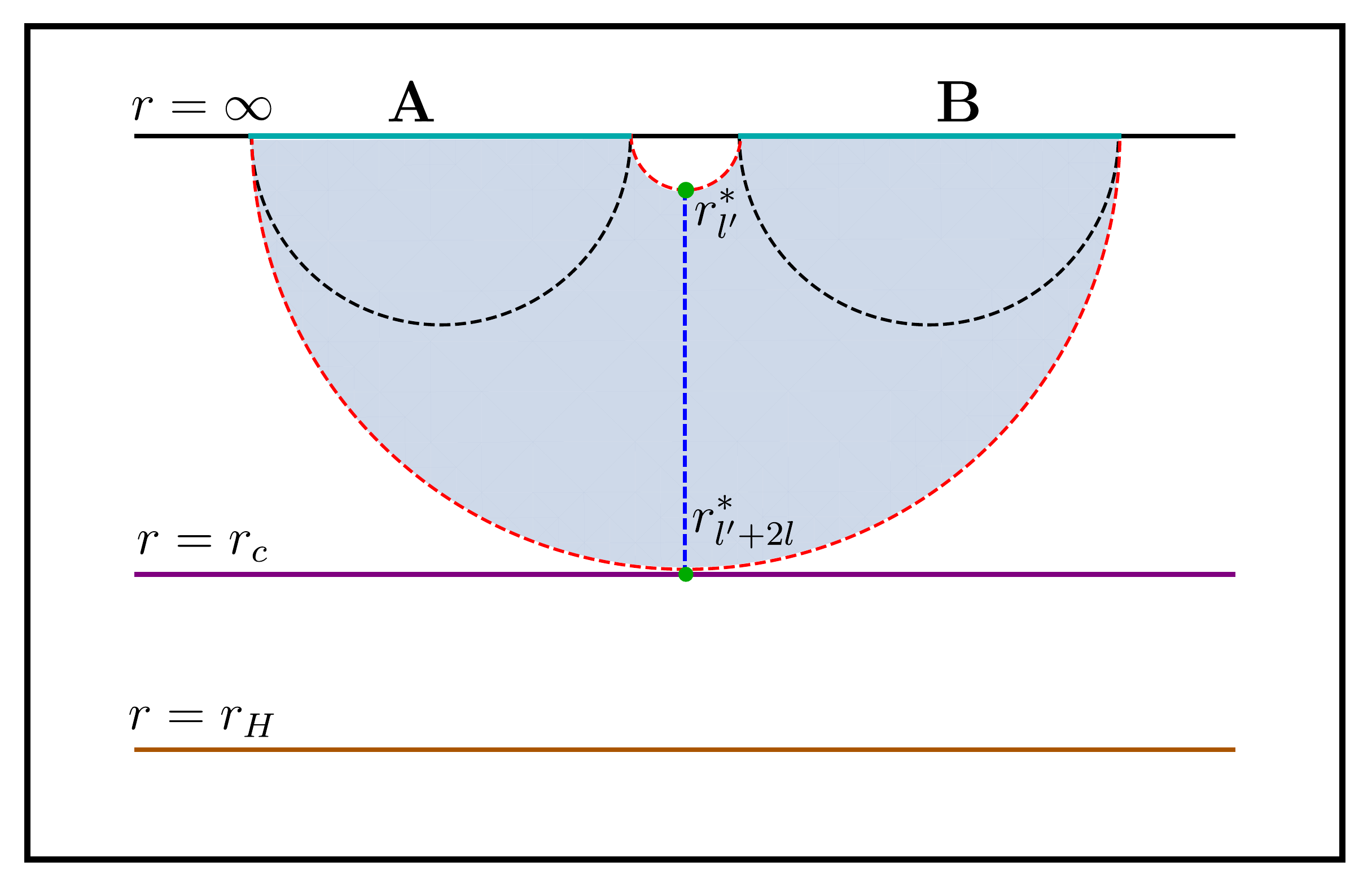}\label{d}}
\subfloat[]{\includegraphics[scale=0.2]{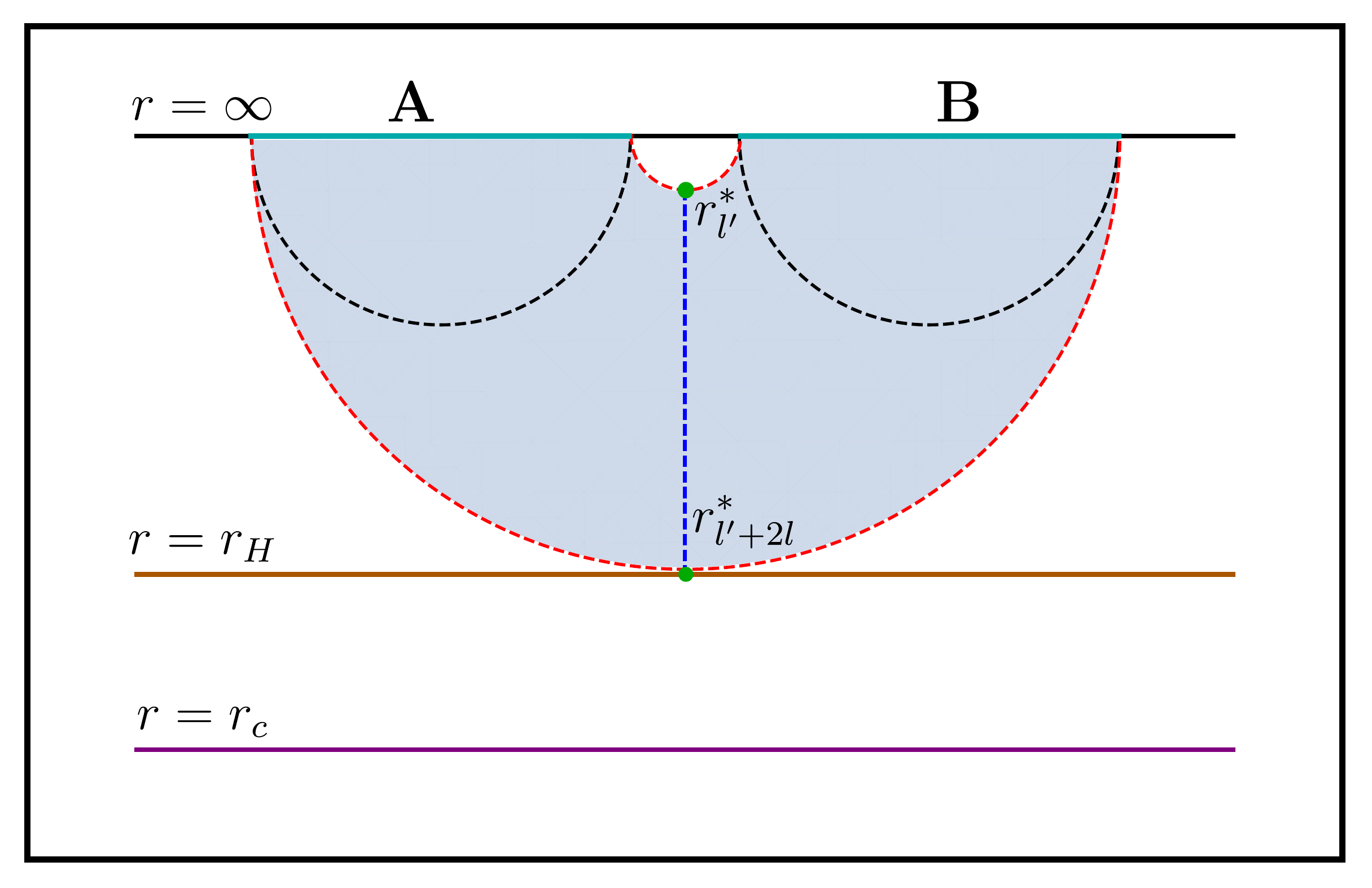}\label{e}}
\subfloat[]{\includegraphics[scale=0.2]{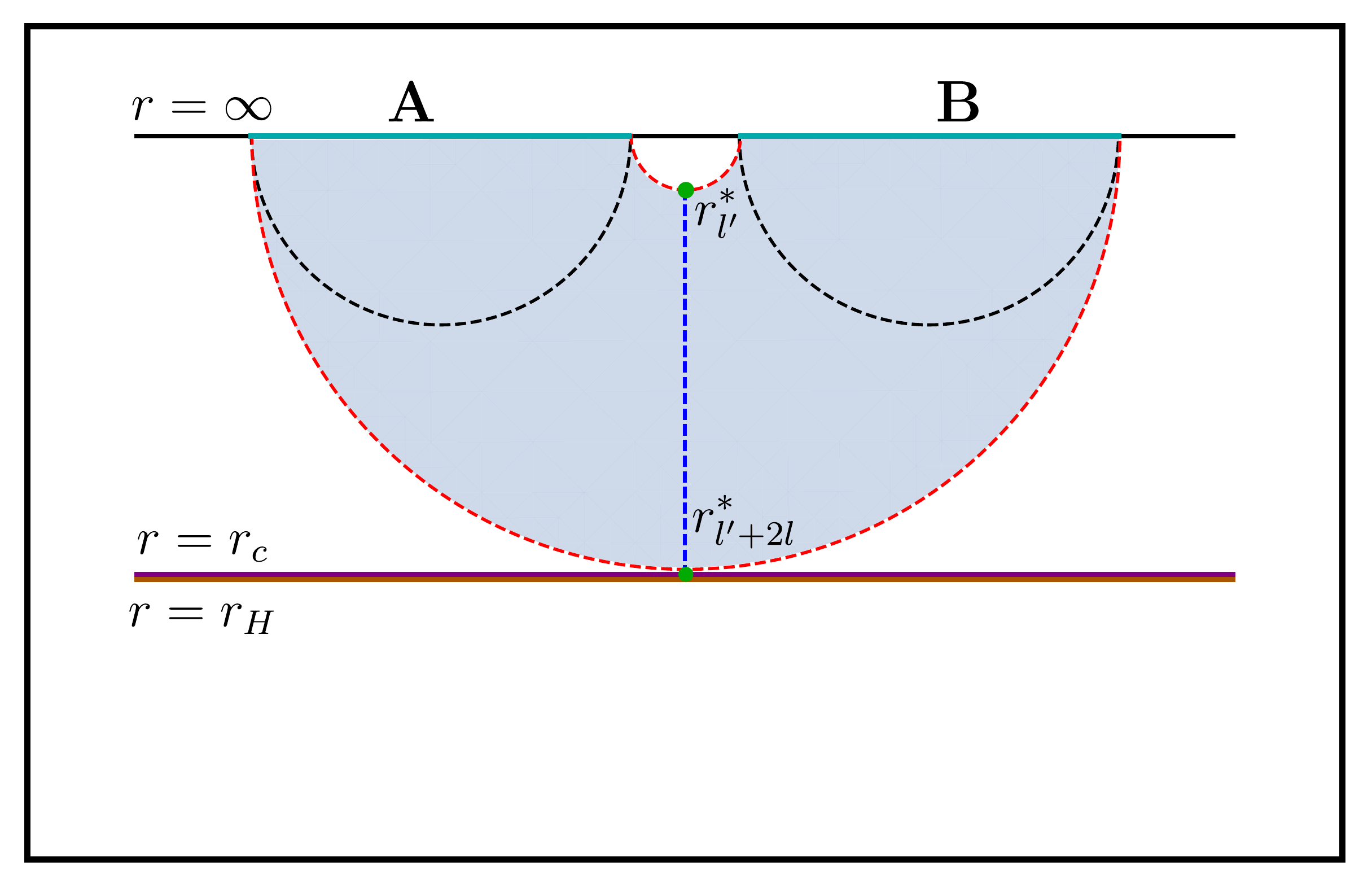}\label{f}}
\caption{(a): The high energy limit i.e. $l\Lambda_c \ \& \ l'\Lambda_c\ll 1$ or equivalently $r_c\ll r^*_{l'} \ \& \ r^*_{l'+2l}$ at zero temperature. (b): The intermediate energy limit i.e. $l'\Lambda_c\ll 1 \ll l\Lambda_c$ or equivalently $ r^*_{l'} \gg r_c \rightarrow r^*_{l'+2l}$ at zero temperature. (c): The low temperature limit i.e. $lT \ \& \ l'T\ll 1$ or equivalently $r_H\ll r^*_{l'} \ \& \ r^*_{l'+2l}$  at high energy. (d): The low temperature limit at intermediate energy. (e): The intermediate temperature i.e. $l'T\ll 1 \ll lT$ or equivalently $r^*_{l'} \gg r_H\rightarrow r^*_{l'+2l}$ at high energy. (f): The intermediate temperature at intermediate energy. This regime is equivalent to the transition limit i.e. $T\rightarrow \frac{\Lambda_c}{\sqrt{2}\pi}$ or equivalently $r_H\rightarrow r_c$.}
\label{limitsT1}
\end{figure}
\begin{align}\label{EoP1}
E_p&=\frac{L^2}{4G_N^{(5)}}\int _{r^*_{l'+2l}} ^{r^*_{l'}} r e^{\frac{3}{2}\left(\frac{r_c}{r}\right)^2}dr \cr
&=\frac{L^2}{16G_N^{(5)}}\Bigg\lbrace 2 {r^*_{l'}}^2 e^{\frac{3}{2}\left(\frac{r_c}{r^*_{l'}}\right)^2}-2 {r^*_{l'+2l}}^2 e^{\frac{3}{2}\left(\frac{r_c}{r^*_{l'+2l}}\right)^2}\cr
&\ \ \ \ \ \ \ \ \ + 3r_c^2Ei\Bigg(\frac{3}{2}\left(\frac{r_c}{r^*_{l'+2l}}\right)^2\Bigg)- 3r_c^2 Ei\Bigg(\frac{3}{2}\left(\frac{r_c}{r^*_{l'}}\right)^2\Bigg)\Bigg\rbrace ,
\end{align}
where $r^*_{l'}$ and $r^*_{l'+2l}$ in eq.\eqref{EoP1} denote the turning point of $\Gamma_{l'}$ and $\Gamma_{2l+l'}$, respectively, and $Ei(x)$ is the exponential integral defined for a real non-zero values of $x$
\begin{align}
Ei(x)=-\int_{-x}^\infty dt \frac{e^{-t}}{t}.
\end{align}
Using eq.\eqref{Lengthh1}, the relation between $r^*_{l'}$ and $l'$ is given by
\begin{align}\label{Length1}
l'(r^*_{l'})=\frac{2}{r^*_{l'}}\sum\limits_{n=0}^{\infty}\frac{\Gamma(n+\frac{1}{2})}{\sqrt{\pi}\Gamma(n+1)}\int_0^1 u^{6n+3} e^{3(n+\frac{1}{2})\left(\frac{r_c}{r^*_{l'}}\right)^2(1-u^2)} du,
\end{align}
where $u=\frac{r^*_{l'}}{r}$. Similar equation can be obtained for $r^*_{l'+2l}$ by replacing $l'$ and $r^*_{l'}$ with $l'+2l$ and $r^*_{l'+2l}$ in above equation, respectively. In order to find the EoP as a function of $l$ and $l'$, we should solve equation \eqref{Length1} for $r^*_{l'}$ and similarly find $r^*_{l'+2l}$ and then substitute these in equation \eqref{EoP1}. Since we can not analytically solve the equations to find $r^*_{l'}$ and $r^*_{l'+2l}$  as a function of $l'$ and $l$ we need to focus on the high and  intermediate energy limit. Note that, in the low energy limit, i.e. $1 \ll l\Lambda_c \ \& \ l' \Lambda_c$, we have disconnected configuration where the HMI and consequently the EoP vanishes for this case.  The high and intermediate energy limits in terms of bulk parameters is depicted in figure \ref{limitsT1}.
\subsubsection{High energy limit}
In the high energy limit, i.e. $l\Lambda _c\ \&\ l'\Lambda _c\ll 1$ or equivalently $r_c \ll r^*_{l'}\ \&\  r^*_{l'+2l}$, the extremal surfaces $\Gamma_{l'}$ and $\Gamma_{l'+2l}$ are restricted to be near the boundary, see figure \ref{a}. Therefore, the leading contribution to the EoP comes from the AdS boundary and the non-conformality effects appear as sub-leading terms which correspond to the deviation of the bulk geometry from pure AdS. 
We can easily obtain $r^*_{l'}$ and $r^*_{l'+2l}$ from eq.\eqref{turning1} and then substitute back these in eq.\eqref{EoP1}. Keeping up to second order in $l\Lambda_c $ and $l'\Lambda_c $ we finally obtain
\begin{align}\label{EoPhighE}
\hat{E}_p(l\Lambda_c,l'\Lambda_c)&\equiv\frac{4G_N^{(5)}E_p(l,l',l\Lambda_c,l'\Lambda_c)}{L^2\Lambda_c^2}\cr
&=2 a_1^2\left(\frac{1}{\left( l'\Lambda_c \right)^2}-\frac{1}{\left(l'\Lambda_c +2l\Lambda_c \right)^2}\right)+ C_1 l\Lambda_c(l\Lambda_c+l'\Lambda_c) \cr
&+\frac{3}{4}\log\left(\frac{l'\Lambda_c+2l\Lambda_c}{l'\Lambda_c}\right), \ \ \   C_1>0
\end{align}
where $\hat{E}_p(l\Lambda_c,l'\Lambda_c)$ is the redefined $E_p(l,l',l\Lambda_c,l'\Lambda_c)$ which is given by eq.\eqref{Ep-highE}  and $C_1$ is a numerical constant given by eq.\eqref{C1} in appendix \ref{Appendix3}.
The first term in eq.\eqref{EoPhighE} corresponds to the EoP obtained for the pure AdS$_5$ which is positive \cite{Jokela:2019ebz}. The other terms correspond to the non-conformal effects which are always positive and hence the non-conformal effects increase $\hat{E}_p(l\Lambda_c,l'\Lambda_c)$. From eq.\eqref{EoPhighE} we observe that $\hat{E}_p(l\Lambda_c,l'\Lambda_c)$ increase (decrease) by increasing $l\Lambda_c$ ($l'\Lambda_c$) for fixed $l'\Lambda_c$ ($l\Lambda_c$).
\subsubsection{Intermediate energy limit}
The intermediate energy limit is defined by $l'\Lambda_c \ll 1\ll l\Lambda_c$ or equivalently $r_c\ll r^*_{l'}$ and $r^*_{l'+2l}\to r_c$ which is depicted in figure \ref{b}. In this limit,  the extremal surface $\Gamma_{l'}$ is restricted to be near the boundary and the turning point of the extremal surface $\Gamma_{2l+l'}$ approaches $r_c$. Expanding the first and fourth terms in eq.\eqref{EoP1} up to the second order in $\frac{r_c}{r^*_{l'}}$, replacing $r^*_{l'}$ with eq.\eqref{turning1} and considering the limit of $r^*_{l'+2l}\to r_c$ we reach the following expression
\begin{align}\label{EoPE}
\hat{E}_p(l'\Lambda_c)\equiv\frac{4G_N^{(5)}E_p(l',l'\Lambda_c)}{L^2\Lambda_c^2}=\frac{1}{(l'\Lambda_c)^2}\bigg\lbrace 2a_1^2+&\left(C_2-\frac{3}{4}\log(l'\Lambda_c )\right)(l'\Lambda_c)^2\cr
&-\frac{C_1}{4}(l'\Lambda_c )^4\bigg\rbrace , \ \ \ \ \ \  C_1, C_2>0,
\end{align}
where $\hat{E}_p(l'\Lambda_c)$ is the redefined $E_p(l',l'\Lambda_c)$ which is given by eq.\eqref{Ep-intE} and $C_2$ is a numerical constant given by eq.\eqref{C2} in appendix \ref{Appendix3}. As we expect in the limit $l'\Lambda_c \ll 1\ll l\Lambda_c$, $\hat{E}_p(l'\Lambda_c)$ does not depend on the length of the subsystems $l$ which is coincides with \cite{BabaeiVelni:2019pkw,Amrahi:2020jqg}. The first term is the leading term diverges in the limit $l'\rightarrow 0$, where the two subsystems touch each other. The other two terms indicate the non-conformal effects which are always positive and hence the non-conformality increases $\hat{E}_p(l'\Lambda_c)$. From eq.\eqref{EoPE} we observe that $\hat{E}_p(l'\Lambda_c)$ decrease by increasing $l'\Lambda_c $.
\subsection{EoP at finite temperature}
In order to investigate the thermal behavior of the EoP in the MBH background, we use background \eqref{MBH} and develop a systematic expansion at low ($lT \ \&\ l' T \ll 1$) and intermediate($l'T \ll 1\ll lT$) temperature  in the high energy limit ($l\Lambda _c\ \&\ l'\Lambda _c\ll 1$)  and  in the intermediate energy limit ($l'\Lambda_c  \ll 1\ll l\Lambda_c $). Using eqs. \eqref{EWCS} and \eqref{eop} and following the same calculations in the previous section we finally reach the following expression
\begin{align}\label{EopT}
E_p=\frac{L^2}{4G_N^{(5)}}\int _{r^*_{l'+2l}} ^{r^*_{l'}} r e^{\frac{3}{2}\left(\frac{r_c}{r}\right)^2}\left(1-\frac{r_H^4}{r^4}\right)^{-\frac{1}{2}} dr,
\end{align}
where $r^*_{l'}$ and $r^*_{l'+2l}$ are given by eq.\eqref{LengthhT} by replacing $l$ with $l'$ and $l'+2l$, respectively. Unfortunately, the above integral can not be solved analytically. In order to calculate eq.\eqref{EopT}, we use eq.\eqref{expansionn} and the expansion of the exponential function $e ^y=\sum\limits_{m=0}^{\infty}\frac{y^m}{\Gamma(m+1)}$ by identifying $x=-\frac{r_H^4}{r^4}$ and  $y=\frac{3}{2}\left(\frac{r_c}{r^*_{l'}}\right)^2$ and we finally obtain
\begin{align}\label{EoPT1}
E_p=\frac{L^2}{4G_N^{(5)}}\sum\limits_{m=0}^{\infty}\sum\limits_{n=0}^{\infty}\frac{(\frac{3}{2})^m\Gamma(n+\frac{1}{2})r_c^{2m}r_H^{4n}}{\sqrt{\pi}\Gamma(n+1)\Gamma(m+1)}\int_{r^*_{l'+2l}} ^{r^*_{l'}} r^{1-4n-2m}dr .
\end{align}
One can check that $\vert x\vert < 1$ for any allowable values of background parameters, $r_H<  r^*_{l'}$ and $r_c < r^*_{l'}$. Using eq.\eqref{lengthhT1} the relation between $r^*_{l'}$ and $l'$ is given by
\begin{align}\label{lengthT1}
l'(r^*_{l'})=\frac{2}{r^*_{l'}}\sum\limits_{m=0}^{\infty}\sum\limits_{n=0}^{\infty}\frac{\Gamma(n+\frac{1}{2})\Gamma(m+\frac{1}{2})}{\pi \Gamma(n+1)\Gamma(m+1)}\left(\frac{r_H}{r^*_{l'}}\right)^{4m}\int_0^1 u^{6n+4m+3} e^{3(n+\frac{1}{2})\left(\frac{r_c}{r^*_{l'}}\right)^2(1-u^2)} du, 
\end{align}
where $u=\frac{r^*_{l'}}{r}$. The relation between $r^*_{l'+2l}$ and $l'+2l$ can be obtained easily by replacing $r^*_{l'}$ and $l'$ in the above equation with $r^*_{l'+2l}$ and $l'+2l$, respectively. Now, we should solve eq.\eqref{lengthT1} for $r^*_{l'}$ and similar equation for $r^*_{l'+2l}$ and then substitute back them in eq.\eqref{EoPT1} to get the EoP in terms of $l$ and $l'$. In order to do analytical calculations we consider the low and intermediate temperature  in the high energy limit and  in the intermediate energy limit, see figure \ref{limitsT1}. In the high temperature limit i.e. $1 \ll lT \ \&\ l' T$ and in the low energy limit i.e. $1 \ll l\Lambda_c \ \&\ l' \Lambda_c$ we have a disconnected configuration and hence the HMI and consequently the EoP become zero in this limits. 
\subsubsection{Low temperature at high energy}
Here, we focus on the limit of low temperature i.e. $lT \ \&\ l' T\ll 1$ at the high energy i.e. $l\Lambda _c\ \&\ l'\Lambda _c\ll 1$, see figure \ref{c}. In this regime, the leading contribution to the EoP comes from the near boundary expansion and the thermal and non-conformal corrections appear as sub-leading terms. We want to get the EoP in terms of $l$ and $l'$ and hence we should have $r^*_{l'}$ and $r^*_{l'+2l}$ in terms of $l'$ and $l'+2l$ , respectively, which can be obtained from eq.\eqref{rstar}.
By taking the integral in eq.\eqref{EoPT1} and keeping up to the second order in $l\Lambda_c$, $l'\Lambda_c$, $lT$ and $l'T$, we obtain
\begin{align}\label{EoPlowThighE}
\tilde{E}_p(l\Lambda_c &,l'\Lambda_c ,lT,l'T)\equiv\frac{4G_N^{5)}E_p(l,l',l\Lambda_c,l'\Lambda_c ,lT,l'T)}{L^2\Lambda_c T}\cr
&=2 a_1^2\left(\frac{1}{( l'\Lambda_c )(l'T)}-\frac{1}{(l'\Lambda_c +2l\Lambda_c )(l'T+2lT)}\right)+C_3\frac{(lT)^2(lT+l'T)}{l\Lambda_c}\cr
&+\frac{3(l\Lambda_c)}{4(lT)}\log\left(\frac{l\Lambda_c+2l'\Lambda_c}{l'\Lambda_c}\right)+C_1 \frac{(l\Lambda_c)^2(l\Lambda_c+l'\Lambda_c)}{lT} , \ \ \ C_1>0, C_3<0,
\end{align}
where $\tilde{E}_p(l\Lambda_c,l'\Lambda_c ,lT,l'T)$ is the redefined $E_p(l,l',l\Lambda_c,l'\Lambda_c ,lT,l'T)$ which is given by eq.\eqref{Ep-lowThighE} and numerical constant $C_3$ is given by eq.\eqref{C3} in appendix \ref{Appendix3}.
The first two terms, including $a_1$ and $C_3$ correspond to the AdS and AdS black hole, respectively and the next two terms are the non-conformal effects which are always positive and hence the non-conformality increase $\tilde{E}_p(l\Lambda_c,l'\Lambda_c ,lT,l'T)$. Since $C_3$ is a negative constant the second term is always negative and hence the thermal fluctuations decrease $\tilde{E}_p(l\Lambda_c,l'\Lambda_c ,lT,l'T)$. Since the EoP is a measure of total correlation between two subsystems, the thermal fluctuations promote disentangling between them in this regime. In the following we represent the results corresponding to the mentioned limits:
\begin{itemize}
\item $\Lambda_c\neq 0$ and $T=0$: In this case, we reproduce the results obtained for MAdS background which is dual to non-conformal field theory at zero temperature. The leading term corresponds to the conformal EoP and the non-conformal corrections appear as the sub-leading terms. These corrections are always positive and hence non-conformality increase $\hat{E}_p(l\Lambda_c,l'\Lambda_c)$.
\item $\Lambda_c=0$ and $T\neq 0$: In this case, we reproduce the previous results obtained for AdS black hole, see for example \cite{BabaeiVelni:2019pkw}. The leading term corresponds to the zero temperature EoP and the finite temperature correction appears as the sub-leading term. The constant $C_3$ is negative and hence the finite temperature corrections decrease $\tilde{E}_p(lT,l'T)$ of our configuration.
\item $\Lambda_c= 0$ and $T=0$:  We reach the previous results obtained for pure AdS$_5$ which is corresponding to the EoP of our configuration in zero temperature conformal field theory \cite{Jokela:2019ebz}. Obviously, this term is positive.
\end{itemize}
Here, we can study  $\tilde{E}_p(l\Lambda_c,l'\Lambda_c ,lT,l'T)$ near the transition point. By taking the transition limit $T\rightarrow \frac{\Lambda_c}{\sqrt{2}\pi}$ from eq.\eqref{EoPlowThighE}, we get 
\begin{align}\label{EoPtra}
\tilde{E}_p(l\Lambda_c,l'\Lambda_c,lT,l'T)\bigg\vert_{T\rightarrow\frac{\Lambda_c}{\sqrt{2}\pi}}&=\sqrt{2}\pi\Bigg\lbrace 2 a_1^2\left(\frac{1}{\left( l'\Lambda_c \right)^2}-\frac{1}{\left(l'\Lambda_c +2l\Lambda_c \right)^2}\right)\cr
&+\left(C_1+\frac{C_3}{4\pi^4}\right)  (l\Lambda _c)(l\Lambda _c+l'\Lambda _c)+\frac{3}{4}\log\left(\frac{l'+2l}{l'}\right)\Bigg\rbrace . \ \ \ \
\end{align}
Similar to the previous sections, we fix $l\Lambda_c$ and $l'\Lambda_c$ and compare $\hat{E}_p(l\Lambda_c,l'\Lambda_c)$ at zero temperature, eq.\eqref{EoPhighE}, with $\tilde{E}_p(l\Lambda_c,l'\Lambda_c,lT,l'T)$ at finite temperature in the transition limit, eq.\eqref{EoPtra}. We get
\begin{align}
\frac{\tilde{E}_p(l\Lambda_c,l'\Lambda_c,lT,l'T)}{\sqrt{2}\pi}\bigg\vert _{T\rightarrow\frac{\Lambda_c}{\sqrt{2}\pi}}- \hat{E}_p(l\Lambda_c,l'\Lambda_c) =\frac{C_3}{4\pi^4 }  (l\Lambda_c)^2(l\Lambda_c+l'\Lambda_c)^2<0.
\end{align}
This result shows that near the transition point, the correlation between two subsystems is larger at zero temperature than the finite temperature and hence the state at zero temperature is the favorable one. In the high energy limit, since the energy of the subsystems is so much bigger than the energy of the phase transition point, see figure \ref{c}, we expect that the zero temperature state is more favorable than the finite temperature one.
\subsubsection{Low temperature at intermediate energy }\label{lowTintE}
In this subsection we consider the limit of low temperature, i.e. $lT \ \&\ l'T\ll 1$ at the intermediate energy i.e. $l'\Lambda_c \ll 1\ll l \Lambda_c $, see figure \ref{d}.  In this limit $T \ll \Lambda _c$ ($r_H \ll r_c$). In this regime, the extremal surface $\Gamma_{l'}$ is restricted to be near the boundary and the turning point of the extremal surface $\Gamma_{2l+l'}$ approaches $r_c$. Using eq. \eqref{EoPT1} and \eqref{rstar}, doing some computations and keeping up to the 4th order in $l'\Lambda_c$ and $l'T$ we reach the following expression (see appendix \ref{Appendix3} for the details of the calculations)
\begin{figure}
\centering
\includegraphics[scale=0.38]{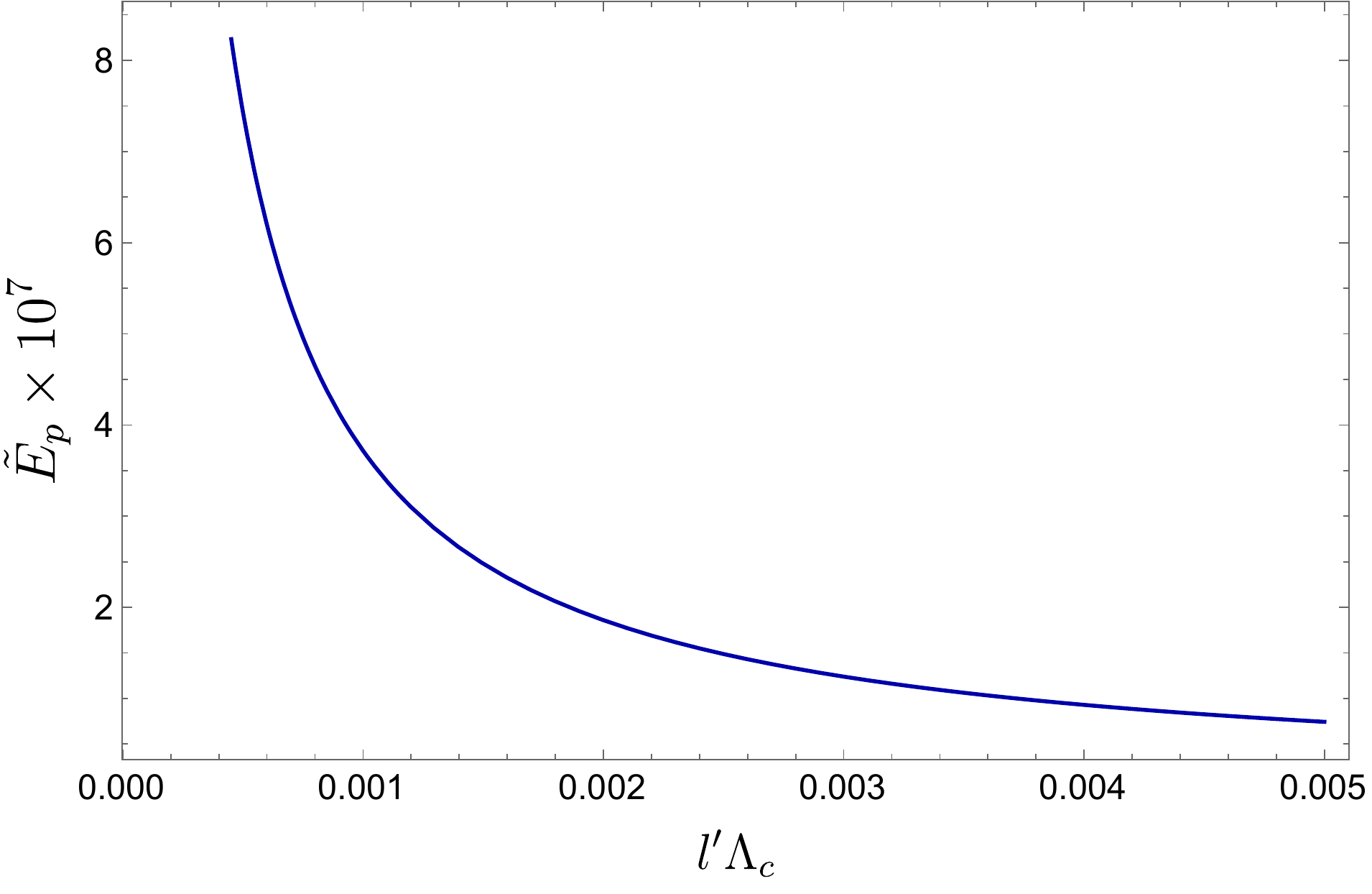}
\includegraphics[scale=0.38]{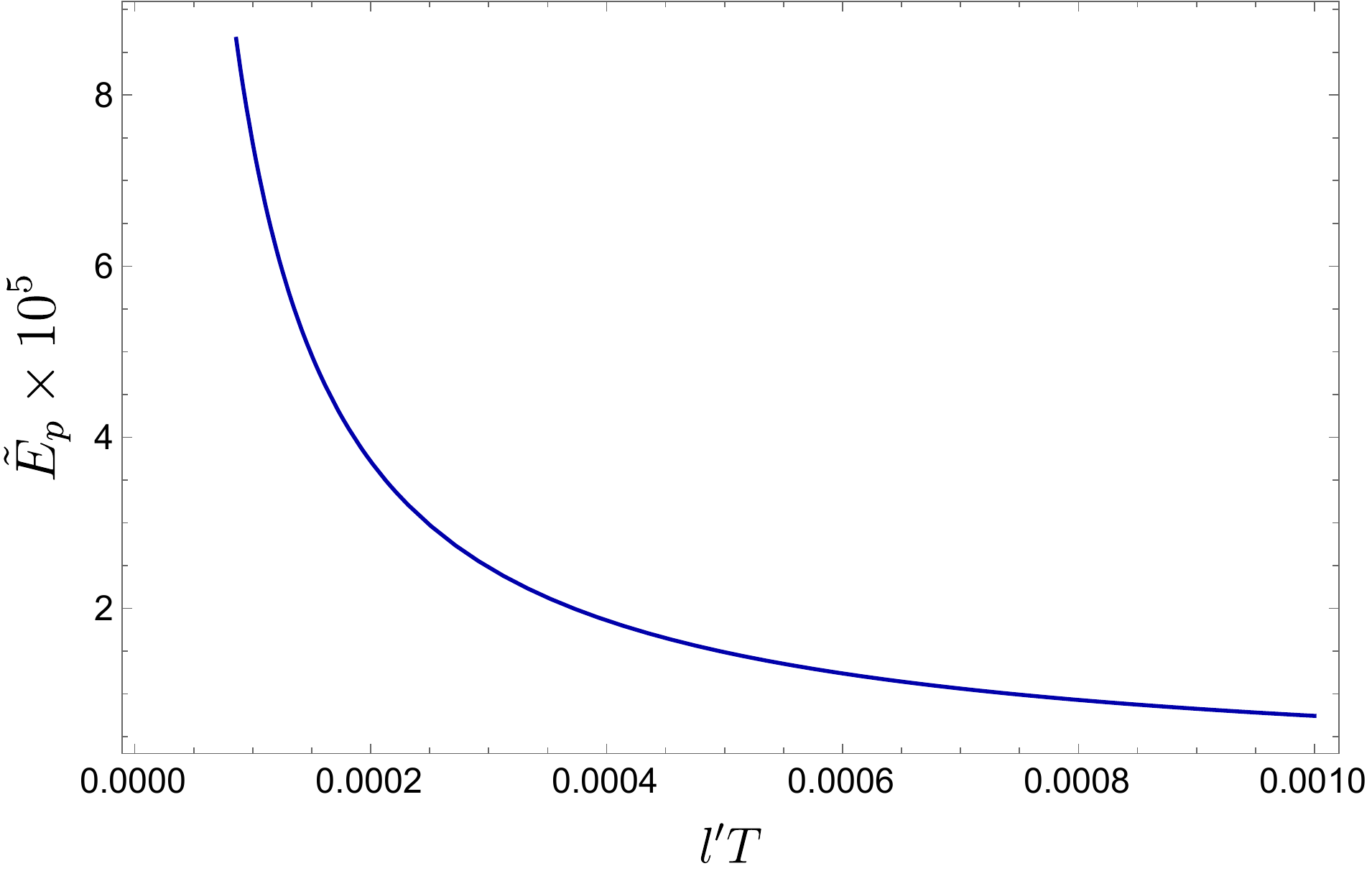}
\caption{Left: $\tilde{E}_p(l'\Lambda_c,l'T)$ in terms of $l'\Lambda _c$ for fixed $l'T=10^{-5}$ in the low temperature limit at intermediate energy. Right: $\tilde{E}_p(l'\Lambda_c,l'T)$ in terms of $l'T$ for fixed $l'\Lambda_c=0.005$ in the low temperature limit at intermediate energy.}
\label{LowTIntermediateE}
\end{figure}
\begin{align}\label{EoPlTiE}
\tilde{E}_p(l'\Lambda_c,l'T)&\equiv\frac{4G_N^{(5)}E_p(l',l'\Lambda_c,l'T)}{L^2\Lambda_c T}=\frac{1}{(l'\Lambda_c)(l'T)}\Bigg\lbrace  2a_1^2-\frac{C_3}{4}(l'T)^4\cr
&+\Bigg[C_2-\frac{3}{4}\log(l'\Lambda_c)+C_6\left(\frac{l'T}{l'\Lambda_c}\right)^4+C_4(l'T)^4
\Bigg](l' \Lambda_c)^2\cr
&-\left[\frac{C_1}{4}+C_5(l'T)^4\right](l'\Lambda _c)^4\Bigg\rbrace , \ \ C_1,C_2,C_4 >0, C_3,C_5,C_6<0
\end{align}
where $\tilde{E}_p(l'\Lambda_c,l'T)$ is the redefined $E_p(l',l'\Lambda_c,l'T)$ which is given by eq.\eqref{Ep-lowTintE} and the numerical constants  $C_4$, $C_5$ and $C_6$ are given by eq.\eqref{C4} in appendix \ref{Appendix3}. As one can see, $l$ does not play any role in our final result which is in complete agreement with the results obtained in \cite{Amrahi:2020jqg}. The first term in eq.\eqref{EoPlTiE} which is the leading term, obeys an area law divergence in $l'\rightarrow 0$ limit, where the two subsystems touch each other. The second term, including $C_3$, is dominant term in the finite temperature corrections which is always positive and hence the finite temperature corrections increase $\tilde{E}_p(l'\Lambda_c,l'T)$ in this regime unlike the low temperature limit at high energy. The term including $C_2$ and the logarithmic term are the dominant terms in the non-conformal effects. Since these terms are always positive the non-conformal effects increase $\tilde{E}_p(l'\Lambda_c,l'T)$. From eq.\eqref{EoPlTiE} we observe that for fixed $l'\Lambda_c$ ($l'T$) $\tilde{E}_p(l'\Lambda_c,l'T)$ decreases by increasing $l'T$ ($l'\Lambda_c$). This result depicted clearly in figure \ref{LowTIntermediateE} where we plot $\tilde{E}_p(l'\Lambda_c,l'T)$ in terms of $l'\Lambda_c$ ($l'T$) for fixed $l'T$ ($l'\Lambda_c$).  If we set $T=0$, we reproduce the previous results which is obtained for MAdS background in the intermediate energy limit eq.\eqref{EoPE}.

From eqs. \eqref{EoPlowThighE} and \eqref{EoPlTiE} one can see that there is a crossover regime between high and intermediate energy in the low temperature where the sign of the thermal effect in the EoP changes. In the low temperature limit at high energy the sign of the thermal effects in the EoP and HMI are similar and these effects decrease the EoP and HMI. In the regime of low temperature at intermediate energy we have $l'\ll l$ and $T\ll \Lambda_c$. If we consider $l'$ to be very small the EoP approach to the HEE and hence it is reasonable that the thermal effects are positive in this regime.
\subsubsection{Intermediate temperature at high energy}\label{intThighE}
In this subsection we consider the intermediate  temperature limit i.e. $l'T \ll 1\ll lT $ at the high energy limit i.e  i.e. $l\Lambda _c\ \&\ l'\Lambda _c\ll 1$. In this limit $\Lambda_c \ll T$ ($r_c \ll r_H$), see figure \ref{e}. In this regime, the extremal surface $\Gamma_{l'}$ is restricted to be near the AdS boundary and the turning point of the extremal surface $\Gamma_{l'+2l}$ approaches $r_H$. Doing some calculations and keeping up to the 4th order in $l'\Lambda_c$ and $l'T$, we obtain the following expression (the details of the calculation can be found in appendix \ref{Appendix3})
\begin{align}\label{EoPIntT}
\tilde{E}_p(l'\Lambda_c,l'T)&\equiv \frac{4G_N^{(5)}E_p(l',l'\Lambda_c,l'T)}{L^2\Lambda_c T}=\frac{1}{(l'\Lambda_c)(l'T)}\Bigg\lbrace  2a_1^2-\frac{C_3}{4}(l'T)^4\cr
&\ \ \ \ \ \ \ \ +\Bigg[C_7-\frac{3}{4}\log(l'T)+\frac{9}{128\pi}\left(\frac{l'\Lambda_c}{l'T}\right)^2-C_4(l'T)^4
\Bigg](l' \Lambda_c)^2\cr
&\ \ \ \ \ \ \ \  -\left[\frac{C_1}{4}+C_5(l'T)^4\right](l'\Lambda _c)^4\Bigg\rbrace, \ \ C_1,C_4>0, C_3,C_5,C_7<0, \ \ \
\end{align}
where $\tilde{E}_p(l'\Lambda_c,l'T)$ is the redefined $E_p(l',l'\Lambda_c,l'T)$ which is given by eq.\eqref{Ep-intThighE} and the numerical constant $C_7$ is given by eq.\eqref{C7} in appendix \ref{Appendix3}. The first and second terms are the same as in the previous subsection. It is seen that $\tilde{E}_p(l'\Lambda_c,l'T)$ does not depend on the length of the subsystems and diverges in $l'\rightarrow 0$ limit. 
Since $C_3$ is a negative constant the second term, the dominant term, is always positive and hence the finite temperature correction increases $\tilde{E}_p(l'\Lambda_c,l'T)$ unlike the HMI which the temperature effects decrease the HMI in this regime. This behavior is coincided with the results obtained in \cite{BabaeiVelni:2019pkw}. The logarithmic term is the dominant term in the non-conformal effects. This terms is always positive and hence the non-conformal effects increase $\tilde{E}_p(l'\Lambda_c,l'T)$. For $\Lambda_c=0$  we reproduce the previous results obtained for AdS block holes \cite{BabaeiVelni:2019pkw}. 
From eq.\eqref{EoPIntT} we observe that $\tilde{E}_p(l'\Lambda_c,l'T)$  decreases by increasing $l'\Lambda_c $ ($l'T$) for fixed  $l'T$ ($l'\Lambda_c $).
\subsubsection{Intermediate temperature at intermediate energy }
In this subsection we consider the intermediate  temperature limit i.e. $l'T \ll 1\ll lT$  at the intermediate energy  i.e. $l'\Lambda_c \ll 1\ll l\Lambda_c $, see figure \ref{f}. In other words, in the intermediate energy limit  we take $r_c\rightarrow r_H$ ($T\rightarrow\frac{\Lambda_c }{\sqrt{2}\pi}$) which is  the transition limit. Doing some calculations and keeping up to the 4th order in $l'\Lambda_c$ we obtain (the details of calculations can be found in appendix \ref{Appendix3})
\begin{align}\label{EoPinttra}
\tilde{E}_p(l'\Lambda_c,l'T)\bigg\vert_{T\rightarrow \frac{\Lambda_c}{\sqrt{2}\pi}}&\equiv\frac{4G_N^{(5)}E_p(l',l'\Lambda_c,l'T)}{L^2\Lambda_c T}\bigg\vert_{T\rightarrow \frac{\Lambda_c}{\sqrt{2}\pi}}\cr
&=\frac{\sqrt{2}\pi}{(l'\Lambda_c)^2}\bigg\lbrace 2a_1^2+\left(C_7-\frac{3}{4}\log(l'\Lambda_c)\right)(l'\Lambda_c)^2 \cr
& \ \ \ \ \ \ \ \ \ \ \ \ \ \ \ \ +C_8(l'\Lambda_c)^4\bigg\rbrace ,\ \  C_7 >0,\ C_8<0, \ \ \ \ \
\end{align}
where $\tilde{E}_p(l'\Lambda_c,l'T)$ is the redefined $E_p(l',l'\Lambda_c,l'T)$ in the transition limit  which is given by eq.\eqref{Ep-intTintE} and numerical coefficients $C_7$ and $C_8$ are given by eq.\eqref{C5} in appendix \ref{Appendix3}.
Similar to the two previous subsections, $\tilde{E}_p(l'\Lambda_c,l'T)$ does not depend on the length of the subsystems. The first term, including $a_1$, is the leading term which diverges in $l'\rightarrow 0$ limit and the second term, including $C_7$ and logarithmic term, is dominant term in the non-conformal effects which is always positive and hence the non-conformality increases $\tilde{E}_p(l'\Lambda_c,l'T)$.
To get a better understanding of the transition limit of the $\tilde{E}_p(l'\Lambda_c,l'T)$  in the intermediate temperature limit at intermediate energy, we fixed $(l'\Lambda_c)$ and compare eq.\eqref{EoPinttra} with the result obtained in eq.\eqref{EoPE}. We have
\begin{align}
 \tilde{E}_p(l'\Lambda_c,l'T)\big\vert _{T\rightarrow\frac{\Lambda_c}{\sqrt{2}\pi}}- \hat{E}_p (l'\Lambda_c)=0.773+0.025(l'\Lambda_c)^2>0 .
\end{align}
From the above equation we observe that in the intermediate energy limit, unlike the high energy limit, near the transition point the two subsystems are less correlated at zero temperature and hence the state at finite temperature is favorable. In the intermediate energy limit, since the energy of the subsystems is close to the energy of the phase transition point, see figure \ref{f}, we expect that in this limit the state at finite temperature is more favorable than the zero temperature.
\section{Conclusion}\label{section6}
In this paper, we consider a non-conformal field theory at zero and finite temperature which has holographic dual. Using gauge/gravity duality, we study the HEE, HMI and EoP in the backgrounds which are dual to the mentioned model. In order to obtain analytical expressions for these observables, we use a systematic expansion and consider some specific limits such as high and intermediate energy at zero, low, intermediate and high temperature. In the following we present the results  obtained for the HEE, HMI and EoP:
\begin{itemize}
\item Entanglement entropy: We calculate the redefined HEE using the RT-prescription for a strip with width $l$ and length $L$ in the high energy limit at the zero, low and high temperature. \textit{In the all of these regimes the non-conformal (thermal) effects decrease (increase) the HEE}. Due to the non-conformality a logarithmic divergent term appeares in the HEE. At high temperature, two terms proportional to $T^2$ and $T^3$ appear in the HEE which the first term scales with the area $L^2$ and the second term scales with the volume of the strip $lL^2$ and hence the first term corresponds to the entanglement entropy and the second term corresponds to the thermal entropy. In the high energy limit and near the phase transition point ($T\rightarrow \frac{\Lambda_c}{\sqrt{2}\pi}$), the subsystems $A$ and $\bar{A}$ are less entangled at zero temperature than the finite temperature and hence  the state at zero temperature is the favorable one.
\item Mutual information: We calculate the redefined HMI for a symmetric configuration including two disjoint strip with equal width $l$ which are separated by the distance $l'$ in the high energy limit at zero, low and intermediate temperature. \textit{In the all of these regimes the non-conformal (thermal) effects increase(decrease) the redefined HMI}. Near the phase transition point, the HMI of the two subsystems is larger at zero temperature than the finite temperature case. Therefore, at high energy the state at zero temperature is the favorable one.
\item Entanglement of purification: We calculate the redefined EoP for a symmetric configuration including two disjoint strip with equal width $l$ which are separated by the distance $l'$ at zero and finite temperature using the holographic proposal which gives the EoP in terms of entanglement wedge cross-section. \textit{In the all studied regimes the non-conformal effects increase the redefined EoP}. At low temperature the thermal fluctuations decrease the redefined EoP in the the high energy limit while they increase the redefined EoP in the intermediate energy limit. In the intermediate temperature limit, the effects of temperature increase the redefined EoP in the high and intermediate energy limit. Near the phase transition point, in the high energy limit, the redefined EoP between two subsystems is larger at zero temperature than the finite temperature one while, in the intermediate energy limit, the redefined EoP between two subsystems is smaller at zero temperature than the finite temperature one. Therefore, near the transition point,  in the high energy (intermediate) limit the state at zero (finite) temperature is the favorable one.
\end{itemize}
The non-conformal and thermal effects in the studied regimes are summarized in tables \ref{non-conformal} and \ref{thermal}, respectively. 
\begin{table}[ht]
\caption{Non-conformal effect ($\Lambda_c$)}
\begin{scriptsize}
{\sffamily%
\begin{tabular}{*{13}{c|}}
\cline{2-13}
\multirow{4}{*}{} & \multicolumn{4}{c|}{high Energy} & \multicolumn{4}{c|}{intermediate energy} & \multicolumn{4}{c|}{low energy}\\
\cline{2-13}
 & \multicolumn{1}{c|}{$T=0$} & \multicolumn{1}{c|}{low T} & \multicolumn{1}{c|}{int T} & \multicolumn{1}{c|}{high T} & \multicolumn{1}{c|}{$T=0$} & \multicolumn{1}{c|}{low T} & \multicolumn{1}{c|}{int T} & \multicolumn{1}{c|}{high T}  & \multicolumn{1}{c|}{$T=0$} & \multicolumn{1}{c|}{low T} & \multicolumn{1}{c|}{int T} & \multicolumn{1}{c|}{high T} \\
\hline
\multicolumn{1}{|c|}{HEE} & $\downarrow$ & $\downarrow$  & $-$& $\downarrow$  & $-$ & $-$  & $-$ & $-$ & $\times$ & $\times$ & $\times$ & $\times$\\
\hline
\multicolumn{1}{|c|}{HMI} & $\uparrow$ & $\uparrow$  & $\uparrow$ & 0  & $\times$ & $\times$  & $\times$ & 0 & 0 & 0 & 0 & 0\\
\hline
\multicolumn{1}{|c|}{EoP} &  $\uparrow$ & $\uparrow$  & $\uparrow$ & 0  & $\uparrow$ & $\uparrow$  & $\uparrow$ & 0 & 0 & 0 & 0 & 0\\
\hline
\end{tabular}
}%
\end{scriptsize}
\label{non-conformal}
\end{table}
\begin{table}[ht]
\centering
\caption{Thermal effect (T)}
\begin{scriptsize}
{\sffamily%
\begin{tabular}{*{10}{c|}}
\cline{2-10}
\multirow{4}{*}{} & \multicolumn{3}{c|}{Low temperature} & \multicolumn{3}{c|}{Intermediate temperature} & \multicolumn{3}{c|}{High temperature}\\
\cline{2-10}
 & \multicolumn{1}{c|}{high E} & \multicolumn{1}{c|}{int E} & \multicolumn{1}{c|}{low E}  & \multicolumn{1}{c|}{high E} & \multicolumn{1}{c|}{int E} & \multicolumn{1}{c|}{low E}& \multicolumn{1}{c|}{high E} & \multicolumn{1}{c|}{int E} & \multicolumn{1}{c|}{low E} \\
\hline
\multicolumn{1}{|c|}{HEE}  & $\uparrow$  & $-$ & $\times$  & $-$  & $-$ & $-$  & $\uparrow$ & $-$ & $\times$\\
\hline
\multicolumn{1}{|c|}{HMI}  & $\downarrow$  & $\times$ & 0   & $\downarrow$  & $\times$  & 0 & 0 & 0 & 0\\
\hline
\multicolumn{1}{|c|}{EoP}  & $\downarrow$   & $\uparrow$ & 0   & $\uparrow$  & $\uparrow$ & 0 & 0 & 0 & 0\\
\hline
\end{tabular}
}%
\end{scriptsize}
\label{thermal}
\begin{flushleft}
$\uparrow$: Non-conformal and thermal effects increase the mentioned quantity. \\
$\downarrow$: Non-conformal and thermal effects decrease the mentioned quantity.\\
$-$: These regimes does not exist for the mentioned quantity.\\
$\times$ : We can not reach the analytical results in these regimes.\\
$0$ : The mentioned quantity vanishes in these regimes.
\end{flushleft}
\end{table}
\section*{Acknowledgement}
We would like to kindly thank M. Lezgi for useful comments and discussions on related topics.
\appendix
\section{Entanglement entropy}\label{Appendix1}
\begin{itemize}
\item Zero temperature: Here we review the computations of HEE at zero temperature in the high energy limit $l\Lambda _c\ll 1$. By expanding eq.\eqref{Lengthh1} up to the 4th order in $\frac{r_c}{r^*}$ we obtain
\begin{align}
l(r^*)=\frac{2}{r^*}\sum\limits_{n=0}^{\infty}\frac{\Gamma(n+\frac{1}{2})}{\sqrt{\pi}\Gamma(n+1)}\int _0^1 u^{6n+3}\bigg\lbrace &1+ \frac{3}{2}(2n+1)(1-u^2)(\frac{r_c}{r^*})^2\cr
&+\frac{9}{8} (2n+1)^2(1-u^2)^2(\frac{r_c}{r^*})^4\bigg\rbrace du.
\end{align}
Integrating the above equation, we get 
\begin{align}\label{lengthh2}
l(r^*)=\frac{2}{r^*}\left\lbrace a_1+a_2 \left(\frac{r_c}{r^*}\right)^2+a_3 \left(\frac{r_c}{r^*}\right)^4\right\rbrace , \ \ \ \ \ \ \ \ a_1, a_2, a_3>0,
\end{align}
where numerical coefficients $a_1$, $a_2$ and $a_3$ are given by
\begin{align}\label{qeoff1}
a_1&=\sum\limits_{n=0}^{\infty}\frac{\Gamma(n+\frac{1}{2})}{\sqrt{\pi}\Gamma(n+1)(4+6n)}=\frac{\sqrt{\pi}\Gamma(\frac{5}{3})}{4\Gamma(\frac{7}{6})}\simeq 0.431,\cr
a_2&=\sum\limits_{n=0}^{\infty}\frac{\Gamma(n+\frac{1}{2})(1+2n)}{\sqrt{\pi}\Gamma(n+1)(8+20n+12n^2)}=\frac{1}{2}-\frac{\sqrt{\pi}\Gamma(\frac{5}{3})}{8\Gamma(\frac{7}{6})}\simeq 0.284,\cr
a_3&=\sum\limits_{n=0}^{\infty}\frac{3\Gamma(n+\frac{1}{2})(1+2n)^2}{8\sqrt{\pi}\Gamma(n+1)(8+26n+27n^2+9n^3)}=0.285\left[2\Gamma(\frac{5}{3})\Gamma(\frac{11}{6})+\Gamma(\frac{7}{6})\Gamma(\frac{7}{3})\right]\cr
&+\frac{3}{140} \left[-35+{}_{3}F_{2}\left(\frac{3}{2},\frac{5}{3},\frac{7}{3};\frac{8}{3},\frac{10}{3};1\right)\right]\simeq 0.180 .
\end{align}
Solving eq.\eqref{lengthh2} for $r^*$ perturbatively up to 4th order in $l\Lambda_c $, we obtain
\begin{align}\label{turning1}
r^*(l)=\frac{1}{l}\left\lbrace \bar{a}_1+\bar{a}_2(l\Lambda_c )^2+\bar{a}_3(l\Lambda_c )^4\right\rbrace , \ \ \ \ \ \ \ \ \bar{a}_1, \bar{a}_2>0,\  \bar{a}_3<0,
\end{align}
where $\bar{a}_1$, $\bar{a}_2$ and $\bar{a}_3$ are numerical constants given by
\begin{align}\label{qeoff2}
\bar{a}_1&=\frac{\sqrt{\pi}\Gamma(\frac{5}{3})}{2\Gamma(\frac{7}{6})}\simeq 0.862,\cr
\bar{a}_2&=\frac{4\Gamma(\frac{7}{6})^2\left(\frac{1}{2}-\frac{\sqrt{\pi}\Gamma(\frac{5}{3})}{8\Gamma(\frac{7}{6})}\right)}{\pi \Gamma(\frac{5}{3})^2}\simeq 0.382,\cr
\bar{a}_3&=3\left(\frac{2\Gamma(\frac{7}{6})}{\sqrt{\pi}\gamma(\frac{5}{3})}\right)^5\Bigg\lbrace -\left(\frac{1}{2}-\frac{\sqrt{\pi}\Gamma(\frac{5}{3})}{8\Gamma(\frac{7}{6})}\right)^2\cr
& +\frac{\Gamma(\frac{5}{3})\left[189\sqrt{\pi}\left(2\Gamma(\frac{5}{3})\Gamma(\frac{11}{6})+\Gamma(\frac{7}{6})\Gamma(\frac{7}{3})\right)+8\pi\left(-35+{}_{3}F_{2}\left(\frac{3}{2},\frac{5}{3},\frac{7}{3};\frac{8}{3},\frac{10}{3};1\right)\right) \right]}{4480\sqrt{\pi}\Gamma(\frac{7}{6})}\Bigg\rbrace\cr
& \simeq -0.346.
\end{align}
We do the same calculations for the area. Using eq.\eqref{HEE2} and keeping up to $\left(\frac{r_c}{r^*}\right)^4$, we obtain
\begin{align}
\mathcal{A}=2{r^*}^2L^2\sum\limits_{n=0}^{\infty}\frac{\Gamma(n+\frac{1}{2})}{\sqrt{\pi}\Gamma(n+1)}\int _{r^*\epsilon}^1 u^{6n-3}\Bigg\lbrace &1+ \left(3n+(\frac{3}{2}-3n)u^2\right)\left(\frac{r_c}{r^*}\right)^2\cr
&+\frac{1}{2} \left(3n+(\frac{3}{2}-3n)u^2\right)^2\left(\frac{r_c}{r^*}\right)^4\Bigg\rbrace du. \ \ \ \ \ 
\end{align}
By integrating the above equation and using eq.\eqref{HEE} we reach the following expression
\begin{align}
S=\frac{1}{4G_N^{(5)}}\frac{L^2}{\epsilon^2}-\frac{3}{8G_N^{(5)}}\Lambda_c^2 L^2\log (\Lambda_c \epsilon)+S_{finite}
\end{align}
where
\begin{align}\label{HEEhigh}
S_{finite}=\frac{1}{4G_N^{(5)}}L^2{r^*}^2\Big[&K_1+\left(K_2-3\log\left(\frac{r^*}{\sqrt{2}r_c} \right)\right)(\frac{r_c}{r^*})^2\cr
&+K_3(\frac{r_c}{r^*})^4\Big], \ \ \ \ \ \ K_1<0, \ K_2 , K_3>0 ,
\end{align}
where numerical coefficients $K_1$, $K_2$ and $K_3$ are given by
\begin{align}\label{Kqeof}
K_1=&2\sum\limits_{n=0}^{\infty}\frac{\Gamma(n+\frac{1}{2})}{\sqrt{\pi}\Gamma(n+1)}\frac{1}{6n-2}=-\frac{\sqrt{\pi}\Gamma(\frac{2}{3})}{\Gamma(\frac{1}{6})}  \simeq -0.431\cr
K_2=&2\sum\limits_{n=1}^{\infty}\frac{\Gamma(n+\frac{1}{2})}{\sqrt{\pi}\Gamma(n+1)}\frac{1-5n}{4n-12n^2}= {}_{2}F_{1}\left(1,\frac{4}{3};\frac{5}{3};-1\right)+\log(2) \simeq 1.262 \cr
K_3=&2\sum\limits_{n=0}^{\infty}\frac{\Gamma(n+\frac{1}{2})}{\sqrt{\pi}\Gamma(n+1)}\frac{21-87n}{16-144n^2}
=\frac{21}{32\sqrt{\pi}}\bigg[\Gamma(\frac{2}{3})\Gamma(\frac{5}{6})+\Gamma(\frac{1}{6})\Gamma(\frac{4}{3})\cr
&\ \ \ \ \ \ \ \ \ \ \ \ \ \ \ \ \ \ \ \ \ \ \ \ \ \ \ \ \ \ \ \ \  +\frac{29\sqrt{\pi}}{28}{}_{3}F_{2}\left(\frac{2}{3},\frac{4}{3},\frac{3}{2};\frac{5}{3},\frac{7}{3};1\right)\bigg] \simeq 4.166 . \ \ \ \ \ \ 
\end{align}
Replacing $r^*$ from eq.\eqref{turning1} in eq.\eqref{HEEhigh}, we obtain
\begin{align}\label{HEEhigh1}
S_{finite}(l,l\Lambda_c)=\frac{1}{4G_N^{(5)}}\frac{L^2}{l^2}\Big[& \kappa_1 +\left(\kappa_2 +\frac{3}{2}\log(l\Lambda_c)\right) (l \Lambda_c)^2\cr
&+\kappa_3(l\Lambda_c)^4\Big], \ \ \ \kappa _1<0, \ \kappa_2 , \kappa_3>0,
\end{align}
where
\begin{align}\label{Kappa}
\kappa _1&=4a_1^2K_1\simeq -0.321,\cr
\kappa _2&=\frac{a_2K_1}{a_1}+\frac{K_2}{2} -\frac{3}{2}\log(2a_1) \simeq 0.568,\cr
\kappa _3&=\frac{K_3a_1^2-5K_1a_2^2+2K_1a_1a_3-3a_1a_2}{16a_1^4} \simeq 0.93 .
\end{align}
\item Low temperature at high energy: Here,  we review the computations of HEE at low temperature $lT \ll 1$ in the high energy limit $l\Lambda _c\ll 1$. In this limit, we expand \eqref{lengthhT1} up to 4th order in $\frac{r_c}{r^*}$ \begin{align}
l(r^*)&=\frac{2}{r^*}\sum\limits_{n=0}^{\infty}\frac{\Gamma(n+\frac{1}{2})}{\sqrt{\pi}\Gamma(n+1)}\int _0^1 u^{6n+3}\Bigg\lbrace 1+\frac{1}{2}u^4(\frac{r_H}{r^*})^4 \cr
&+\left[\frac{3}{2}(2n+1)(1-u^2)+\frac{3}{4}(2n+1)(1-u^2)u^4(\frac{r_H}{r^*})^4 \right](\frac{r_c}{r^*})^2\cr
&+\left[\frac{9}{8} (2n+1)^2(1-u^2)^2+\frac{9}{6} (2n+1)^2(1-u^2)^2u^4(\frac{r_H}{r^*})^4 \right](\frac{r_c}{r^*})^4\Bigg\rbrace du. \ \ \
\end{align}
Integrating the above equation, we get 
\begin{align}\label{LengthhLT}
l(r^*)=\frac{2}{r^*}\Bigg\lbrace a_1&+b_1\left(\frac{r_H}{r^*}\right)^4+\left[a_2+b_2\left(\frac{r_H}{r^*}\right)^4\right]\left(\frac{r_c}{r^*}\right)^2\cr
&+\left[a_3+b_3\left(\frac{r_H}{r^*}\right)^4\right]\left(\frac{r_c}{r^*}\right)^4\Bigg\rbrace \ \ \ \ \ \ \ b_1,b_2,b_3>0,
\end{align}
where numerical coefficients $b_1$, $b_2$ and $b_3$ are given by
\begin{align}\label{qeoff3}
b_1&=\sum\limits_{n=0}^{\infty}\frac{\Gamma(n+\frac{1}{2})}{\sqrt{\pi}\Gamma(n+1)(16+12n)}=\frac{\sqrt{\pi}\Gamma(\frac{7}{3})}{16\Gamma(\frac{11}{6})}\simeq 0.140,\cr
b_2&=\sum\limits_{n=0}^{\infty}\frac{\Gamma(n+\frac{1}{2})(3+6n)}{\sqrt{\pi}\Gamma(n+1)(160+216n+72n^2)}\cr
&=\frac{3}{2240}\left\lbrace \frac{72}{\sqrt{\pi}}\left(3\Gamma(\frac{13}{6})\Gamma(\frac{7}{3})-2\Gamma(\frac{5}{6})\Gamma(\frac{8}{3})\right)+5 {}_{3}F_{2}\left(\frac{3}{2},\frac{7}{3},\frac{8}{3};\frac{10}{3},\frac{11}{3};1\right)\right\rbrace \simeq 0.081,\cr
b_3&=\sum\limits_{n=0}^{\infty}\frac{3\Gamma(n+\frac{1}{2})(1+2n)^2}{16\sqrt{\pi}\Gamma(n+1)(40+74n+45n^2+9n^3)}\cr
&=\frac{81}{5600\pi}\left[95\Gamma(\frac{16}{6})\Gamma(\frac{7}{3})184\Gamma(\frac{11}{6})\Gamma(\frac{8}{3})\right]\cr
&+\frac{3}{448}\left[168+{}_{3}F_{2}\left(\frac{3}{2},\frac{7}{3},\frac{8}{3};\frac{10}{3},\frac{11}{3};1\right)\right]\simeq 0.054.
\end{align}
Solving eq.\eqref{LengthhLT} for $r^*$ perturbatively up to 4th order in $l\Lambda_c $ and $lT$, we obtain
\begin{align}\label{rstar}
r^*(l)=\frac{1}{l}\left\lbrace \bar{a}_1+\bar{a}_2(l\Lambda_c )^2+\bar{a}_3(l\Lambda_c )^4+\bar{a}_4(lT)^4\right\rbrace , \ \ \ \ \ \ \bar{a}_1,\bar{a}_2,\bar{a}_4>0, \ \bar{a}_3<0
\end{align}
where numerical coefficient $\bar{a}_4$ is given by
\begin{align}\label{bbar}
\bar{a}_4=\frac{2\pi^{5/2}\Gamma(\frac{7}{6})^4\Gamma(\frac{7}{3})}{\Gamma(\frac{5}{3})^4\Gamma(\frac{11}{6})}\simeq 49.392.
\end{align}
Similar to the previous section, we expand eq.\eqref{HEET1} keeping up to 4th order in $\frac{r_c}{r^*}$ and $\frac{r_H}{r^*}$
\begin{align}
\mathcal{A}=&2{r^*}^2L^2\sum\limits_{n=0}^{\infty}\frac{\Gamma(n+\frac{1}{2})}{\sqrt{\pi}\Gamma(n+1)}\int _{r^* \epsilon}^1u^{6n-3}e^{(3n+(\frac{3}{2}-3n)u^2)(\frac{r_c}{r^*})^2} du\cr
&+2{r^*}^2L^2\sum\limits_{n=0}^{\infty}\frac{\Gamma(n+\frac{1}{2})}{2\sqrt{\pi}\Gamma(n+1)}(\frac{r_H}{r^*})^4\int _{r^* \epsilon}^1u^{6n+1}e^{(3n+(\frac{3}{2}-3n)u^2)(\frac{r_c}{r^*})^2} du.
\end{align}
Taking the integrals and using eq.\eqref{HEE} we reach the following expression for HEE in the mentioned limit
\begin{align}
S=\frac{1}{4G_N^{(5)}}\frac{L^2}{\epsilon^2}-\frac{3}{8G_N^{(5)}}\Lambda_c^2 L^2\log (\Lambda_c \epsilon)+S_{finite}
\end{align}
where
\begin{align}\label{HEET2}
S_{finite}=\frac{1}{4G_N^{(5)}}{r^*}^2L^2\bigg\lbrace & K_1+\bar{K}_1(\frac{r_H}{r^*})^4+\left[K_2-3\log\left(\frac{r^*}{\sqrt{2}r_c} \right)+\bar{K}_2(\frac{r_H}{r^*})^4\right](\frac{r_c}{r^*})^2\cr
&+\left[K_3+\bar{K}_3(\frac{r_H}{r^*})^4\right](\frac{r_c}{r^*})^4\bigg\rbrace , \ \ \ \ \ \ \ \bar{K}_1,\bar{K}_2,\bar{K}_3>0, \ \ \ 
\end{align}
where $\bar{K}_1$, $\bar{K}_2$ and $\bar{K}_3$ are numerical constants given by
\begin{align}\label{Kbar}
\bar{K}_1=&2\sum\limits_{n=0}^{\infty}\frac{\Gamma(n+\frac{1}{2})}{\sqrt{\pi}\Gamma(n+1)}\frac{1}{6n+2}=\frac{\sqrt{\pi}\Gamma(\frac{4}{3})}{\Gamma(\frac{5}{6})}\simeq 0.701\cr
\bar{K}_2=&2\sum\limits_{n=0}^{\infty}\frac{\Gamma(n+\frac{1}{2})}{\sqrt{\pi}\Gamma(n+1)}\frac{3+15n}{8+36n+36n^2}\cr
=&\frac{3}{16}\left[\frac{12\left(2 \Gamma(\frac{7}{6})\Gamma(\frac{4}{3})-\Gamma(\frac{5}{6})\Gamma(\frac{5}{3})\right)}{\sqrt{\pi}}+{}_{3}F_{2}\left(\frac{4}{3},\frac{3}{2},\frac{5}{3};\frac{7}{3},\frac{8}{3};1\right)\right]\simeq 0.808 \cr
\bar{K}_3=&2\sum\limits_{n=0}^{\infty}\frac{\Gamma(n+\frac{1}{2})}{\sqrt{\pi}\Gamma(n+1)}\frac{3(2+13n+29n^2)}{16(2+11n+18n^2+9n^3)}\cr
&=\frac{27}{4}+\frac{3}{8\sqrt{\pi}}\left(11\Gamma(\frac{1}{3})\Gamma(\frac{7}{6}-38\Gamma(\frac{2}{3})\Gamma(\frac{5}{6})\right)\cr
&+\frac{87}{320}{}_{3}F_{2}\left(\frac{4}{3},\frac{3}{2},\frac{5}{3};\frac{7}{3},\frac{8}{3};1\right)\simeq 0.707.
\end{align}
Replacing $r^*$ from eq.\eqref{rstar} in eq.\eqref{HEET2} and keeping up to $(l\Lambda_c )^4$ and $(lT)^4$, we obtain
\begin{align}\label{HEElow}
S_{finite}=\frac{L^2}{l^2}\Bigg\lbrace &\kappa_1 +\bar{\kappa}_1(lT)^4+\left[\kappa_2+\frac{3}{2}\log(l\Lambda_c) +\bar{\kappa}_2(lT)^4\right] (l \Lambda_c)^2\cr
&+\left[\kappa_3 +\bar{\kappa}_3(lT)^4\right] (l\Lambda_c)^4\Bigg\rbrace , \ \ \ \bar{\kappa}_1>0, \bar{\kappa}_2,\bar{\kappa}_3<0,
\end{align}
where $\bar{\kappa}_1$, $\bar{\kappa}_2$ and $\bar{\kappa}_3$ is given by
\begin{align}\label{kappabar}
\bar{\kappa}_1&=\frac{\pi^4(2K_1 b_1+\bar{K}_1a_1)}{4a_1^3} \simeq 55.0983,\cr
\bar{\kappa}_2&=\frac{\pi^4(2K_1 a_2 b_1-2\bar{K}_1a_1a_2+\bar{K}_2a_1^2)}{32a_1^6}\simeq -26.5767,\cr
\bar{\kappa}_3&=\frac{\pi^4(2K_1b_1a_1a_3 -6K_1 a_2^2 b_1+9\bar{K}_1a_1a_2^2-2\bar{K}_1a_1^2 a_3 -4\bar{K}_2a_1^2a_2-2K_3a_1^2b_1+\bar{K}_3a_1^3)}{256a_1^9}\cr
&\simeq -159.136.
\end{align}
\item High temperature at high energy: In the subsection \ref{EE-highT}, we performed our calculations in details in the regime of high temperature and high energy. Here, we bring the numerical constants which obtained in the mentioned subsection
\begin{align}\label{F1}
F_1&=-\frac{3 \pi^\frac{5}{2} \Gamma \left(\frac{2}{3}\right)}{ \Gamma \left(\frac{1}{6}\right)}+2\pi^2\left[-\frac{3\sqrt{\pi}\Gamma(\frac{2}{3})}{2\Gamma(\frac{1}{6})}+\sum\limits_{n=1}^{\infty}\frac{3}{(4n-2)(4n+1)}\frac{\Gamma(n+\frac{1}{2})\Gamma(\frac{2n+2}{3})}{\Gamma(n+1)\Gamma(\frac{4n+1}{6})}\right]\cr
&=\frac{\pi^\frac{3}{2}}{60}\Bigg[5 \sqrt{\pi } \, _5F_4\left(\frac{1}{2},\frac{5}{6},1,\frac{7}{6},\frac{3}{2};\frac{5}{4},\frac{4}{3},\frac{5}{3},\frac{7}{4};1\right)\cr
&+3 \Gamma \left(\frac{1}{3}\right) \Gamma \left(\frac{1}{6}\right) \, _4F_3\left(\frac{1}{6},\frac{1}{2},\frac{5}{6},\frac{7}{6};\frac{11}{12},\frac{4}{3},\frac{17}{12};1\right)\cr
&-90 \Gamma \left(\frac{2}{3}\right) \Gamma \left(\frac{5}{6}\right) \left(\, _4F_3\left(-\frac{1}{6},\frac{1}{6},\frac{1}{2},\frac{5}{6};\frac{7}{12},\frac{2}{3},\frac{13}{12};1\right)-1\right)\Bigg] \simeq -6.572.
\end{align}
\begin{align}\label{B0}
B_0&=\frac{510 \pi  \Gamma \left(\frac{2}{3}\right)-495 \pi  \Gamma \left(\frac{5}{3}\right)-162 \pi  \Gamma \left(\frac{8}{3}\right)+20 \sqrt{\pi } \Gamma \left(\frac{1}{6}\right)+40 \sqrt{\pi } \log (4) \Gamma \left(\frac{1}{6}\right)}{160 \sqrt{\pi } \Gamma \left(\frac{1}{6}\right)}\cr 
&-\frac{1}{2} \simeq -0.0284.
\end{align}
\begin{align}\label{F2}
F_2&=B_0-\frac{3}{2}\log(2a_1)+\sum\limits_{m=1}^{\infty}\frac{3\Gamma(m+\frac{1}{2})}{8\sqrt{\pi}\Gamma(m+1)} \cr
&\times\Bigg[\frac{1}{m}\left(\, _4F_3\left(\frac{1}{2},\frac{2 m}{5}+\frac{4}{5},\frac{2 m}{3}-\frac{1}{3},\frac{2 m}{3};\frac{2 m}{5}-\frac{1}{5},\frac{2 m}{3}+\frac{2}{3},\frac{2 m}{3}+1;1\right)\right)\cr
&-\frac{1}{(m+1) (2 m+3)}\left(\, _3F_2\left(\frac{3}{2},\frac{2 m}{3}+\frac{2}{3},\frac{2 m}{3}+1;\frac{2 m}{3}+\frac{5}{3},\frac{2 m}{3}+2;1\right)\right)\Bigg]\cr
&\approx 2.5157.\ \ \ \ \
\end{align}

\end{itemize}
\section{Mutual information}\label{Appendix2}
In this appendix we present the result obtained for the HMI before our redefinitions.
\begin{itemize}
\item Zero temperature: 
\begin{align}\label{I-highE}
 I(l,l',l\Lambda_c,l'\Lambda_c)=&\frac{L^2}{4G_N^{(5)}}\Bigg\lbrace \kappa _1\left[\frac{2}{l^2}-\frac{1}{l'^2}-\frac{1}{(l'+2l)^2}\right]\cr
 &+\frac{3(l\Lambda_c)^2}{2l^2}\log\left(\frac{(l\Lambda_c)^2}{(l'\Lambda_c)(l'\Lambda_c +2l\Lambda_c)} \right)-2\kappa _3\frac{(l'\Lambda _c+ l\Lambda _c)^4}{(l'+l)^2}\Bigg\rbrace. \ \ \
\end{align}
\item Low temperature:
\begin{align}\label{I-lowT}
 I(l,l',l\Lambda_c,l'\Lambda_c ,lT,l'T)
=&\frac{L^2}{4G_N^{(5)}}\Bigg\lbrace \kappa _1\left[\frac{2}{l^2}-\frac{1}{l'^2}-\frac{1}{(l'+2l)^2}\right]\cr
&+\frac{3(l\Lambda_c)^2}{2l^2}\log\left(\frac{(l\Lambda_c)^2}{(l'\Lambda_c)(l'\Lambda_c +2l\Lambda_c)} \right)\cr
&-2\bar{\kappa}_1\frac{(l'T+ lT)^4}{(l'+l)^2}-2\kappa _3\frac{(l'\Lambda _c+ l\Lambda _c)^4}{(l'+l)^2}\Bigg\rbrace .
\end{align}
\item High temperature:
\begin{align}\label{I-intT}
I(l',l'\Lambda_c ,l'T)=\frac{1}{4G_N^{(5)}}\frac{L^2}{l'^2}\Bigg\lbrace & -\kappa_1+F_1(l'T)^2 -\pi^3(l'T)^3-\bar{\kappa}_1 (l'T)^4\cr
&+\left[F_3-\frac{3}{2}\log(l'\Lambda_c)-\bar{\kappa}_2(l'T)^4\right](l'\Lambda_c )^2 \cr
&-\left[\kappa_3+\bar{\kappa}_3(l'T)^4\right] (l'\Lambda_c )^4\Bigg\rbrace , \ \ F_3>0, \ \ 
\end{align}
where $F_3$ is a numerical constant given by
\begin{align}\label{F3}
F_3=F_2-\kappa_2 \approx 1.9477
\end{align}
\end{itemize}
\section{Entanglement of purification}\label{Appendix3}
In this appendix we present the result obtained for the EoP before our redefinitions. We also review the computations of EoP in subsections \ref{lowTintE} and \ref{intThighE}.
\begin{itemize}
\item High energy at zero temperature:
\begin{align}\label{Ep-highE}
E_p(l,l',l\Lambda_c,l'\Lambda_c)=\frac{L^2}{4G_N^{(5)}}\Bigg\lbrace &2 a_1^2\left(\frac{1}{\left( l'\right)^2}-\frac{1}{\left(l'+2l \right)^2}\right)+ C_1 \frac{(l\Lambda_c)^2(l\Lambda_c+l'\Lambda_c)^2}{l(l+l')} \cr
&+\frac{3(l\Lambda_c)^2}{4l^2}\log\left(\frac{l'+2l}{l'}\right)\Bigg\rbrace, \ \ \   C_1>0.
\end{align}
where $C_1$ is a numerical coefficient given by
\begin{align}\label{C1}
C_1=\frac{9a_1^2+40a_2^2-16a_1a_3-24a_1a_2}{64a_1^4} \simeq 0.326.
\end{align}
\item Intermediate energy at zero temperature:
\begin{align}\label{Ep-intE}
E_p(l',l'\Lambda_c)=\frac{1}{4G_N^{(5)}}\frac{L^2}{l'^2}\bigg\lbrace 2a_1^2+\Big(C_2&-\frac{3}{4}\log(l'\Lambda_c )\Big)(l'\Lambda_c)^2\cr
&-\frac{C_1}{4}(l'\Lambda_c )^4\bigg\rbrace , \ \ \ \ \ \  C_1, C_2>0,
\end{align}
where $C_2$ is a numerical constant given by
\begin{align}\label{C2}
C_2\equiv\frac{3}{8}\left[1-\gamma +\log\left(\frac{16a_1^2}{3}\right)-\frac{2}{3}e^\frac{3}{2}+Ei\left(\frac{3}{2}\right)+\frac{4a_2}{3a_1}\right]\simeq 0.6027,
\end{align}
where $\gamma$ is the Euler-Mascheroni constant.
\item low temperature at high energy: 
\begin{align}\label{Ep-lowThighE}
E_p(l,l',l\Lambda_c,l'\Lambda_c ,lT,l'T)
=\frac{L^2}{4G_N^{5)}}\Bigg\lbrace &2 a_1^2\left(\frac{1}{l'^2}-\frac{1}{(l'+2l)^2}\right)+C_3\frac{(lT)^2(lT+l'T)^2}{l(l+l')}\cr
&+\frac{3(l\Lambda_c)^2}{4l^2}\log\left(\frac{l'+2l}{l'}\right)+C_1 \frac{(l\Lambda_c)^2(l\Lambda_c+l'\Lambda_c)^2}{l(l+l')}\Bigg\rbrace , \cr
&\ \ \ C_1>0, C_3<0,
\end{align}
where numerical constant $C_3$ is given by 
\begin{align}\label{C3}
C_3&=\frac{\pi^4(a_1-4b_1)}{4a_1^3}\simeq -39.396.
\end{align}
\item Low temperature at intermediate energy:
 We take the Integral in \eqref{EoPT1} and obtain the following expression for the EoP
\begin{align}\label{EoPT3}
E_p&=\frac{L^2}{4G_N^{(5)}}\Bigg\lbrace \frac{1}{2}\left({r^*_{l'}}^2-{r^*_{l'+2l}}^2\right)+\frac{3r_c^2}{4}\log\left(\frac{r^*_{l'}}{r^*_{l'+2l}}\right)\cr
&+r_c^2\sum\limits_{m=2}^\infty \frac{\left(\frac{3}{2}\right)^mr_c^{2m}}{\Gamma(m+1)(2-2m)}\left[\left(\frac{r_c}{r^*_{l'}}\right)^{2m-2}-\left(\frac{r_c}{r^*_{l'+2l}}\right)^{2m-2}\right]\cr
&+\sum\limits_{n=1}^\infty\sum\limits_{m=0}^\infty\frac{\Gamma(n+\frac{1}{2})\left(\frac{3}{2}\right)^mr_c^{2m}r_H^{4n}}{\sqrt{\pi}\Gamma(m+1)\Gamma(n+1)(2-4n-2m)}{r^*_{l'}}^{2-4n-2m}\cr
&+\frac{3}{4}\sum\limits_{n=1}^\infty\frac{\left(\frac{2}{3}\right)^{2n}r_H^2}{\sqrt{\pi}\Gamma(n+1)}\left(\frac{r_H}{r^*_{l'+2l}}\right)^{4n-2}\left(\frac{r_c}{r^*_{l'+2l}}\right)^{1-2n}\Gamma\left(2n-1,0,-\frac{3}{2}\left(\frac{r_c}{r^*_{l'+2l}}\right)^2\right)\Bigg\rbrace , \cr
\end{align}
We should check carefully the convergence of the infinite series in the above equation. Obviously the series including $\frac{r_c}{r^*_{l'}}$ is convergent because in this limit we have $r_c\ll r^*_{l'}$. For large $n$, the infinite series in the second line in the above equation which contain $\frac{r_c}{r^*_{l'+2l}}$ behaves as $\left(\frac{3e}{2m}\right)^{m+\frac{3}{2}}\left(\frac{r_c}{r^*_{l'+2l}}\right)^{2m-2}$ and hence converges in the $r^*_{l'+2l}\rightarrow r_c$ limit. We numerically check the convergence of the infinite series in the fourth line  in eq.\eqref{EoPT3}. For large $n$ in the limit of $r^*_{l'+2l}\rightarrow r_c$ this series behaves as $n^{-\frac{3}{2}}$ and hence this series are convergent. Therefore, we can safely take limit $r^*_{l'+2l}\rightarrow r_c$. For the other part of the intermediate energy limit i.e $r_H\ll r_c\ll r^*_{l'}$, we expand \eqref{EoPT3} up to second order in  $\frac{r_c}{r^*_{l'}}$, $\frac{r_H}{r^*_{l'}}$ and $\frac{r_H}{r_c}$. Using \eqref{rstar} for $r^*_{l'}$, we obtain
\begin{align}\label{Ep-lowTintE}
E_p(l',l'\Lambda_c,l'T)&=\frac{1}{4G_N^{(5)}}\frac{L^2}{l'^2}\Bigg\lbrace  2a_1^2-\frac{C_3}{4}(l'T)^4\cr
&+\Bigg[C_2-\frac{3}{4}\log(l'\Lambda_c)+C_6\left(\frac{l'T}{l'\Lambda_c}\right)^4-C_4(l'T)^4
\Bigg](l' \Lambda_c)^2\cr
&-\left[\frac{C_1}{4}+C_5(l'T)^4\right](l'\Lambda _c)^4\Bigg\rbrace , \ \ C_1,C_2,C_4>0 , C_3,C_5,C_6<0, \ \ \ 
\end{align}
where numerical constants  $C_4$, $C_5$ and $C_6$ are given by
\begin{align}\label{C4}
C_4&=\frac{\pi^4(-3a_1^2+16a_2b_1+8a_1a_2)}{2^{9}a_1^6}\simeq 31.420,\cr
C_5&=\frac{\pi^4\left(-96a_2^2b_1+32a_1a_3b_1-144a_1^2b_1-96a_2^2a_1^3+48a_2^2a_1-16a_1^2a_3+24a_1^2a_1-3a_1^3\right)}{2^{13}a_1^9}\cr
&\simeq -68.089,\cr
C_6&=\frac{\left(1-e^\frac{3}{2}\right)\pi^4}{3}\simeq -113.049.
\end{align}
\item Intermediate temperature at high energy:
One can take the integral in eq.\eqref{EoPT1} and obtain the following expression for the EoP
\begin{align}\label{EoPT2}
E_p&=\frac{L^2}{4G_N^{(5)}}\Bigg\lbrace\frac{r_H^2}{\sqrt{\pi}}\sum\limits_{n=0}^{\infty}\frac{\Gamma(n+\frac{1}{2})}{\Gamma(n+1)(2-4n)}\left[\left(\frac{r_H}{r^*_{l'}}\right)^{4n-2}-\left(\frac{r_H}{r^*_{l'+2l}}\right)^{4n-2}\right]+\frac{3r_c^2}{2}\log\left(\frac{r^*_{l'}}{r^*_{l'+2l}}\right)\cr
&-\frac{3r_c^2}{8\sqrt{\pi}}\sum\limits_{n=1}^{\infty}\frac{\Gamma(n+\frac{1}{2})}{\Gamma(n+1)n}\left[\left(\frac{r_H}{r^*_{l'}}\right)^{4n}-\left(\frac{r_H}{r^*_{l'+2l}}\right)^{4n}\right]\cr
&+\sum\limits_{m=2}^{\infty}\sum\limits_{n=0}^{\infty}\frac{(\frac{3}{2})^m\Gamma(n+\frac{1}{2})r_c^{2m}r_H^{4n}}{\sqrt{\pi}\Gamma(n+1)\Gamma(m+1)(2-4n-2m)}{r^*_{l'}}^{2-4n-2m}\cr
&+\frac{1}{2}\sum\limits_{m=2}^{\infty}\frac{(\frac{3}{2})^mr_c^2}{\Gamma(m+1)(m-1)}\left(\frac{r_c}{r^*_{l'+2l}}\right)^{2m-2}{}_{2}F_{1}\left(\frac{1}{2},\frac{1}{2}(m-1);\frac{m+1}{2};\left(\frac{r_H}{r^*_{l'+2l}}\right)^4\right)\Bigg\rbrace
\end{align}
 we should check the convergence of the series in the above equation in the intermediate temperature limit. Obviously the series including $\frac{r_H}{r^*_{l'}}$ is convergent because in this limit we have $r_H\ll r^*_{l'}$. For large $n$, the infinite series in the first and second line in the above equation which contain $\frac{r_H}{r^*_{l'+2l}}$ beahves as $\frac{1}{n^{\frac{3}{2}}}\left(\frac{r_H}{r^*_{l'+2l}}\right)^{4n-2}$ and $\frac{1}{n^{\frac{3}{2}}}\left(\frac{r_H}{r^*_{l'+2l}}\right)^{4n}$, respectively. In the limit of $r^*_{l'+2l}\rightarrow r_H$ both of these series behaves as $n^{-\frac{3}{2}}$ and hence these series are convergent. We numerically check that for large $m$ the infinite series in the fourth line in eq.\eqref{EoPT2} in the limit of $r^*_{l'+2l}\rightarrow r_H$ goes as $\left(\frac{3e}{2m}\right)^{m+1}\left(\frac{r_c}{r^*_{l'+2l}}\right)^{2m-2}$ which $\frac{r_c}{r^*_{l'+2l}} \ll 1$  and hence the infinite series is convergent in the mentioned limit. Therefore, we can safely take $r^*_{l'+2l}\rightarrow r_H$ limit. For another part of the intermediate temperature regime i.e. $r_c\ll r_H\ll r^*_{l'}$ we expand eq.\eqref{EoPT2} up to 4th order in  $\frac{r_c}{r^*_{l'}}$, $\frac{r_H}{r^*_{l'}}$ and $\frac{r_c}{r_H}$ and using eq.\eqref{rstar} for $r^*_{l'}$ we reach the following expression
 \begin{align}\label{Ep-intThighE}
E_p(l',l'\Lambda_c,l'T)&=\frac{1}{4G_N^{(5)}}\frac{L^2}{l'^2}\Bigg\lbrace  2a_1^2-\frac{C_3}{4}(l'T)^4\cr
&+\left[C_7-\frac{3}{4}\log(l'T)+\frac{9}{128\pi}\left(\frac{l'\Lambda_c}{l'T}\right)^2-C_4(l'T)^4
\right](l' \Lambda_c)^2\cr
&-\left[\frac{C_1}{4}+C_5(l'T)^4\right](l'\Lambda _c)^4\Bigg\rbrace, \ \ \ \ C_1,C_4>0, C_3,C_5,C_7<0, \ \ \ \ \ \
\end{align}
where $C_7$ is a numerical constant given by
\begin{align}\label{C7}
C_7&=\frac{a_2}{2a_1}+\frac{3}{8}\log\left(\frac{8a_1^2}{\pi^2}\right)\simeq -0.380.
\end{align}
\item Intermediate temperature at intermediate energy: One can take the integral in eq.\eqref{EoPT1} and obtain eq.\eqref{EoPT2}. We checked the convergence of the series in the first and second line in eq.\eqref{EoPT2} in the limit $r^*_{l'+2l}\to r_H$. We numerically check that for large $m$ the series in the fourth line in eq.\eqref{EoPT2} in the limits $r^*_{l'+2l}\to r_c$ and $r^*_{l'+2l}\to r_H$ behaves as $\left(\frac{3e}{2m}\right)^{m+1}$. Hence, the series are convergent in this regime and we can safely take these limits.  For the limits of $ r_H\ll r^*_{l'}$ and $ r_c\ll r^*_{l'}$, we expand eq.\eqref{EoPT2} up to second order in  $\frac{r_c}{r^*_{l'}}$ and $\frac{r_H}{r^*_{l'}}$ and using eq.\eqref{rstar} for $r^*_{l'}$. Taking the transition limit, we finally obtain
\begin{align}\label{Ep-intTintE}
E_p(l',l'\Lambda_c,l'T)\bigg\vert_{T\rightarrow \frac{\Lambda_c}{\sqrt{2}\pi}}
&=\frac{1}{4G_N^{(5)}}\frac{\sqrt{2}\pi L^2}{l'^2}\bigg\lbrace 2a_1^2+\left(C_8-\frac{3}{4}\log(l'\Lambda_c)\right)(l'\Lambda_c)^2 \cr
&\ \ \ \ \ \ \ \ \ \ \ \ \ \ \ +C_9(l'\Lambda_c)^4\bigg\rbrace , \ \ \ \ \ \ \  C_8 >0,\ C_9<0,
\end{align}
where numerical coefficients $C_8$ and $C_9$ are given by
\begin{align}\label{C5}
&C_8=\frac{1}{2}\left[\frac{a_2}{a_1}+\frac{3}{2}\log\left(\frac{4a_1}{\pi^2}\right)+1.2745\right]\simeq 1.3761,\cr
&C_9=\frac{16a_1(b_1+a_3)-40a_2^2-13a_1^2}{256a_1^4}\simeq -0.3889.
\end{align}
\end{itemize}

\end{document}